\documentclass[10pt,journal,compsoc]{IEEEtran}

\usepackage{url}       
\usepackage{amsmath}
\usepackage{amssymb}
\usepackage{graphicx}
\usepackage{amsfonts}
\usepackage{bbm}
\usepackage{algorithm}
\usepackage{algorithmic}
\usepackage{parskip}
\usepackage[labelfont=bf,textfont=it]{caption}
\usepackage[marginclue,footnote,silent,draft]{fixme}
\usepackage{xcolor}

\floatname{algorithm}{Algorithm}

\begin{document}
\title{Meshless Approximation and Helmholtz-Hodge Decomposition of Vector Fields}
\author{Giuseppe Patan\'e 
\IEEEcompsocitemizethanks{\IEEEcompsocthanksitem G. Patan\'e is with CNR-IMATI, Consiglio Nazionale delle Ricerche, Istituto di Matematica Applicata e Tecnologie Informatiche
Genova, Italy.\protect\\ E-mail: patane@ge.imati.cnr.it}}
\IEEEtitleabstractindextext{
\begin{abstract}
The analysis of vector fields is crucial for the understanding of several physical phenomena, such as natural events (e.g., analysis of waves), diffusive processes, electric and electromagnetic fields. While previous work has been focused mainly on the analysis of 2D or 3D vector fields on volumes or surfaces, we address the meshless analysis of a vector field defined on an arbitrary domain, without assumptions on its dimension and discretisation. The meshless approximation of the Helmholtz-Hodge decomposition of a vector field is achieved by expressing the potential of its components as a linear combination of radial basis functions and by computing the corresponding conservative, irrotational, and harmonic components as solution to a least-squares or to a differential problem. To this end, we identify the conditions on the kernel of the radial basis functions that guarantee the existence of their derivatives. Finally, we demonstrate our approach on 2D and 3D vector fields measured by sensors or generated through simulation.
\end{abstract}
%
%
}
\maketitle
\IEEEdisplaynontitleabstractindextext
\IEEEpeerreviewmaketitle
\IEEEraisesectionheading{\section{Introduction}}
Vector fields are commonly used to model several phenomena, such as natural events, diffusive processes, electric and electromagnetic fields; for instance, vector fields represent the velocity and direction of an object or the magnitude and direction of a force. Measuring and simulating a flow field typically generate a large amount of unstructured, sparse, and noisy vector data, which are a challenging input for the reconstruction of a global representation. Indeed, it is generally complex to convey 3D directional information through stream lines that visualise the flow patterns. Furthermore, the components of the reconstructed vector field must satisfy additional properties such as being conservative (e.g., incompressible fluid flow), irrotational (e.g., magnetic field), harmonic, and invariant with respect to a set of transformations.


In these examples, we typically have heterogeneous data (e.g., scalar values, vectors) in terms of data structures (e.g., vector fields, scalar functions), spatial distribution or resolution (e.g., regular grids, meshes, sparse samples), and data values of any dimension. Then, recovering a common tessellation is generally difficult and time-consuming. As a result of the heterogeneity of the input data, previous work has been focused mainly (i) on the analysis of 2D or 3D vector fields on volumes or surfaces, depending on the discretisation of the input domain and of the differential operators, and (ii) on the computation of the potential of the conservative component of an arbitrary field, without constraints on the values of its potential. In particular, the Helmholtz-Hodge Decomposition (HHD, for short) splits a vector field into its conservative, irrotational, and harmonic component fields, which provide a concise representation of the underlying flow through sources, sinks, and vortices. Then, the HHD is applied to simplify or edit the structure of the input vector field in a coherent and admissible manner, to analyse and visualise the behaviour and properties of the underlying phenomenon. 

In this context, we present a unified approach to (i) the \emph{meshless approximation of heterogeneous data} (Sects.~\ref{sec:THEO-BACKGROUND},~\ref{sec:MESHLESS-APPROXIMATION-VF}), such as scalar values and vectors, measured at sparsely sampled points (e.g., without a regular structure or an underlying grid), and (ii) the \emph{meshless HHD of arbitrary vector fields}, e.g., generated by particle-based fluid simulation or experimental measurements (e.g., wind fields). 

Given a set of function values and vectors sampled at a set of sparse points, we compute a meshless and smooth approximation of the input vector field. Then (Sect.~\ref{sec:MESHLESS-HH}), we introduce three variants of the meshless HHD with Radial Basis Functions (RBFs, for short), which are based on a direct, a least-squares, and a Laplace-based approach, respectively. The meshless approach allows us to (i) exactly compute differential operators by evaluating the 1D derivatives of the kernel generating the RBFs, (ii) impose  different types of interpolating, least-squares, or smoothness constraints, and (iii) approximate data values of any dimension and structure (e.g., scalar values, vectors). In particular, the smoothness order is determined by the regularity of the generating kernel, e.g., a poly-harmonic kernel for~$\mathcal{C}^{k}$ regularity, \mbox{$k\geq 2$}.

With respect to~\cite{PETRONETTO2010}, we provide a \emph{continuous formulation of the HHD}, which is based on the differentiability of the meshless approximation and is independent of the discretisation of differential operators (e.g., gradient, divergence, Laplace-Beltrami operator) involved in the Poisson-based formulation of the HHD. In this way, we improve the approximation accuracy and the resulting performances have the same order of complexity of previous work, in terms of computational cost and storage overhead. While the compactness of the input domain is a standard hypothesis of previous work~\cite{BHATIA2014,BHATIA2014-HODGE} to guarantee the unicity of the HHD, we neither assume that the input domain is compact nor restrict our approximation and decomposition to a compact domain. 

While the meshless HHD~\cite{PETRONETTO2010} applies to 2D SPH (\emph{Smoothed Particle Hydrodynamics}) flows and the mesh-based decomposition~\cite{POLTHIER2000} holds for vector fields on surfaces discretised as triangle meshes, our meshless HHD applies to~$n$D vector fields, is based entirely on a continuous approach, and is independent of the discretisation of partial derivatives with finite differences. As a theoretical analysis of the properties of the resulting HHD, we determine the conditions on the existence of the meshless potential of the conservative and solenoidal components of the decomposition. Then, we discuss the \emph{numerical accuracy and stability of the meshless decomposition}, in terms of the selected centres and of the generating kernel, and its \emph{computational aspects}, in terms of cost, memory requirements, and reduction of the memory footprint for the representation of discrete data (Sect.~\ref{sec:DISCUSSION-GRAD}).

Since the meshless HHD involves the evaluation of the gradient and of the  Laplace operator, we identify the \emph{hypothesis on the generating kernel for the existence of the first and second order derivatives of the corresponding RBFs}. The meshless potentials of the conservative and irrotational components are uniquely defined in terms of the RBFs. In fact, the coefficients of their representations solve the corresponding least-squares systems. Since our meshless HHD is based on the evaluation of differential operators, the residual divergence and the residual rotor are null. On the contrary, previous work~\cite{PETRONETTO2010,RIBEIRO2016} involves a residual divergence and rotor, as differential operators are discretised with finite differences.

Through the meshless approximation, the involved differential operators (i.e., gradient, rotor, Laplace-Beltrami) are computed in linear time by applying their continuous definition, and are not tailored to a specific  (e.g., vertex-, or edge-, or face-based) discretisation~\cite{TONG2003} of the operator itself or of the input domain. The meshless approach guarantees the \emph{consistency of the approximation and decomposition} of the input vector field. It also provides a \emph{compact representation} in terms of memory footprint, by encoding the potentials as the set of the corresponding centres of the RBFs and coefficients. To summarise, our main contributions are
\begin{itemize}
\item a unified approach to the meshless approximation and HHD of vector fields with null residual divergence and rotor;
\item a meshless approach, based on the differentiability of RBFs and independent of discrete derivatives;
\item the identification of the hypothesis on the generating kernel for the existence of RBFs' derivatives;
\item  a generalisation of previous work on the meshless HHD of 2D vector fields~\cite{PETRONETTO2010,RIBEIRO2016} to~$n$D vector fields.
\end{itemize} 
We demonstrate the main features of the proposed approach on 2D/3D vector fields acquired by sensors and generated through simulation; for the evaluation of the approximation accuracy and numerical stability of the meshless decomposition, we consider analytic vector fields. Finally (Sect.~\ref{sec:FUTURE-WORK}), we outline future work on the classification of critical points and centres' selection.

\section{Previous work\label{sec:THEO-BACKGROUND}}
Let us assume that the vector field \mbox{$\mathbf{v}:\Omega\rightarrow\mathbb{R}^{d}$}, defined on a compact domain~$\Omega$ of~$\mathbb{R}^{d}$, assigns a vector to the points \mbox{$\mathcal{P}:=\{\mathbf{p}_{i}\}_{i=1}^{n}$} or cells (e.g., triangles, tetrahedra) of~$\Omega$. We present previous work on the HHD (Sect.~\ref{sec:HH-DECOMPOSITION}), the meshless approximation (Sect.~\ref{sec:MESHLESS-APPROX}), the analysis of potentials and vector fields (Sect.~\ref{sec:VECTOR-ANALYSIS}).

\subsection{Helmolthz-Hodge decomposition\label{sec:HH-DECOMPOSITION}}
A vector field \mbox{$\mathbf{v}:\Omega\rightarrow\mathbb{R}^{d}$} is \emph{conservative} if there exists a \emph{potential function} \mbox{$u:\Omega\rightarrow\mathbb{R}$} such that \mbox{$\mathbf{v}=\nabla u$}. If~$\mathbf{v}$ is conservative, then it is irrotational (i.e., \mbox{$\nabla\wedge\mathbf{v}=\mathbf{0}$}) and the vice versa holds if the input domain is simply connected. A vector field~$\mathbf{v}$ is \emph{solenoidal} if there exists a field~$\mathbf{w}$ such that \mbox{$\mathbf{v}=\nabla\wedge\mathbf{w}$}, or equivalently if its divergence is null (i.e., \mbox{$\nabla\cdot\mathbf{v}=0$}). If~$\mathbf{v}$ is not conservative, then we consider its HHD \mbox{$\mathbf{v}=\nabla u+\nabla\wedge\mathbf{w}+\mathbf{h}$} in terms of a conservative \mbox{$\nabla u$}, a solenoidal \mbox{$\nabla\wedge\mathbf{w}$}, and a harmonic~$\mathbf{h}$ component. To guarantee the uniqueness of the decomposition, we impose that the conservative and solenoidal components are orthogonal and tangential to the boundary of the input domain, respectively. The potentials of the conservative and solenoidal components solve the Poisson equations  \mbox{$\Delta u=\nabla\cdot\mathbf{v}$} and \mbox{$\overline{\Delta}\mathbf{w}=\nabla\wedge\mathbf{v}$}, where~$\Delta$ and \mbox{$\overline{\Delta}:=(\nabla\nabla\cdot)-(\nabla\wedge\nabla\wedge)$} are the standard Laplace-Beltrami and vector Laplace-Beltrami operator, respectively. For 3D, the solution \mbox{$u(\mathbf{p})=\int_{\Omega}G(\mathbf{p},\mathbf{q})f(\mathbf{q})\textrm{d}\mathbf{q}$} to the Poisson equation \mbox{$\Delta u=f$} is computed by convolving the function~$f$ with the \emph{free-space Green's function} \mbox{$G(\mathbf{p},\mathbf{q}):=-(4\pi\|\mathbf{p}-\mathbf{q}\|_{2})^{-1}$} and is approximated as \mbox{$u(\mathbf{p})\approx\sum_{i=1}^{k}f(\mathbf{p}_{i})G(\mathbf{p}_{i},\mathbf{p})\vert V_{i}\vert$}~\cite{BHATIA2014-HODGE,BHATIA2013-SURVEY}, where \mbox{$\vert V_{i}\vert$} is the volume of the 1-star Voronoi cells associated with~$\mathbf{p}_{i}$.

Previous work has addressed the computation of the HHD on triangular meshes~\cite{POLTHIER2000,ZHAO2019} through a variational approach; on tetrahedral meshes~\cite{TONG2003} through a least-squares formulation of its components; and on regular grids~\cite{YU2004} through a decomposition of the grid into a regular triangulation. The \emph{natural HHD}~\cite{BHATIA2014-HODGE} is achieved by separating the vector field on the domain~$\Omega$ and its complement \mbox{$\mathbb{R}^{d}\backslash\Omega$}. Then, the corresponding potentials are computed by solving the Poisson equation and convolving the input functions with the free-space Green's function. The HHD on a domain with boundary is not unique and two decompositions differ by a harmonic function, which is both irrotational and divergence-free. This decomposition becomes unique by either imposing boundary conditions, or restricting the vector field on an open, bounded, and connected sub-domain. In the former case, normal-parallel conditions impose that the conservative and solenoidal components are normal and parallel to the boundary, respectively. In the latter case~\cite{WIEBEL2007}, the \emph{localised flow} in the sub-domain~$\Omega^{\star}$ of~$\Omega$ has the same divergence and curl components of the input vector field but they are isolated from the field on the boundary of~$\Omega^{\star}$.

For the HHD of 2D vector fields, the meshless approach~\cite{PETRONETTO2010,RIBEIRO2016} approximates the gradient, divergence, and Laplace operators through discrete derivatives of the approximation of the input vector field with RBFs generated by a compactly-supported polynomial kernel. Indeed, these approximated differential operators have a fixed accuracy, which is generally linear. In~\cite{MACEBO}, the decomposition is achieved by expressing its components as a linear combination of matrix-valued kernels and computing its coefficients through a learning process. For the choice of the boundary conditions (e.g.,  normal boundary flow on the curl-free and a tangential flow on the divergence-free component) that  guarantee the uniqueness and orthogonality of the HHD, we refer the reader to the work of Bhatia \emph{et al.} ~\cite{BHATIA2013-TVCG-COMMENT}.
%

We briefly recall that the boundary conditions determine the unicity of the HHD and the orthogonality of the conservative and irrotational components. Traditional boundary conditions impose a normal boundary flow for the solenoidal component (i.e.,~$\nabla u$ is normal to the boundary: \mbox{$\nabla u\wedge\mathbf{v}=\mathbf{0}$}) and a tangential flow for the conservative components (i.e., \mbox{$\nabla\wedge\mathbf{w}$}) is parallel to the boundary (i.e., \mbox{$\nabla\wedge\mathbf{w}\cdot\mathbf{n}=0$}). Weaker or different boundary conditions (e.g., the normal flow in the conservative component and the tangential flow in the conservative component) can be applied without guarantees on the uniqueness and/or orthogonality of the decomposition~\cite{BHATIA2013-SURVEY,PETRONETTO2010,BHATIA2013-TVCG-COMMENT}.

\subsection{Meshless approximation\label{sec:MESHLESS-APPROX}}
Meshless approximations have been studied extensively in Computer Graphics and applied to volumetric modelling~\cite{ADAMS2009} and topological analysis~\cite{HART1997}. Choosing a kernel \mbox{$\phi:\mathbb{R}^{+}\rightarrow\mathbb{R}$}, the meshless approximation \mbox{$F:\mathbb{R}^{3}\rightarrow\mathbb{R}$} of \mbox{$f:\mathcal{M}\rightarrow\mathbb{R}$} is defined as a linear combination \mbox{$F(\mathbf{p}):=\sum_{i=1}^{n}\alpha_{i}\phi_{i}(\mathbf{p})$}, where \mbox{$\phi_{i}(\mathbf{p}):=\phi(\|\mathbf{p}-\mathbf{p}_{i}\|_{2})$} is the radial basis function (RBF) centred at~$\mathbf{p}_{i}$. Then, the coefficients \mbox{$\alpha:=(\alpha_{i})_{i=1}^{n}$}, which uniquely satisfy the interpolating conditions \mbox{$F(\mathbf{p}_{i})=f_{i}$}, \mbox{$i=1,\ldots,n$}, are the solutions of the \mbox{$n\times n$} linear system \mbox{$\Phi\alpha=\mathbf{f}$}, where the entries of the Gram matrix~$\Phi$ associated with the RBFs  are \mbox{$\Phi(i,j):=\phi(\|\mathbf{p}_{i}-\mathbf{p}_{j}\|_{2})$}. For the approximation of a vector field~$\mathbf{v}$,  we either apply the meshless approximation to each component or consider the approximating field \mbox{$\mathbf{v}(\mathbf{p}):=\frac{\sum_{\mathbf{p}_{i}\in\mathcal{N}_{\mathbf{p}}}\mathbf{v}(\mathbf{p}_{i})\phi(\|\mathbf{p}-\mathbf{p}_{i}\|_{2})}{\sum_{\mathbf{p}_{i}\in\mathcal{N}_{\mathbf{p}}}\phi(\|\mathbf{p}-\mathbf{p}_{i}\|_{2})}$}, defined as a weighted average~\cite{PRAUN2000,TURK2001} (e.g., Gaussian weights) of the values of~$\mathbf{v}$ in a neighbourhood~$\mathcal{N}_{\mathbf{p}}$ of~$\mathbf{p}$. As neighbourhood of~$\mathbf{p}$, we refer to the~$k$ points of the input point set~$\mathcal{P}$ that are nearest to~$\mathbf{p}$, or to the points belonging to a sphere centred at~$\mathbf{p}$ and with radius proportional to the local sampling density. 

\textbf{RBFs' Centres' selection\label{sec:CENTRES-SELECTION}}
To reduce the amount of memory storage and computation time, a set of centres is selected through clustering, kernel and sampling methods, or sparsificiation. \emph{Clustering} (e.g., \mbox{$k$-means} clustering~\cite{LLOYD82}, PCA - Principal Component Analysis~\cite{JOLLIFFE86}) is applied to group those points that satisfy a common ``property'' (e.g., planarity, closeness) and each basis function is centred at a representative point of each cluster. \emph{Kernel methods}~\cite{CORTES95} (e.g., kernel PCA) evaluate the correlation among points with respect to the scalar product induced by a positive-definite kernel. \emph{Sampling methods} approximate a signal as a linear combination of Gaussian kernels with a fixed support, whose centres (i.e., the samples) are computed through the minimisation of a least-squares energy functional~\cite{zhong2016kernel,sheather2004density}, or satisfy the blue-noise~\cite{fattal2011blue} or spectral~\cite{oztireli2010spectral} properties. \emph{Sparsification} selects the basis functions through a basis pursuit de-noising~\cite{CHEN1998}, standard and orthogonality matching pursuit methods~\cite{CHEN1989,MALLAT1993}, or regularised logistic regression~\cite{NG2004}.
%
%
\begin{figure}[t]
\centering
\begin{tabular}{ll}
\includegraphics[height=100pt]{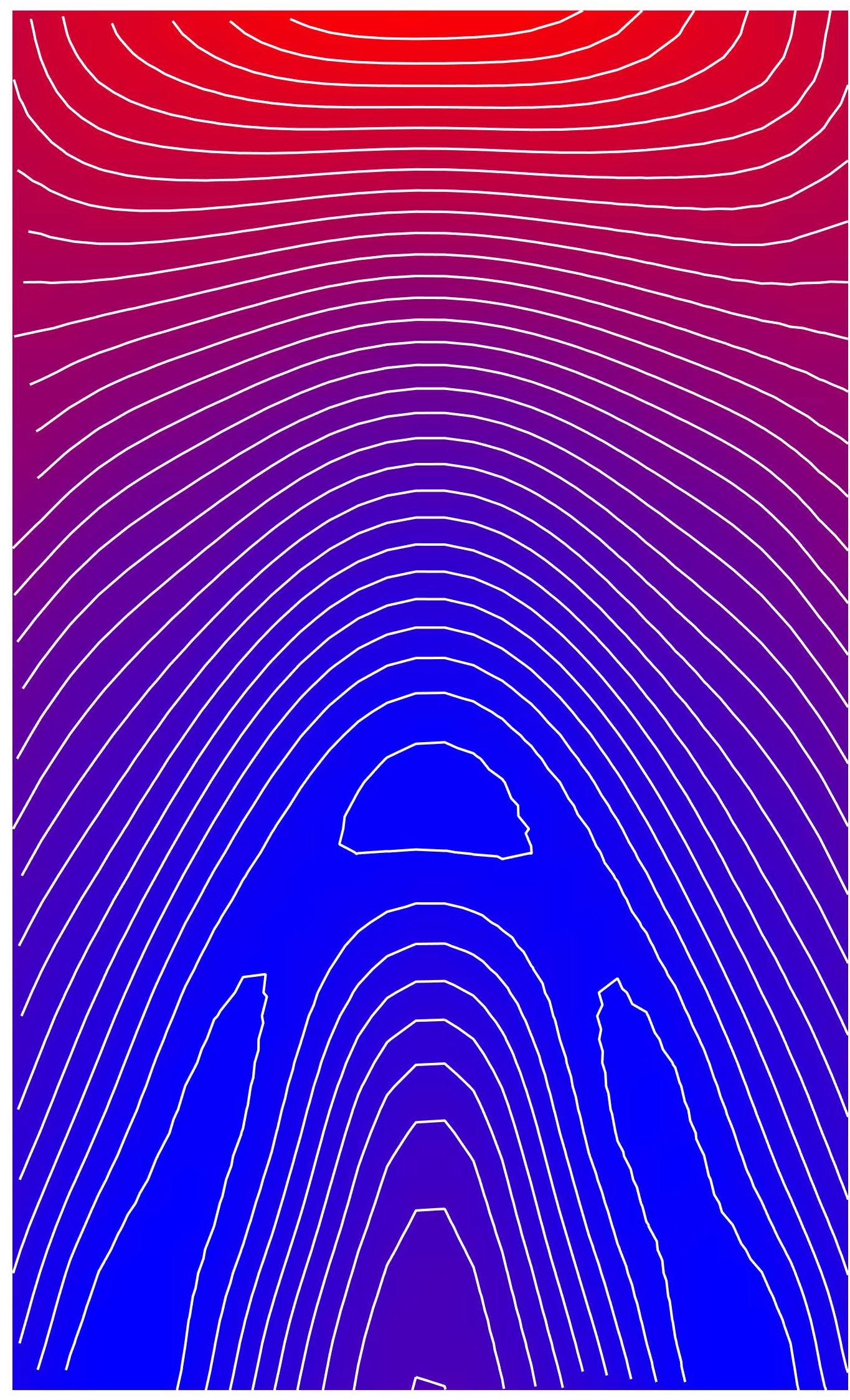}
&\includegraphics[width=160pt]{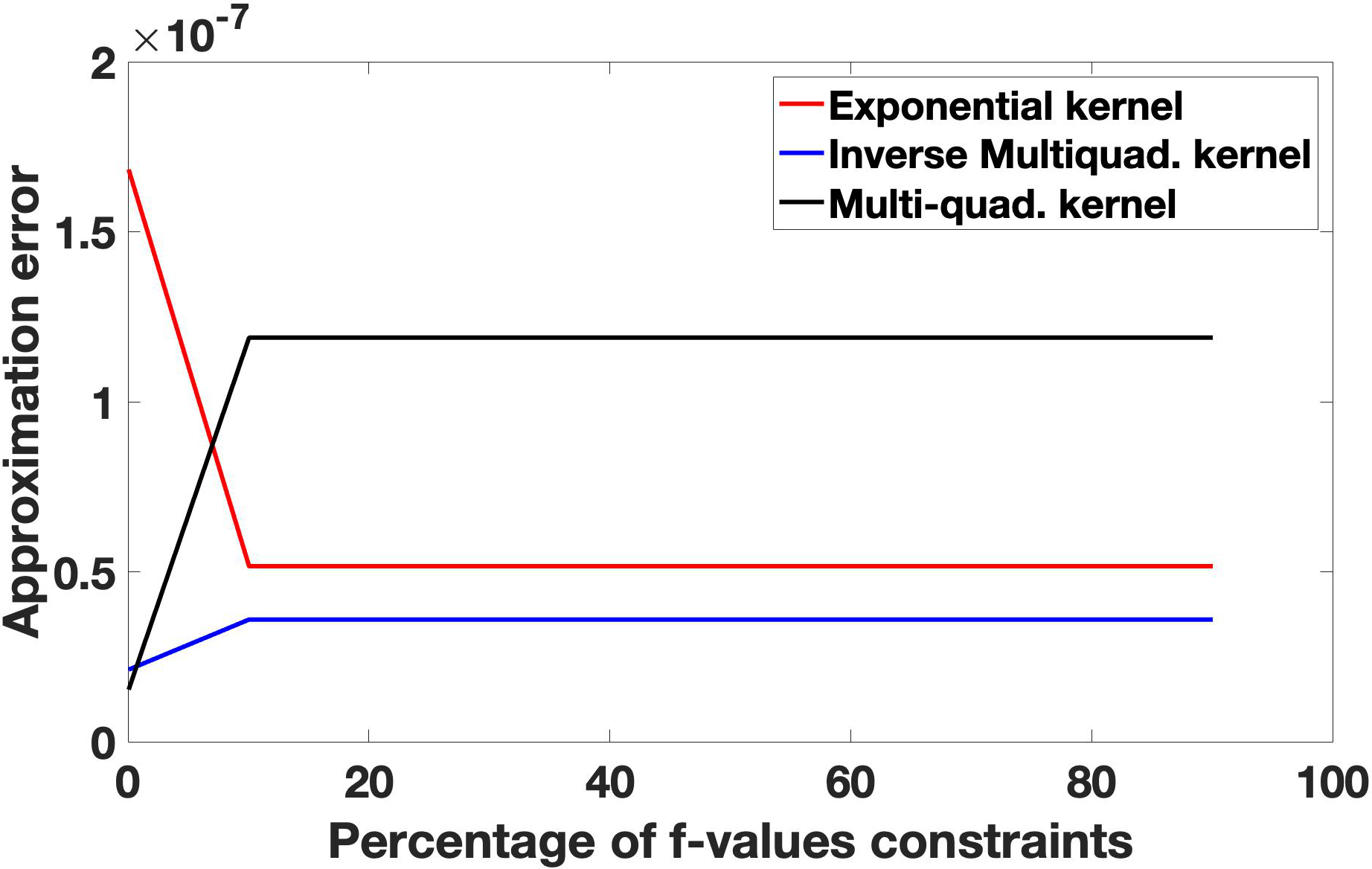}\\
(a) Potential~$u$ &(b) Approximation accuracy\\
\includegraphics[height=100pt]{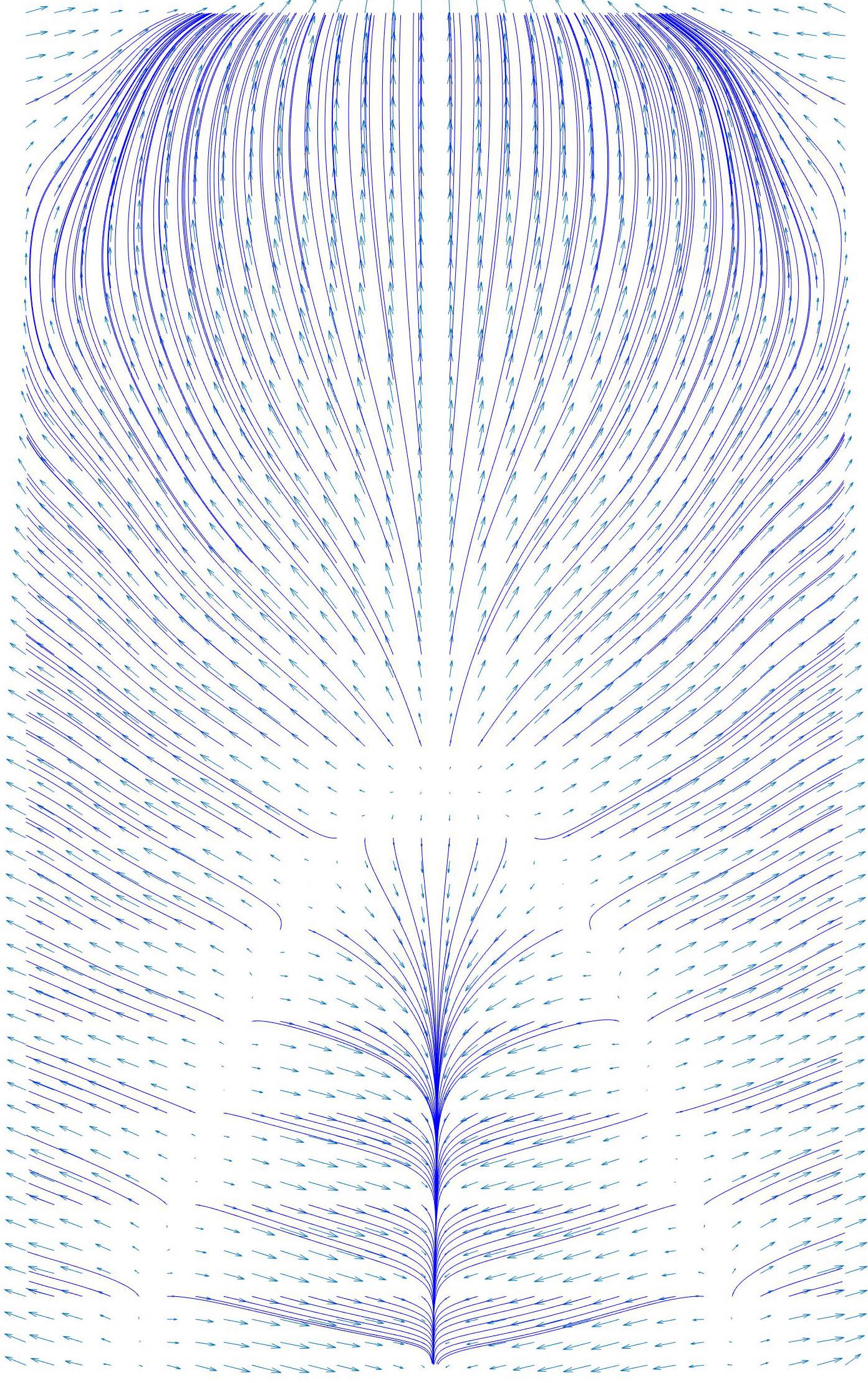}
&\includegraphics[width=160pt]{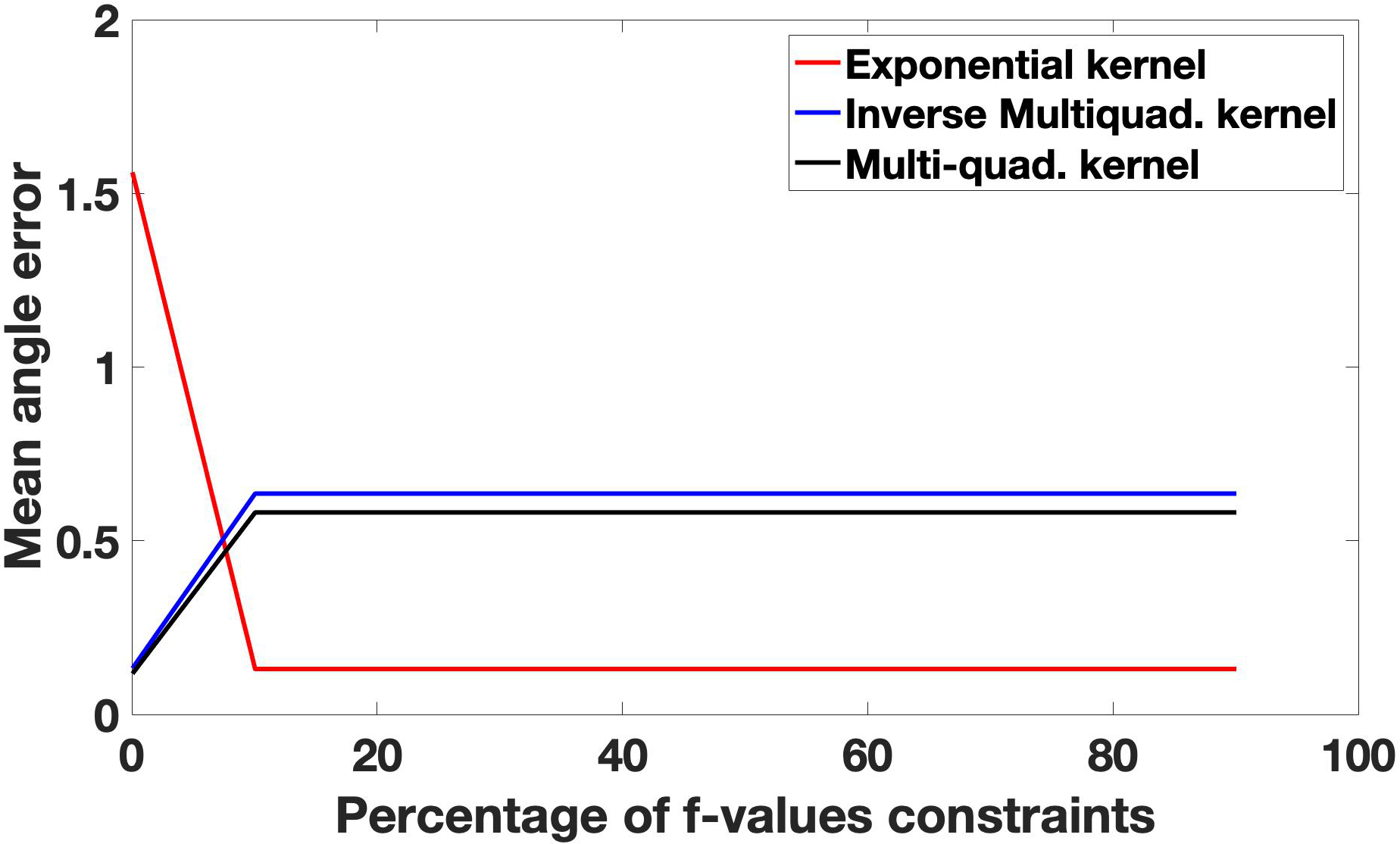}\\
(c) Gradient~$\nabla u$ &(d) Mean angle \mbox{$\angle(\mathbf{v},\nabla u)$}
\end{tabular}
\caption{(a) Meshless potential~$u$, (b) approximation accuracy, (c) gradient field~$\nabla u$, ($y$-axis), (d) mean angle \mbox{$\angle(\mathbf{v},\nabla u)$} between the input vector field~$\mathbf{v}$ and the least-squares meshless gradient field \mbox{$\nabla u$}, induced by different kernels and percentages ($x$-axis) of~$p_{1}\%$ random constraints on~$u$-values ($x$-axis) ($10\%\leq p_{1}\leq 90\%$), \mbox{$(100-p_{1})\%$} random constraints on~$\nabla u$-values. Function and vector constraints are both applied to an additional~$10\%$ of the same input points (i.e., overlapped constraints), thus imposing a total of~$110\%$ least-squares constraints. See also Fig.~\ref{fig:NONANALYTIC-EXAMPLE-ZOOM}.\label{fig:NONANALYTIC-EXAMPLE}}
\end{figure}
\subsection{Analysis of potentials and vector fields\label{sec:VECTOR-ANALYSIS}}
\textbf{Computation of the potential fields}
For the conservative and solenoidal components on discrete surfaces and volumes, the Poisson equation reduces to a sparse linear system, whose solution is computed by applying iterative solvers (e.g., gradient conjugate) or building a multi-resolution mesh pyramid~\cite{YU2004,GUSKOV1999}. In this last case, the Poisson equation is solved at the coarsest resolution and its solution is then mapped back to the finer resolution by collapsing the pyramid and adding the corresponding details to the current approximation of the solution. This approach is efficient but depends on the discretisation and resolution of the input grid. An alternative approach is to consider a moving least-squares approximation~\cite{FARWIG1986,LEVIN1998} with a polynomial basis, which improves the accuracy and smoothness of the gradient of a given scalar function~\cite{CORREA2011}. 
\begin{figure*}[t]
\centering
\begin{tabular}{l|l||l|l||l|l}
\hline
\textbf{SC}~$\mathcal{I}:90\%$	
&\textbf{SC}~$\mathcal{I}:20\%$
&\textbf{SC}~$\mathcal{I}:90\%$	
&\textbf{SC}~$\mathcal{I}:20\%$				
&\textbf{SC}~$\mathcal{I}:90\%$
&\textbf{SC}~$\mathcal{I}:20\%$\\
\textbf{VC}~$\mathcal{J}:20\%$			
&\textbf{VC}~$\mathcal{J}:90\%$			
&\textbf{VC}~$\mathcal{J}:20\%$			
&\textbf{VC}~$\mathcal{J}:90\%$				
&\textbf{VC}~$\mathcal{J}:20\%$			
&\textbf{VC}~$\mathcal{J}:90\%$\\
\hline
\multicolumn{2}{c||}{\textbf{Gaussian kernel}}			
&\multicolumn{2}{c||}{\textbf{Inverse multi-quadratic kernel}}		
&\multicolumn{2}{c}{\textbf{Multi-quadratic kernel}}\\
\includegraphics[height=110pt]{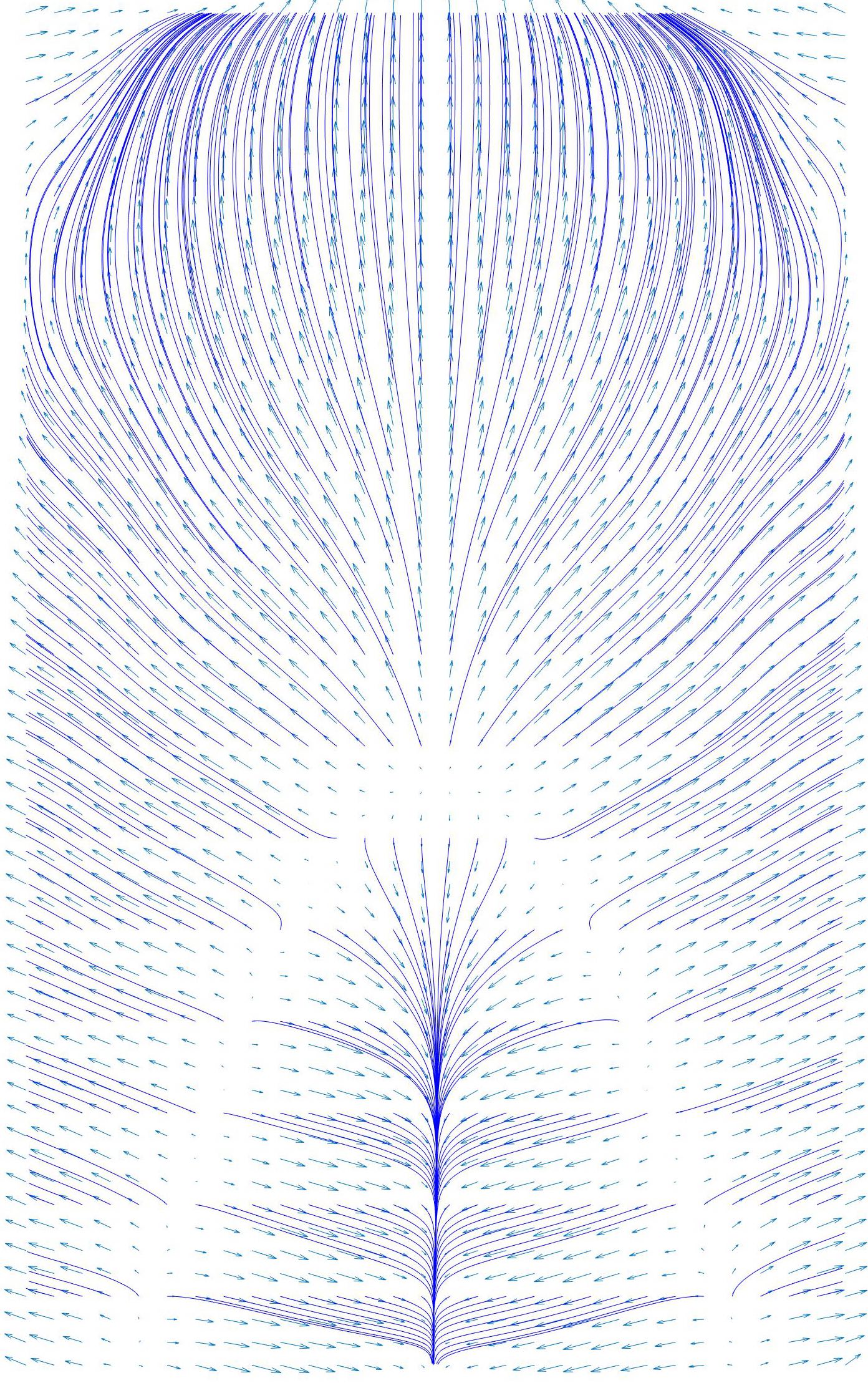}
&\includegraphics[height=110pt]{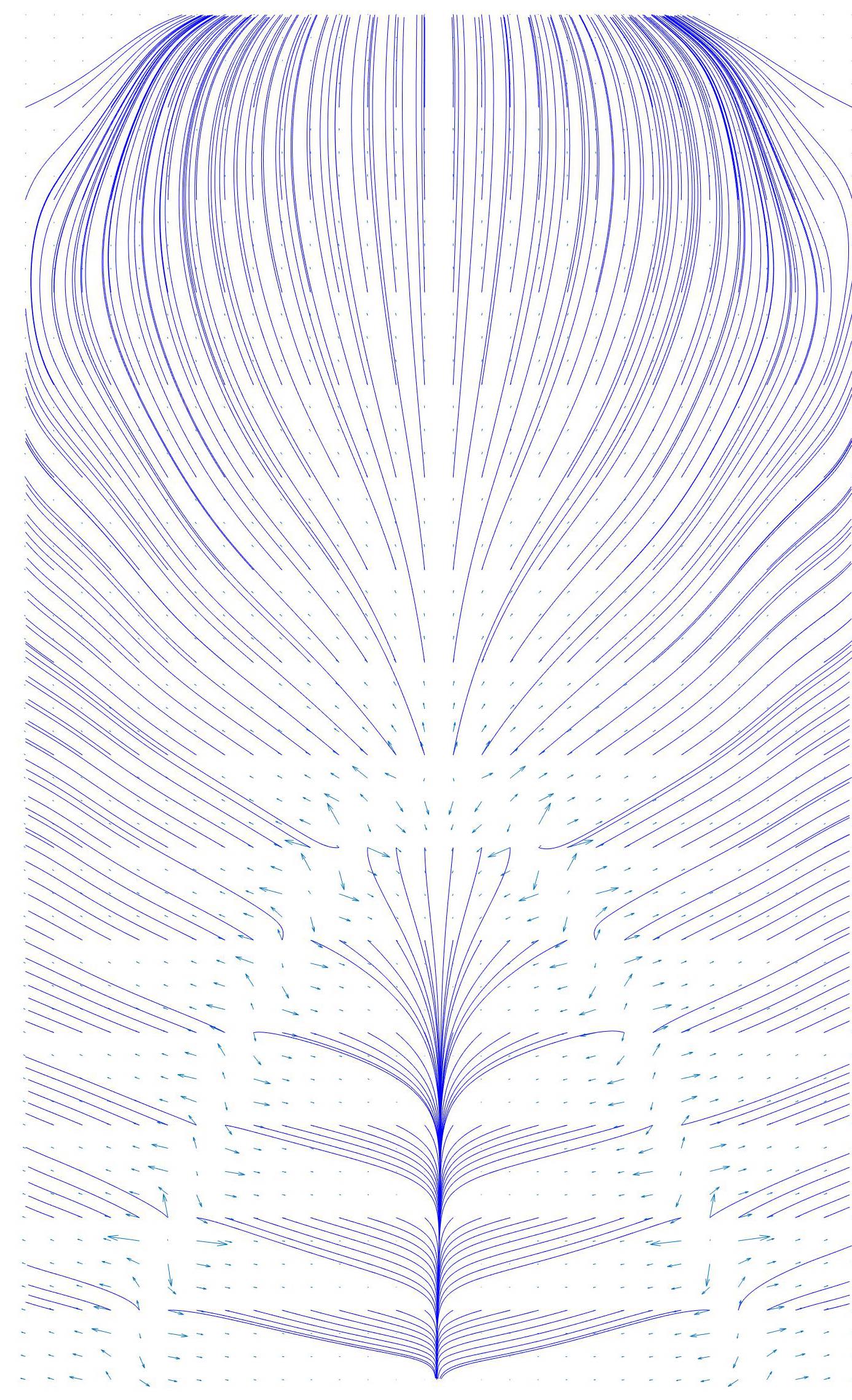}
&\includegraphics[height=110pt]{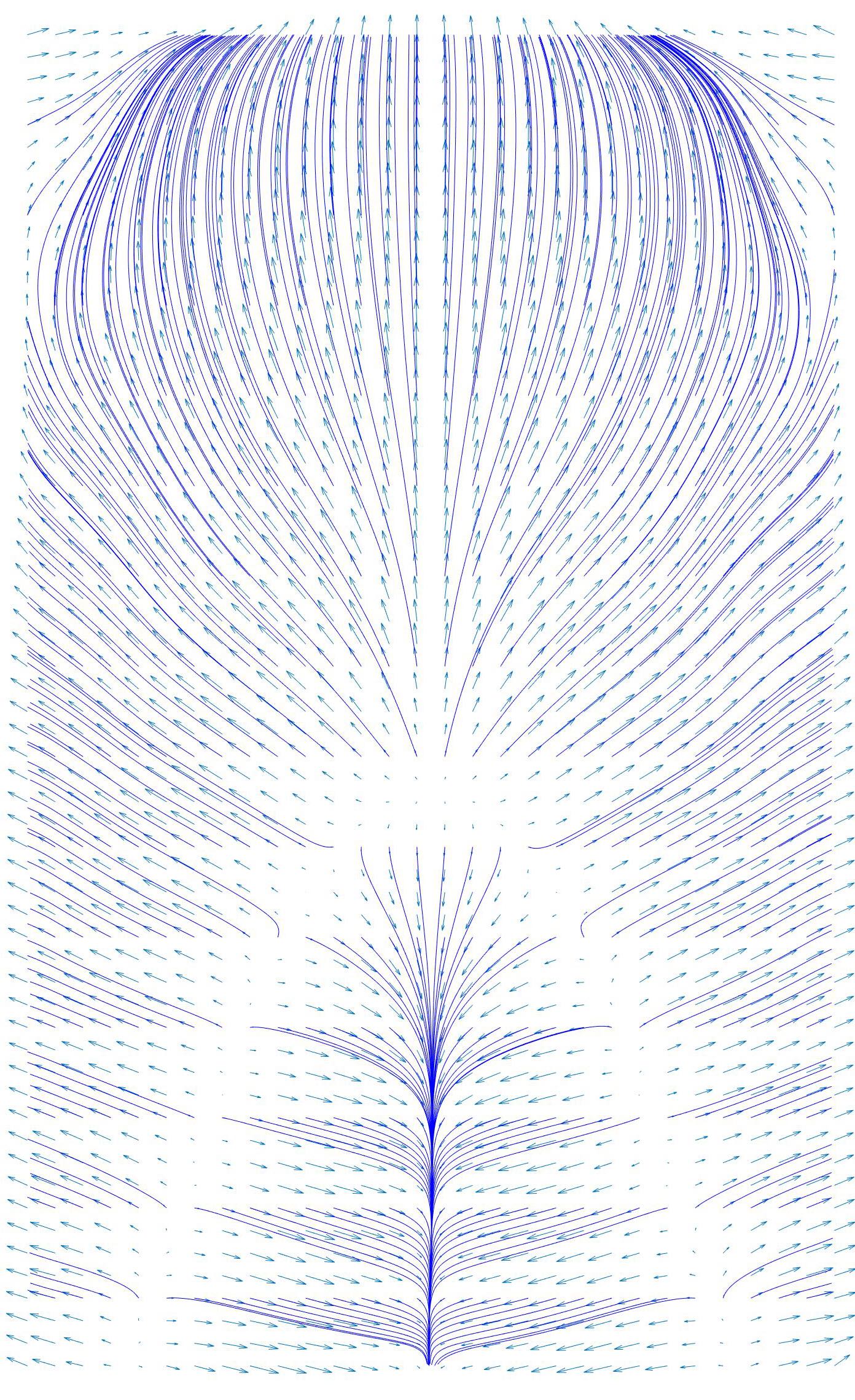}
&\includegraphics[height=110pt]{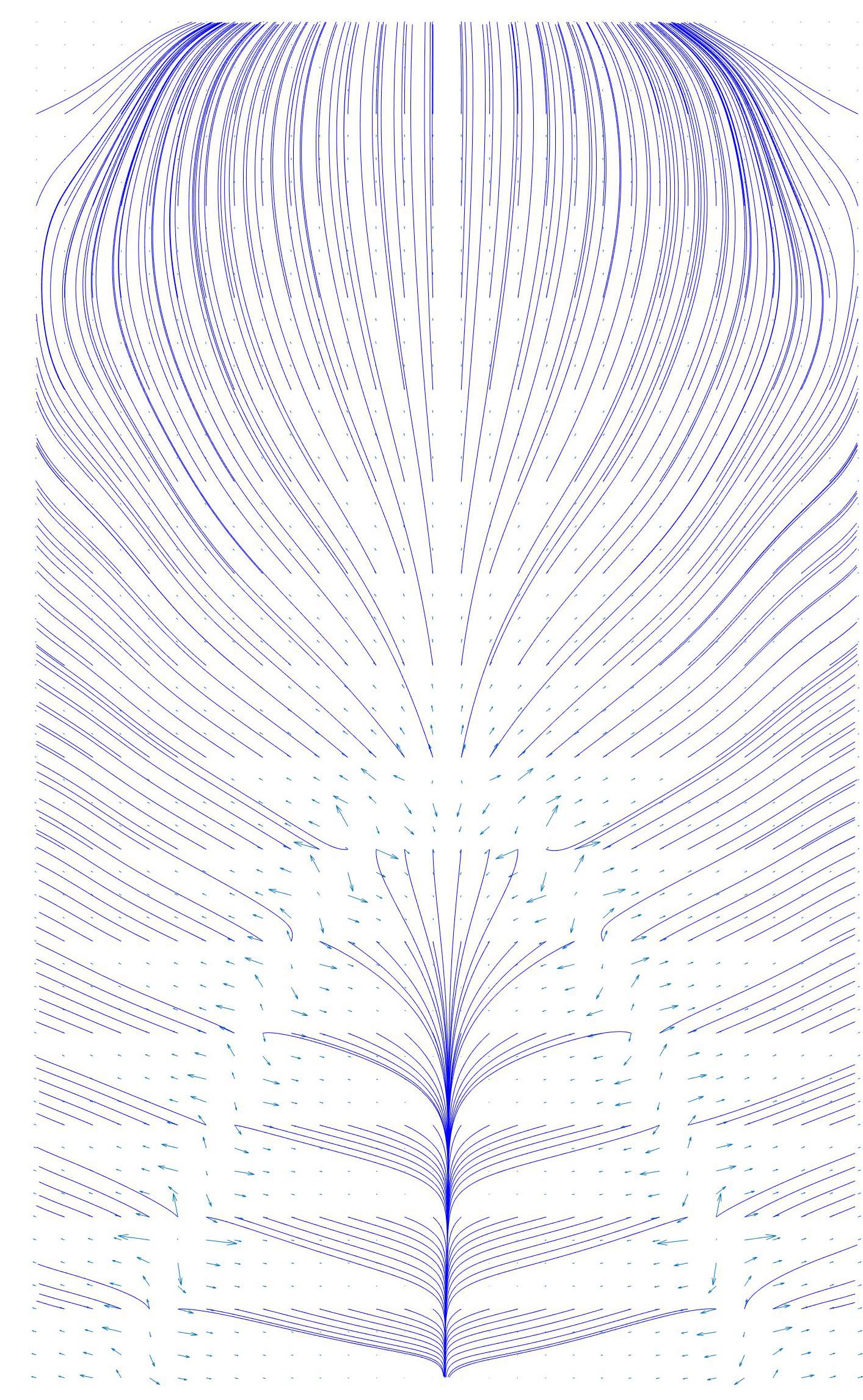}
&\includegraphics[height=110pt]{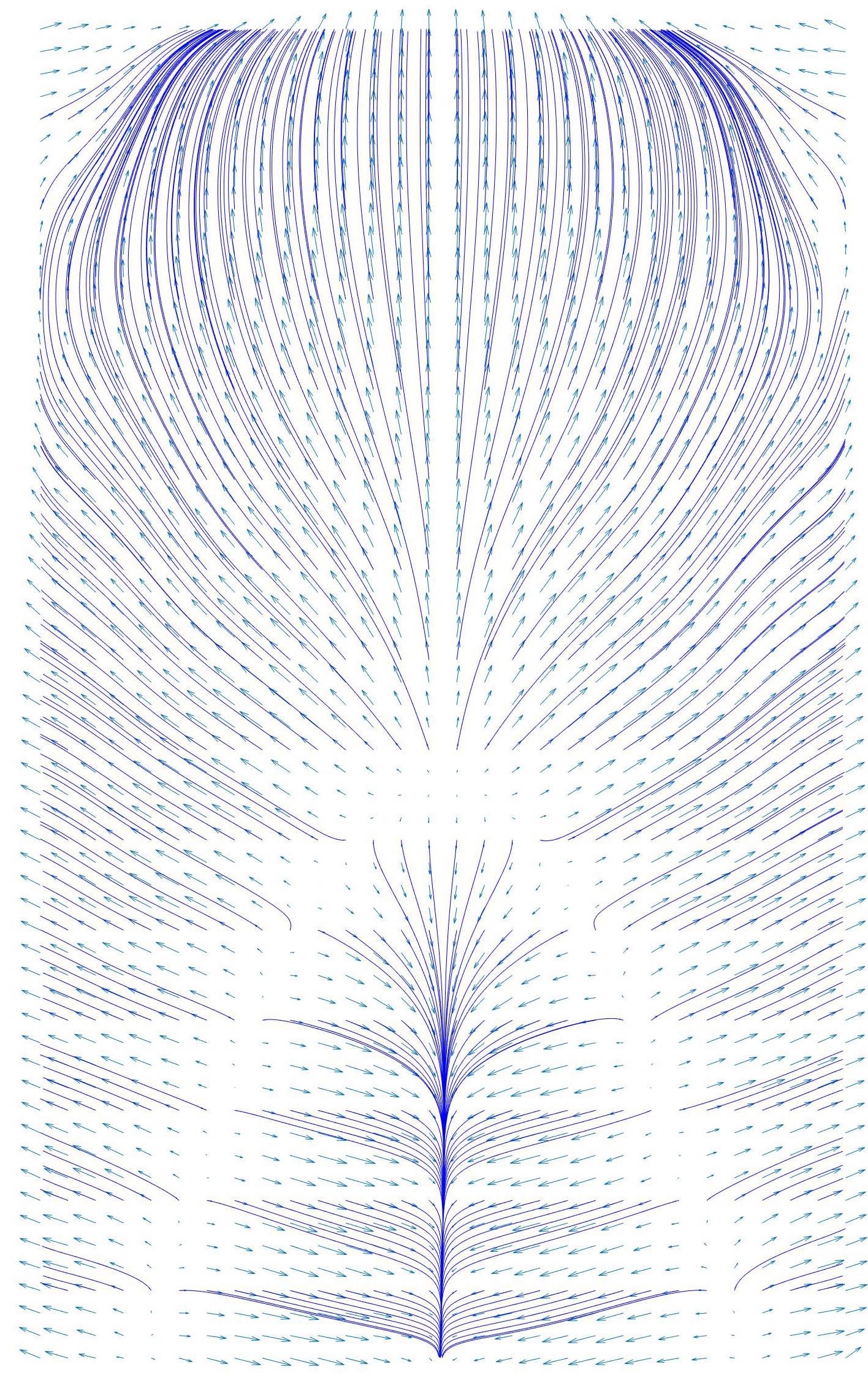}
&\includegraphics[height=110pt]{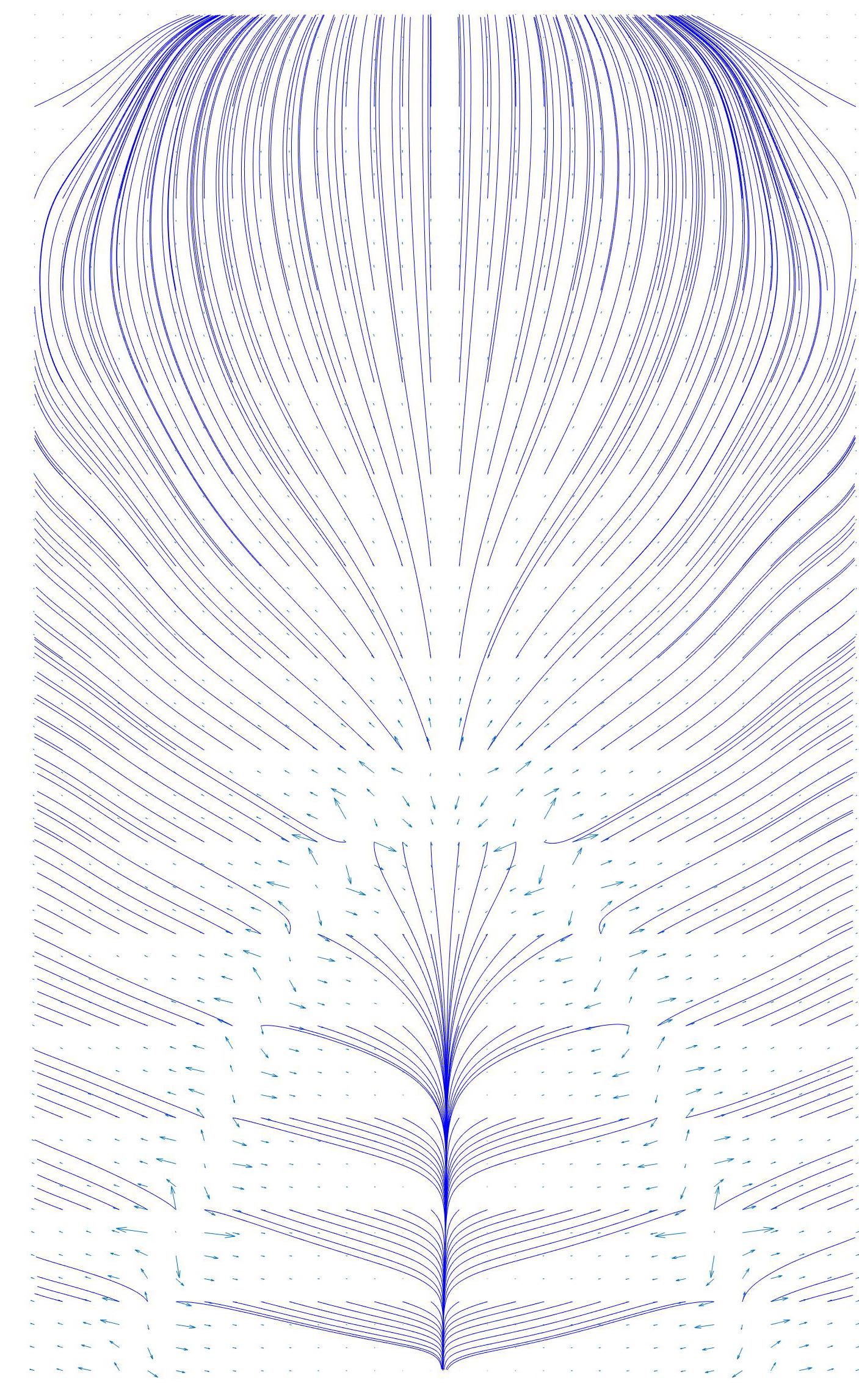}\\
\hline
\end{tabular}
\begin{tabular}{ccc|ccc}
\includegraphics[height=80pt]{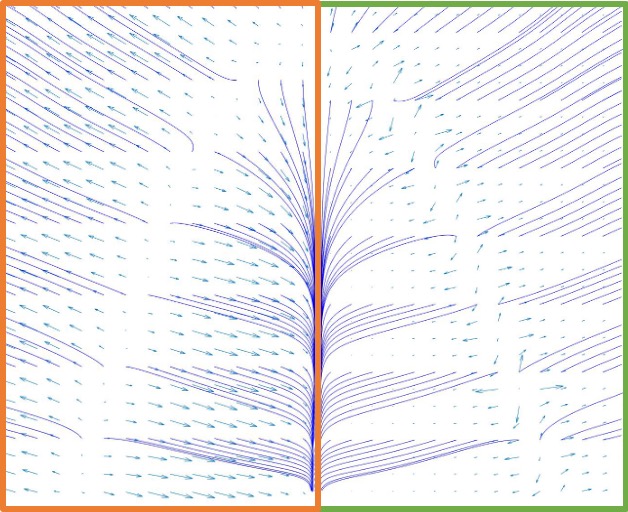}
&\includegraphics[height=80pt]{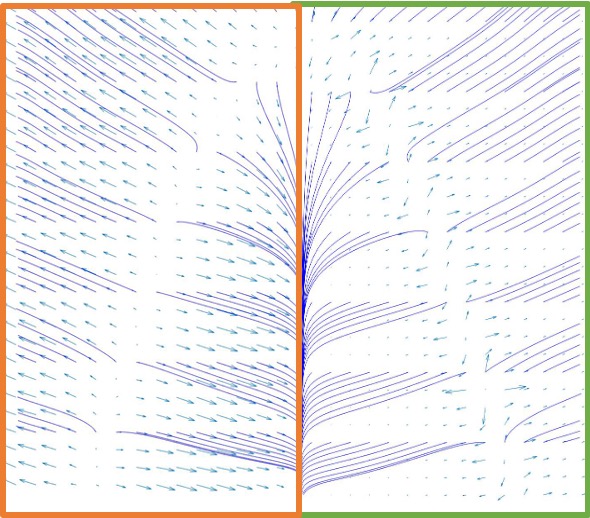}
&\includegraphics[height=80pt]{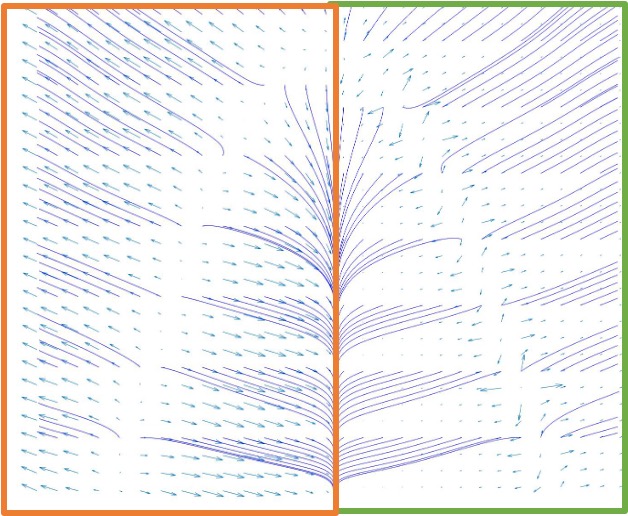}
&\includegraphics[height=85pt]{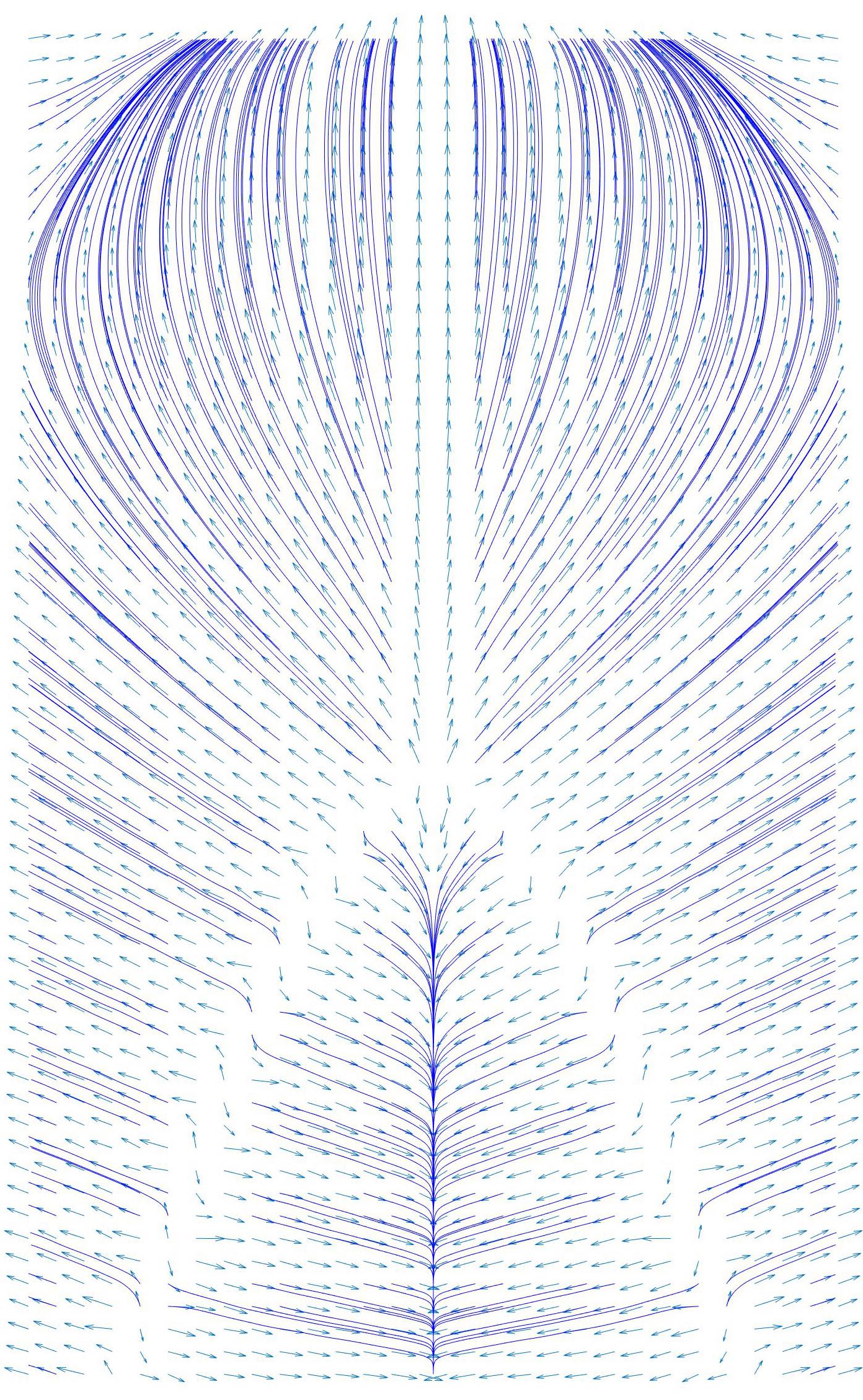}
&\includegraphics[height=85pt]{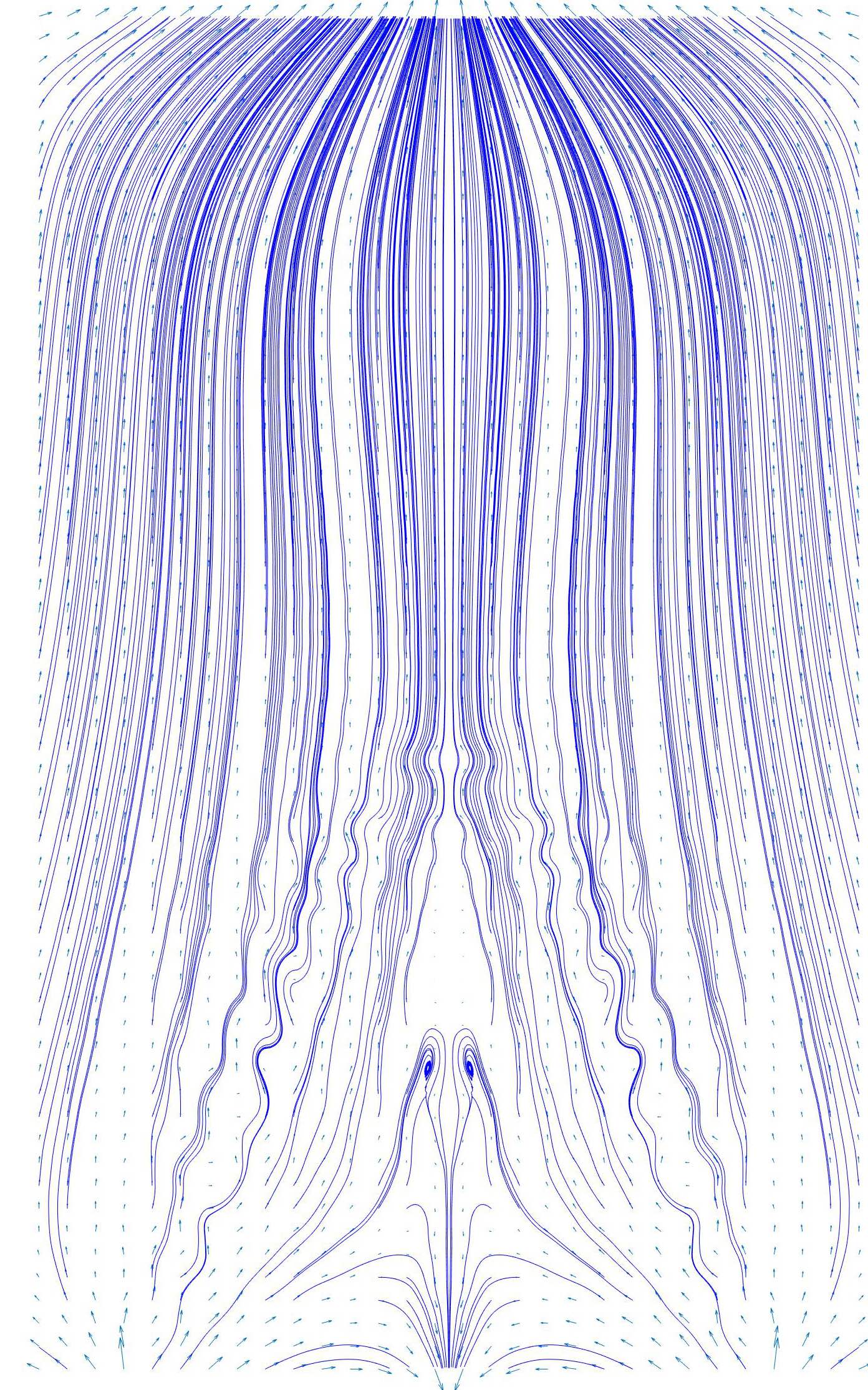}
&\includegraphics[height=85pt]{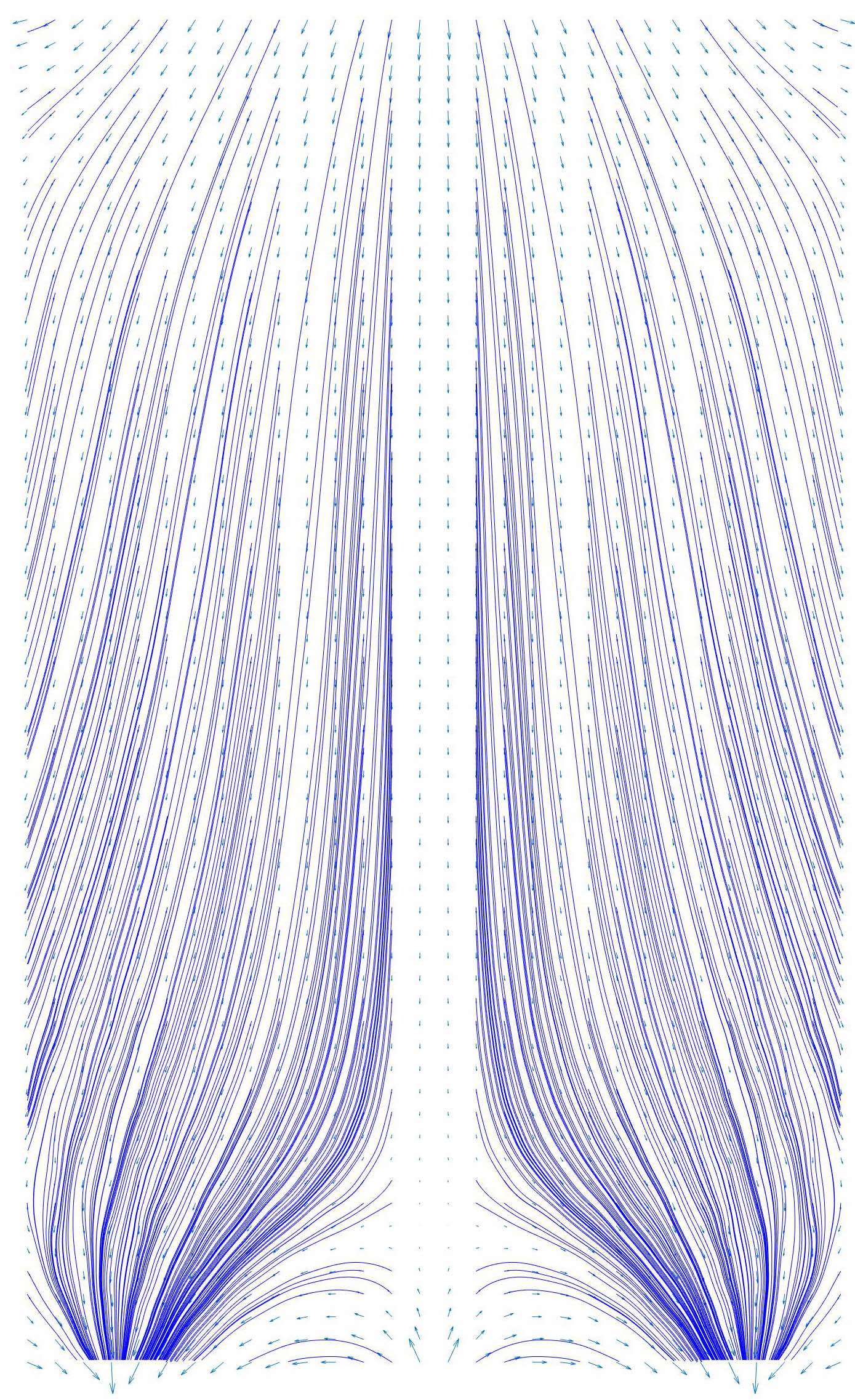}\\
(a) Zoom: Gaussian ker.&Inv. multi-quad. ker. &Multi-quad. ker. &(b)~$\nabla u$ &$\nabla\wedge\mathbf{w}$ &$\mathbf{h}$
\end{tabular}
\caption{1st Row: With reference to Fig.~\ref{fig:NONANALYTIC-EXAMPLE}, streamlines of the approximated meshless vector field, induced by different kernels and percentages of randomly selected function (SC) and vector (VC) constraints. We impose~$p_{1}\%$ constraints on~$u$-values ($x$-axis) (\mbox{$10\%\leq p_{1}\leq 90\%$}), \mbox{$(100-p_{1})\%$} constraints on~$\nabla u$-values. Function and vector constraints are both applied to an additional~$10\%$ of the same input points (i.e., overlapped constraints), thus imposing a total of~$110\%$ least-squares constraints. 2nd Row: (a) zoom on the input (left box) and meshless (right box) approximation, and (b) HHD.\label{fig:NONANALYTIC-EXAMPLE-ZOOM}}
\end{figure*}

\textbf{Analysis of vector and potential fields}
Vector fields are typically visualised through level-set methods~\cite{WESTERMANN2000}, local reference frames~\cite{BHATIA2014,GUNTER2018}, and flow regions~\cite{WINKAUF2004} with a similar geometric, topological, or physical behaviour~\cite{BAUER2002,LEEUW1999,TRICOCHE2000}. Another efficient approach is to compute the flow streamlines~\cite{ROESSL2012}, which are defined as curves whose tangential direction is equal to the velocity of the input field, and partition the input domain in such a way that two streamlines are either disjoint or equal. For the analysis of the potential of the conservative component, we mention the classification of the critical points~\cite{BAUER2012,EDELSBRUNNER2006} and the contour tree~\cite{CARR2000,CARR2004}, the Morse-Smale complex~\cite{BREMER2004,GYULASSY2008}, and the Reeb graph~\cite{PASCUCCI2007,TIERNY2009,PATANE2009-TVCG}. The input vector field and/or the potential of the irrotational component can be analysed by classifying its singularities and streamlines, which are smoothed in order to preserve only the persistent ones~\cite{MANN2002} in case of noise. For more details on these topics, we refer the reader to survey papers on topology-based visualisation~\cite{MCLOUGHLIN2010,PROBITZER2010,PROBITZER2011} and on vortex extraction~\cite{GUNTER2018}.

%
\begin{figure*}[t]
\centering
\begin{tabular}{ccc}
$u_{1}=x^2-y^2$ &$u_{2}=x^2-y^3$ &$u_{3}=x^2+\exp(-xy)+y^3$\\
\includegraphics[height=96pt]{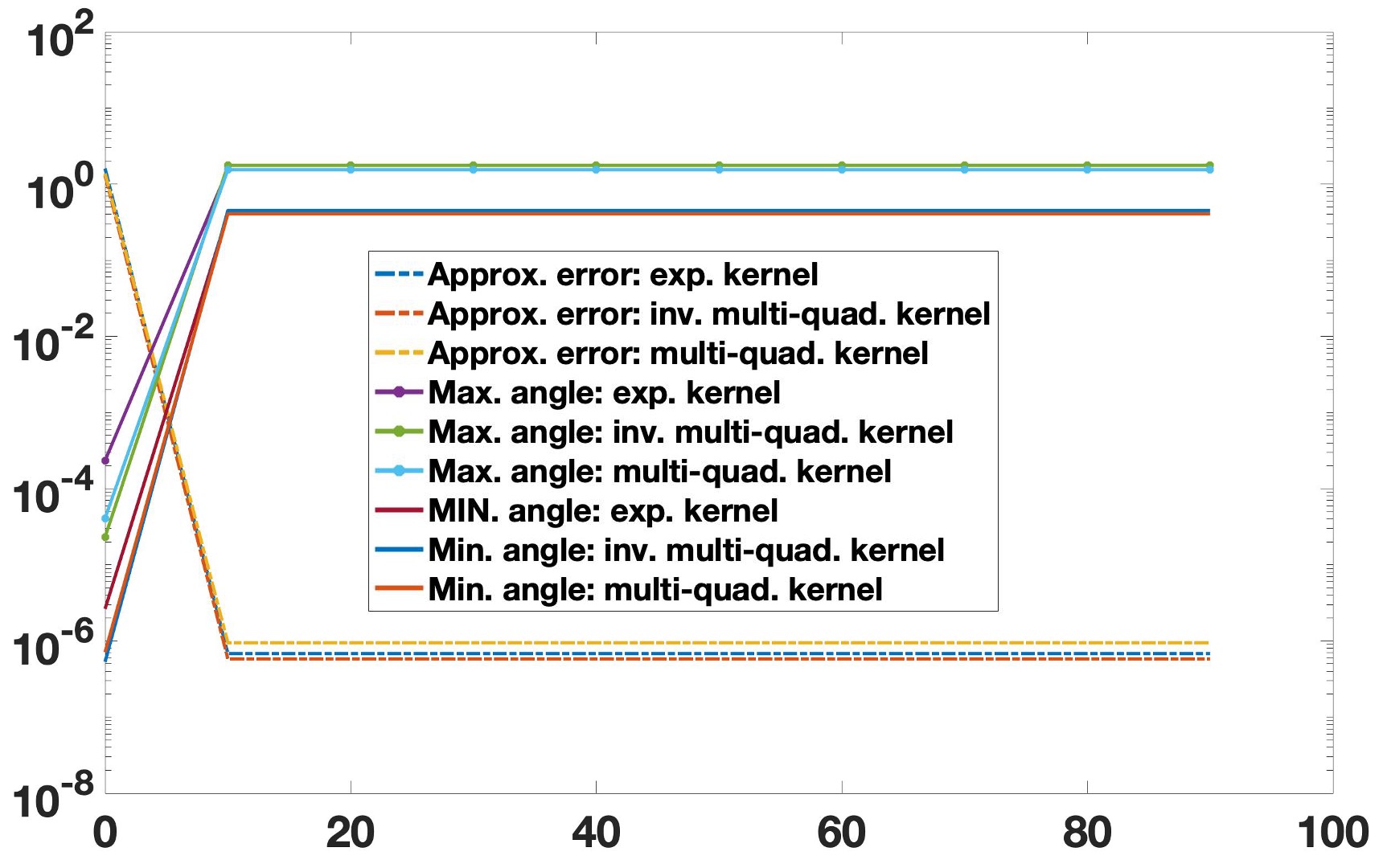}
&\includegraphics[height=96pt]{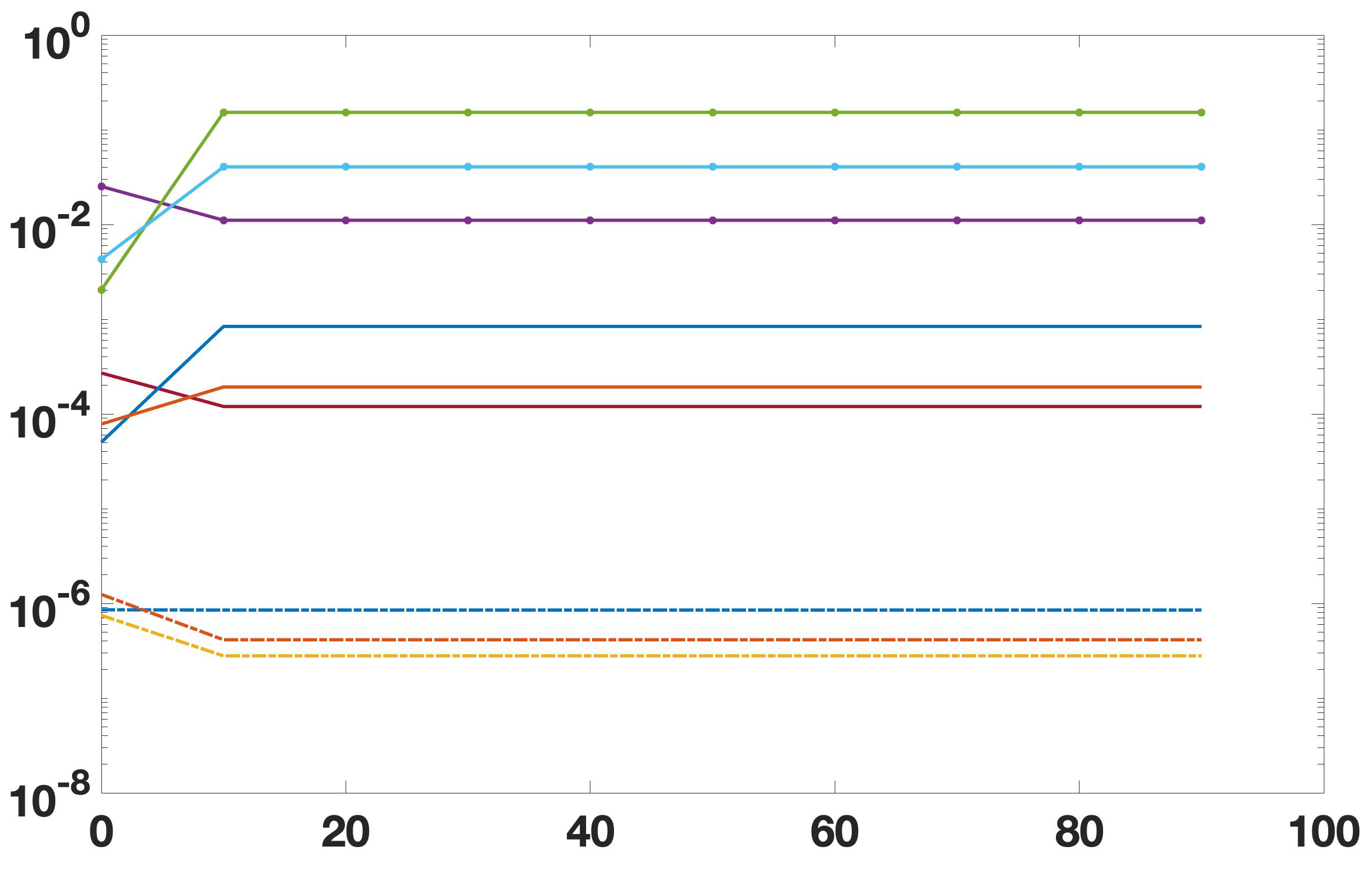}
&\includegraphics[height=96pt]{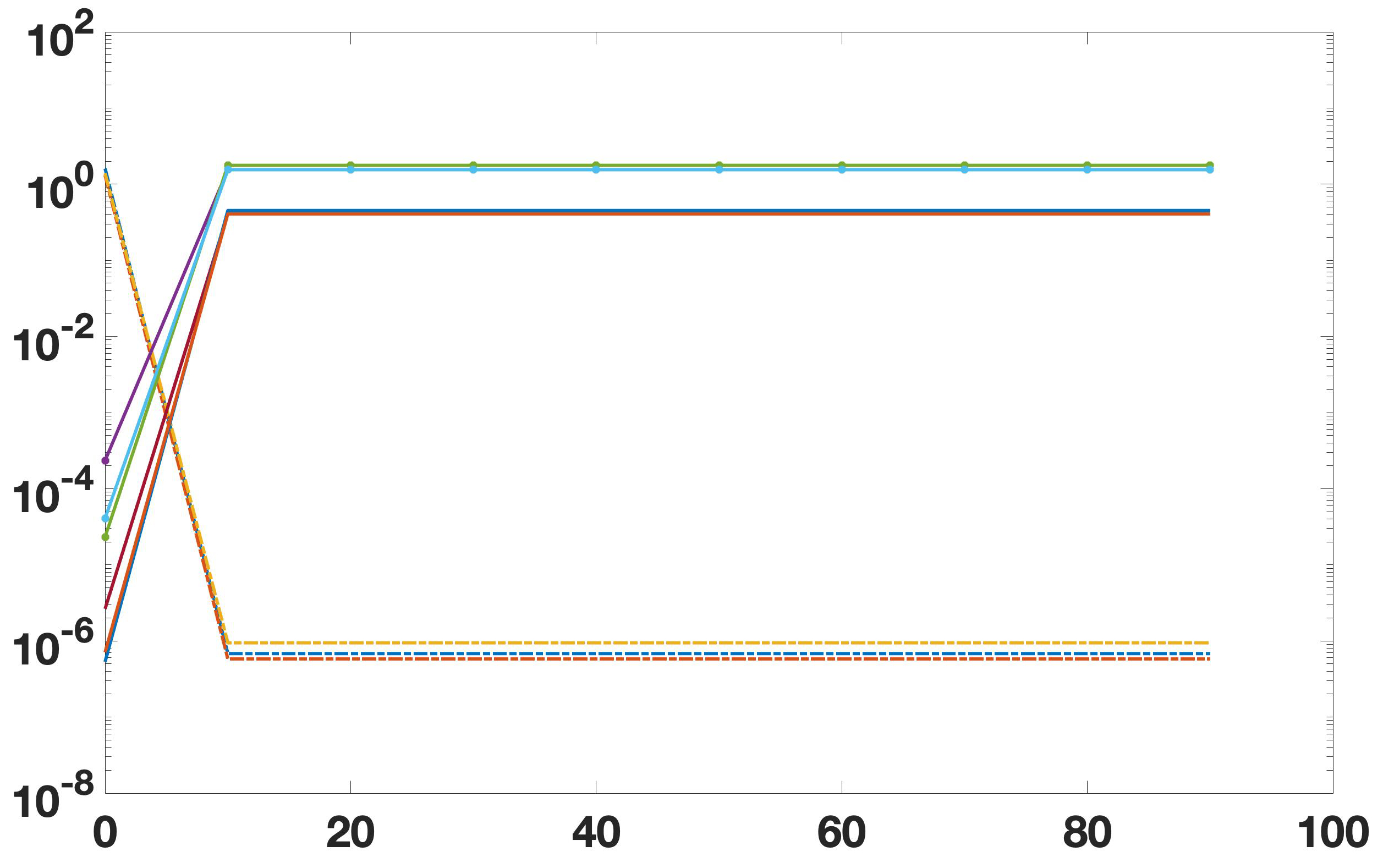}
\end{tabular}
\caption{Analytic vector fields sampled on an irregular 2D point set (c.f., Fig.~\ref{fig:ANALYTIC-ANGLE-ERROR}(a)). ($y$-axis)~$\mathcal{L}_{\infty}$ approximation error and maximum angles \mbox{$\angle(\mathbf{v},\nabla u)$} between the input~$\mathbf{v}$ and meshless \mbox{$\nabla u$} gradient fields, induced by different kernels and percentages ($x$-axis) of constraints on the ($u$,~$\nabla u$) values. We impose~$p_{1}\%$ constraints on~$u$-values ($x$-axis) ($10\%\leq p_{1}\leq 90\%$), \mbox{$(100-p_{1})\%$} constraints on~$\nabla u$-values. Function and vector constraints are both applied to an additional~$10\%$ of the input points, thus imposing a total of~$110\%$ of least-squares constraints (i.e., overlapped constraints). For the potential~$u_{2}$, the distribution of the angle error is shown in Fig.~\ref{fig:ANALYTIC-ANGLE-ERROR}.\label{fig:ANALYTIC-STATISTICS}}
\end{figure*}
\section{Meshless approximation\label{sec:MESHLESS-APPROXIMATION-VF}}
We address  the approximation of the potential of the conservative component of a vector field from a set of discrete vector values, and eventually integrated with a set of scalar values of its potential at the same or different samples. To this end, we apply a meshless least-squares approach, which enforces the consistency of the meshless potential with respect to the input samples and allows us to analytically evaluate its derivatives for the computation of the HHD. In Sect.~\ref{sec:KERNEL-SELECTION}, we discuss the selection of the kernel and of the centres of the RBFs for the meshless approximation. 

\textbf{Meshless potential of mixed scalar and vector fields}
We address the computation of the \emph{meshless potential} \mbox{$\mathbf{v}:\Omega\rightarrow\mathbb{R}$} \emph{of scalar values \mbox{$(f_{i})_{i\in\mathcal{I}}$} and vectors \mbox{$(\mathbf{v}_{i})_{i\in\mathcal{J}}$} measured at sparse sampled points}, i.e., 
\begin{equation}\label{eq:MIXED-CONDITIONS}
u(\mathbf{p}_{i})\approx f_{i},\, i\in\mathcal{I},\qquad
\nabla u(\mathbf{p}_{j})\approx \mathbf{v}_{j},\, j\in\mathcal{J},
\end{equation}
with~$k_{1}$,~$k_{2}$ numbers of indices in~$\mathcal{I}$,~$\mathcal{J}$, respectively. For the mixed interpolation problem, the potential function \mbox{$u(\mathbf{p})=\sum_{i\in\mathcal{I}}\alpha_{i}\phi_{i}(\mathbf{p})+\sum_{j\in\mathcal{J}}\beta_{j}\phi_{j}(\mathbf{p})$} is defined as a linear combination of the RBFs \mbox{$(\phi_{i})_{i}$} centred at \mbox{$(\mathbf{p}_{k})_{k\in\mathcal{I}\cup\mathcal{J}}$}, with unknown coefficients \mbox{$(\alpha_{i})_{i\in\mathcal{I}}$} and  \mbox{$(\beta_{j})_{j\in\mathcal{J}}$}. Imposing the conditions in Eq.~(\ref{eq:MIXED-CONDITIONS})
\begin{equation}\label{eq:LS-HETEROGENEOUS-CONSTRAINTS}
\left\{
\begin{array}{ll}
u(\mathbf{p}_{i})
=\sum_{k\in\mathcal{I}\cup\mathcal{J}}\alpha_{k}\phi_{k}(\mathbf{p}_{i})
\approx f_{i},	&i\in\mathcal{I},\\
\nabla u(\mathbf{p}_{j})=\sum_{k\in\mathcal{I}\cup\mathcal{J}}\alpha_{k}\nabla\phi_{k}(\mathbf{p}_{j})
\approx \mathbf{v}_{j}, &j\in\mathcal{J},
\end{array}
\right.
\end{equation}
we minimise the least-squares energy functional
\begin{equation}\label{eq:MESHLESS-LS}
\mathcal{E}(\alpha)
:=\sum_{i\in\mathcal{I}}\vert u(\mathbf{p}_{i})-f(\mathbf{p}_{i})\vert^{2}+\sum_{j\in\mathcal{J}}\|\nabla u(\mathbf{p}_{j})-\mathbf{v}_{j}\|_{2}^{2}.
\end{equation}
Then, the normal equation is
\begin{equation}\label{eq:LINEAR-SYSTEM}
\left[
\begin{array}{c}
\tilde{\Phi}\\
\Phi
\end{array}
\right]\alpha=
\left[
\begin{array}{c}
\tilde{\mathbf{f}}\\
\tilde{\mathbf{v}}
\end{array}
\right],
\end{equation}
with \mbox{$\tilde{\Phi}\in\mathbb{R}^{k_{1}\times (k_{1}+k_{2})}$}, \mbox{$\Phi\in\mathbb{R}^{3k_{2}\times (k_{1}+k_{2})}$}, \mbox{$(k_{1}+k_{2})$} unknowns, and right hand side vectors \mbox{$\tilde{\mathbf{f}}:=(f(\mathbf{p}_{i}))_{i\in\mathcal{I}}\in\mathbb{R}^{k_{1}}$} and \mbox{$\tilde{\mathbf{v}}\in\mathbb{R}^{3k_{2}}$}. Here, \mbox{$\tilde{\Phi}:=(\phi_{k}(\mathbf{p}_{j}))_{j\in\mathcal{J}}^{k\in\mathcal{I}\cup\mathcal{J}}$} is the Gram matrix associated with the generating kernel and the matrix~$\Phi$ is defined as \mbox{$\Phi:=
[\partial_{x}\phi_{k}(\mathbf{p}_{j}), \partial_{y}\phi_{k}(\mathbf{p}_{j}), \partial_{z}\phi_{k}(\mathbf{p}_{j})]^{\top}_{k,j}$}. If the scalar and vector terms have very different value ranges, it is enough to consider a positive trade-off parameter~$\delta$ as coefficient of the second term in Eq. (\ref{eq:MESHLESS-LS}) and to include~$\delta$ in the corresponding parts $\Phi$, $\tilde{\mathbf{v}}$ of Eq. (\ref{eq:LINEAR-SYSTEM}). Alternatively, the function and vector values are normalised before computing the meshless potential.

\textbf{``Overlapped'' conditions on scalar/vector values}
The least-squares formulation is valid also for those cases where we have both scalar and vector constraints at the same points, or arbitrary vector fields (e.g., not necessarily conservative). In this case, the approximation scheme and solver remain unchanged and each centre~$\mathbf{p}_{i}$, \mbox{$i\in\mathcal{I}\cap\mathcal{J}$}, is counted only once. The number of approximating constraints is greater than the number of unknowns and the corresponding linear system in Eq.~(\ref{eq:LINEAR-SYSTEM}) admits a unique solution. The conservative and meshless vector field \mbox{$\mathbf{v}=\nabla u$} provides the best least-squares approximation of the constrains in Eq.~(\ref{eq:LS-HETEROGENEOUS-CONSTRAINTS}).

\textbf{Examples}
We consider an input potential~$u$ (Fig.~\ref{fig:NONANALYTIC-EXAMPLE}(a)) and vector field \mbox{$\mathbf{v}:=\nabla u$} (Fig.~\ref{fig:NONANALYTIC-EXAMPLE}(b)), which are sampled with a different percentage of constraints on function ($\mathcal{I}$) and vector ($\mathcal{J}$) values. In this case, constraints on the~$u$-values and~$\nabla u$-vectors are both applied to the~$10\%$ of the same input points (i.e., overlapped constraints), thus considering a total of~$110\%$ least-squares constraints. Varying the number of~$u$-values and \mbox{$\nabla u$}-vectors, we compute the meshless potential~$\tilde{u}$ (c.f., Eq.~(\ref{eq:MIXED-CONDITIONS})) and the approximation error \mbox{$\epsilon_{\infty}:=\|u-\tilde{u}\|_{\infty}/\|\tilde{u}\|_{\infty}$} ($y$-axis) between~$u$ and the ground-truth solution. We notice that the~$\mathcal{L}^{2}$-error is bounded as \mbox{$\|u-\tilde{u}\|_{2}\leq\epsilon_{\infty}\vert\Omega\vert$}, where~$\vert\Omega\vert$ is the measure (e.g., area, volume) of~$\Omega$. Then, we report the approximation accuracy~$\epsilon_{\infty}$ (Fig.~\ref{fig:NONANALYTIC-EXAMPLE}(c)) for the potential and the mean of the angle \mbox{$\angle(\nabla u,\nabla\tilde{u})$} (Fig.~\ref{fig:NONANALYTIC-EXAMPLE}(d)) between the input and the approximated vector fields, for a different percentage of constraints on function and vector values. Increasing the number of constraints on the~$u$-values from~$10\%$ to~$90\%$ (Fig.~\ref{fig:NONANALYTIC-EXAMPLE}(b),~$x$-axis), the approximation error ($y$-axis) remains below~$10^{-6}$, with an analogous behaviour with respect to the inverse multi-quadratic and exponential kernels and a slightly higher error with the multi-quadratic kernel. Increasing the number of constraints on the~$\nabla u$-values from~$10\%$ to~$90\%$ (Fig.~\ref{fig:NONANALYTIC-EXAMPLE}(d),~$x$-axis), the maximum angle ($y$-axis) remains below~$1.6$ degree, with an analogous behaviour with respect to the multi-quadratic and inverse multi-quadratic kernels and a slightly higher error with the Gaussian kernel. 

\begin{figure}[t]
\centering
\begin{tabular}{cc}
Irregular sampling of~$\Omega$ &Level-sets of~$u_{3}$\\
(a)\includegraphics[height=60pt]{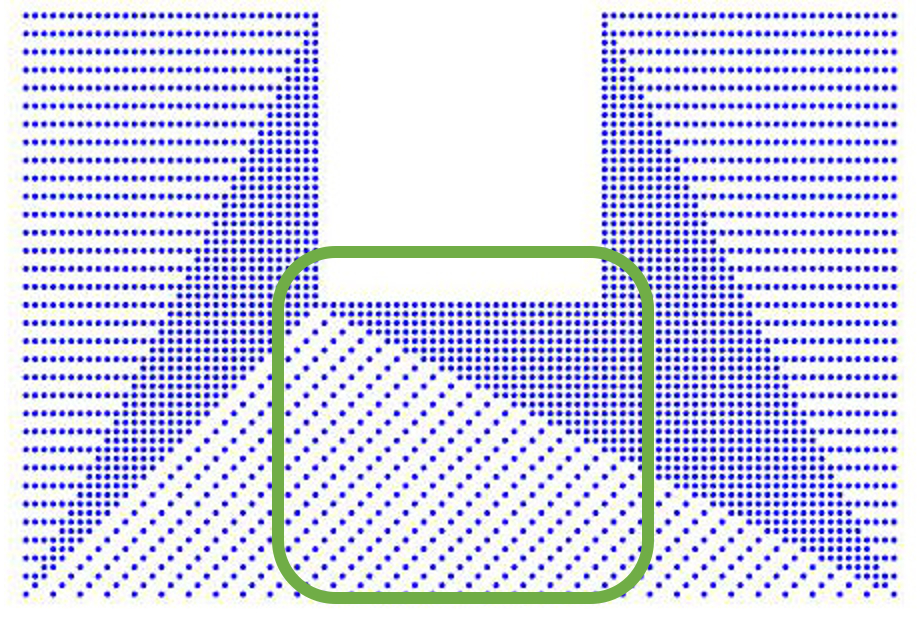}
&(b)\includegraphics[height=60pt]{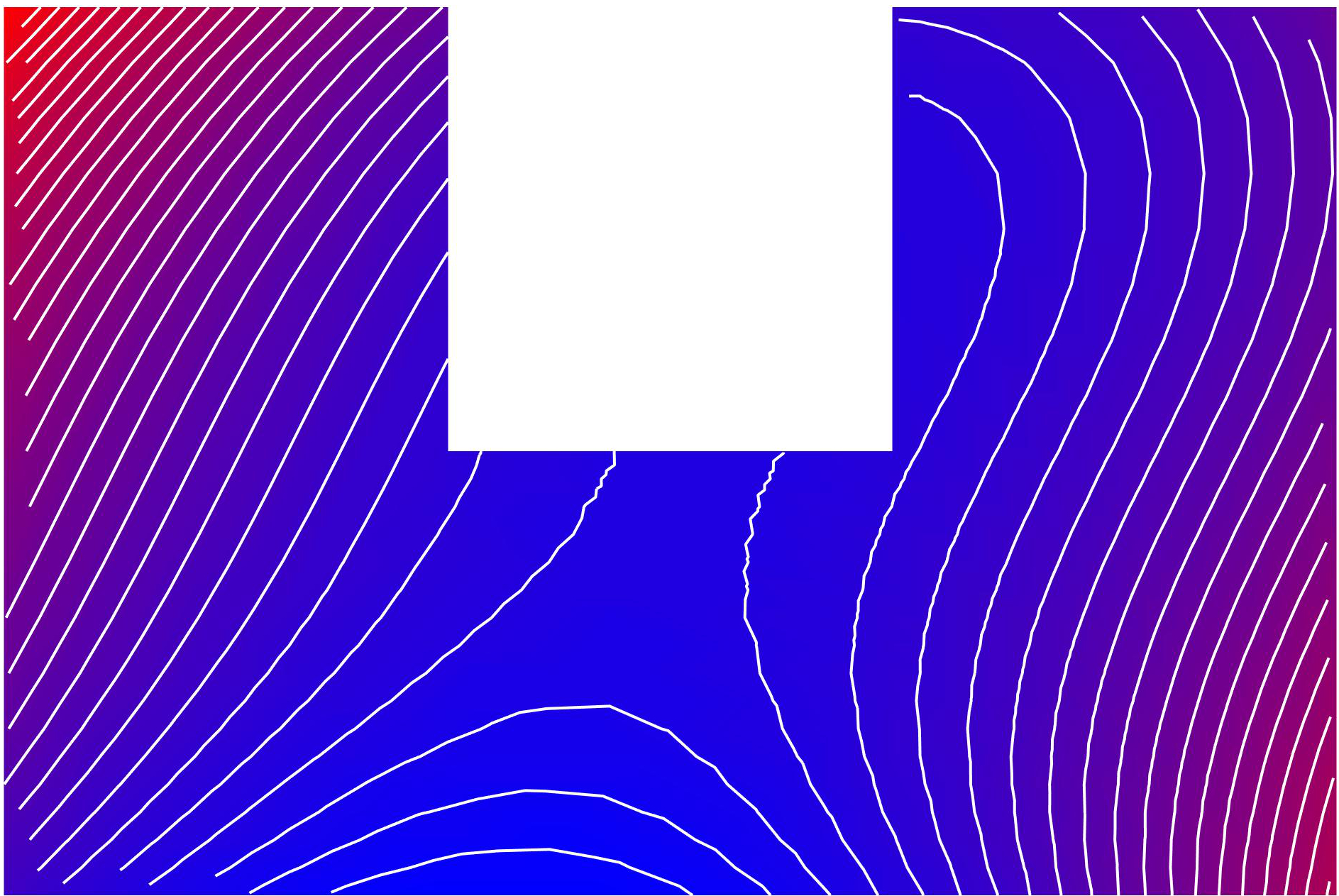}
\end{tabular}
\begin{tabular}{c|c|c}
\hline
\textbf{Gaussian Kern.} &\textbf{Mquad. Kern.} &\textbf{Inv. Mquad. Kern.}\\
\hline
\includegraphics[height=60pt]{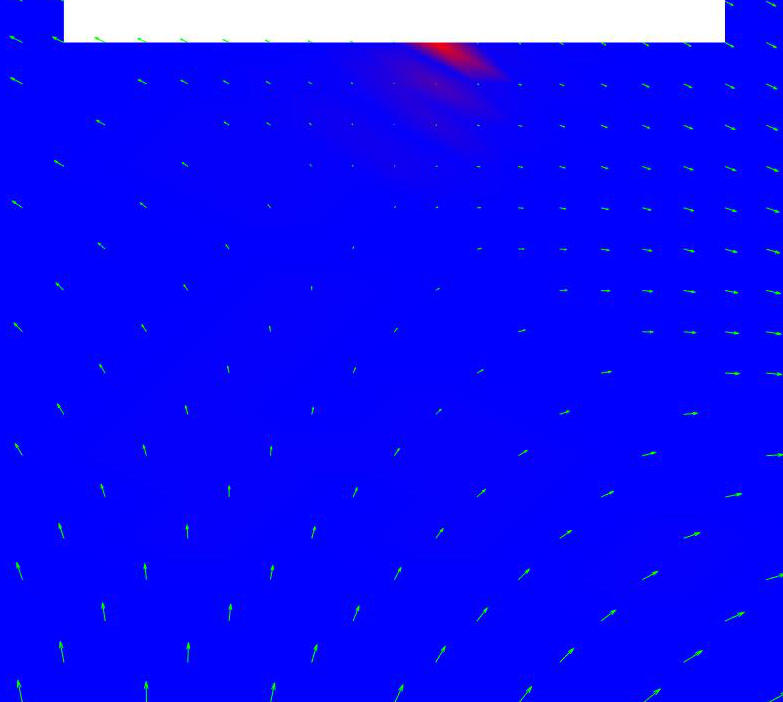}
&\includegraphics[height=60pt]{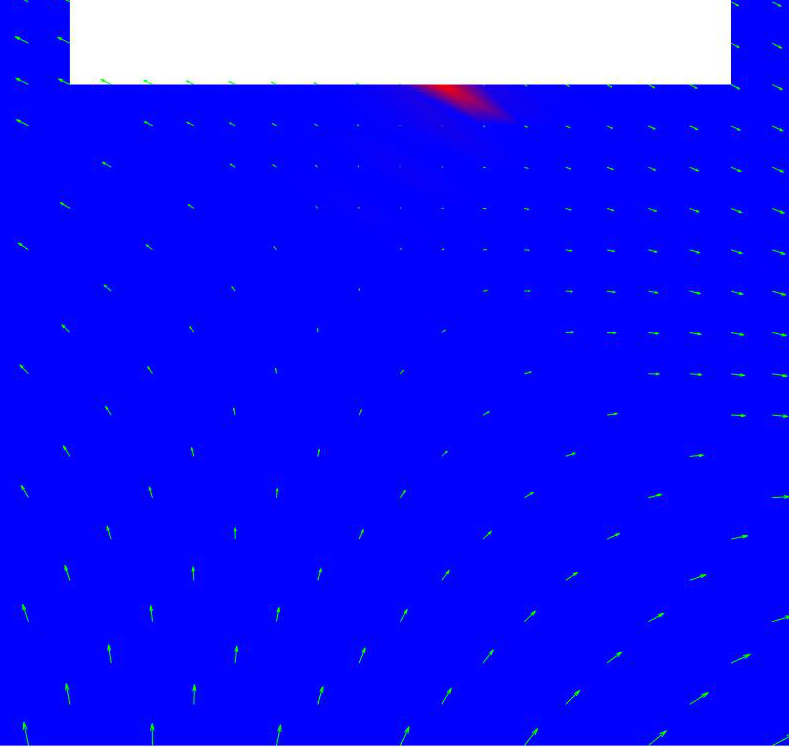}
&\includegraphics[height=60pt]{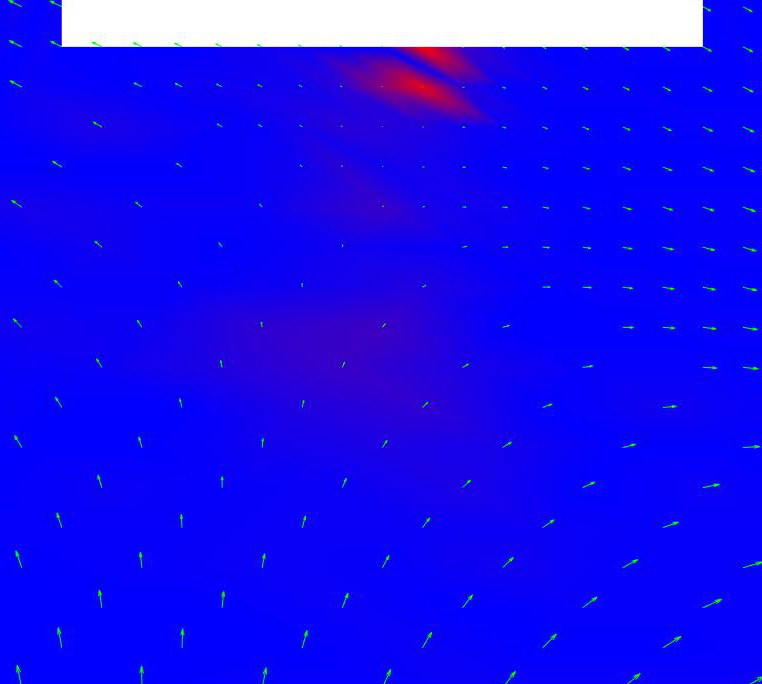}\\	
\multicolumn{3}{c}{$90\%$ Function constr.,\qquad~$20\%$ Vector constr.}\\
\hline
\includegraphics[height=60pt]{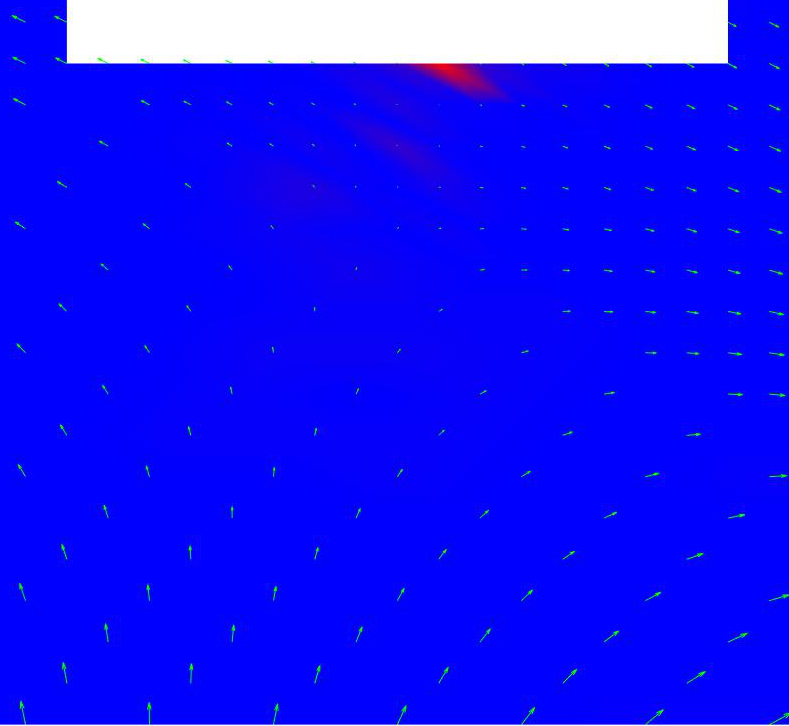}
&\includegraphics[height=60pt]{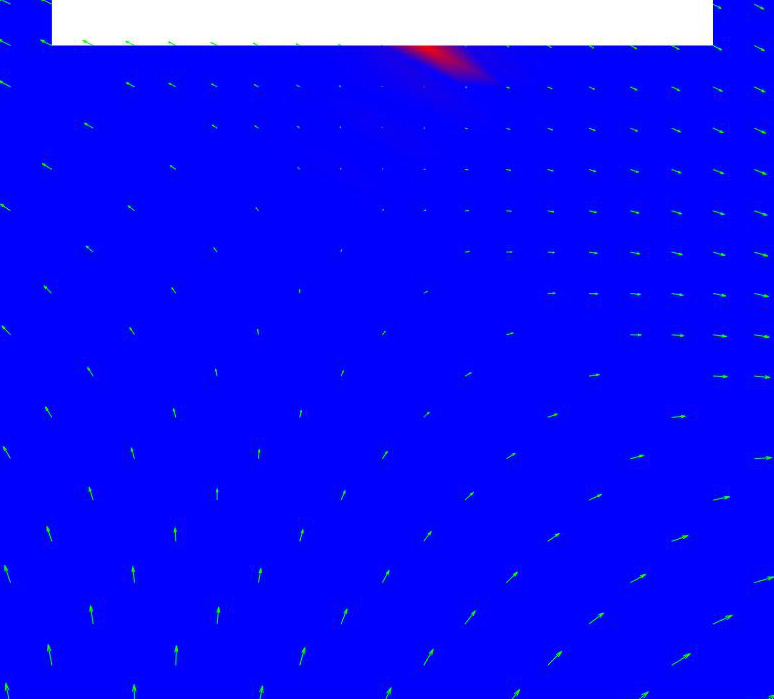}
&\includegraphics[height=60pt]{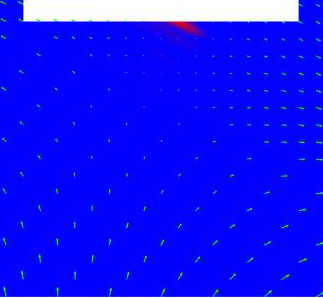}\\
\multicolumn{3}{c}{$20\%$ Function constr.,\qquad~$90\%$ Vector constr.}
\end{tabular}
\caption{1st Row: (a) input domain~$\Omega$ with an irregular distribution of samples and (b) level-sets of the potential~$u_{3}$ in Fig.~\ref{fig:ANALYTIC-STATISTICS}. (2nd, 3rd Rows) Distribution of the angle error of the meshless approximation of \mbox{$\nabla u_{3}$} induced by the Gaussian, multi-quadratic (mquad), and inverse multi-quadratic (inv. mquad) kernels, with a different percentage of randomly selected function and vector constrains. The error is localised in the (red) area close to the upper part of the boundary (green box) of~$\Omega$ and covers a larger area for the inverse multi-quadratic kernel. Blue identifies a null error and red corresponds to the maximum approximation error ($1.53\%$).\label{fig:ANALYTIC-ANGLE-ERROR}}
\end{figure}

In Fig.~\ref{fig:NONANALYTIC-EXAMPLE-ZOOM}, we show the streamlines of the gradient field and the distribution of the angle error, with~$90\%$ of constraints on~$u$-values and~$20\%$ of \mbox{$\nabla u$} constraints, or vice versa with~$20\%$ of constraints on~$u$-values and~$90\%$ of \mbox{$\nabla u$} constraints. Indeed, we impose a total of~$110\%$ least-squares constraints. Comparing the ground-truth vector field in Fig.~\ref{fig:NONANALYTIC-EXAMPLE}(b) with its meshless approximation in Fig.~\ref{fig:NONANALYTIC-EXAMPLE-ZOOM}, the approximation error and the maximum angle \mbox{$\angle(\mathbf{v},\nabla u)$} between the input~$\mathbf{v}$ and meshless \mbox{$\nabla u$} vector fields (Fig.~\ref{fig:NONANALYTIC-EXAMPLE}(c,d)) remain low and only small undulations of the streamlines are visible along the parabolic area where the input vector field is discontinuous. These local artifacts are present where the number of constraints on the gradient is maximum ($\mathcal{J}$:~$90\%$), and disappear as we increase the number of constraints on the values of the potential. Analogously to the tests in Figs.~\ref{fig:NONANALYTIC-EXAMPLE},~\ref{fig:NONANALYTIC-EXAMPLE-ZOOM}, in Fig.~\ref{fig:ANALYTIC-STATISTICS} we select a set of analytic functions and their gradient fields evaluated at a set of irregularly distributed samples (Fig.~\ref{fig:ANALYTIC-ANGLE-ERROR}(a)). The meshless potential induced by these three kernels has a good accuracy in terms of approximation error and maximum angles between the input and the meshless gradient fields. The Gaussian kernel generally provides the most accurate results. According to the angle variation (Fig.~\ref{fig:ANALYTIC-ANGLE-ERROR}, 2nd and 3rd rows), the maximum error is localised in those regions of~$\Omega$ where \mbox{$(u,\nabla u)$} have a complex behaviour and a partial information, e.g., at the domain boundary.

\section{Meshless decomposition\label{sec:MESHLESS-HH}}
Let us assume that the input vector field is known at an arbitrary set \mbox{$\mathcal{B}:=\{\mathbf{b}_{i}\}_{i=1}^{t}$} of points of the input domain. For instance, the vector field on each triangle or tetrahedron~$T_{i}$ is associated with the corresponding barycentre~$\mathbf{b}_{i}$. The output is a smooth approximation of the input vector field and its HHD, which can be re-sampled at any point, or used to analyse the behaviour of the underlying phenomenon. To this end, we introduce a direct (Sect.~\ref{sec:APPROX-DECOMP}), least-squares (Sect.~\ref{WEIGHTED-HH-DECOMPOSIION}), and Laplace-based (Sect.~\ref{eq:MESHLESS-HH-LAPLACE}) HHD with RBFs.
\begin{algorithm}[t]
\caption{Meshless HHD.\label{al:HH-DECOMPOSITION}}
{\small{\begin{algorithmic}[1]
\REQUIRE A discrete vector field \mbox{$\mathbf{v}:\mathcal{P}\rightarrow\mathbb{R}^{3}$}, with~$\mathcal{P}$ point set, a positive-definite kernel \mbox{$\phi:\mathbb{R}\rightarrow\mathbb{R}$}, and a set of centres \mbox{$\mathcal{C}:=\{\mathbf{c}_{i}\}_{i=1}^{k}$} with RBFs \mbox{$\phi_{i}(\mathbf{p}):=\phi(\|\mathbf{p}-\mathbf{c}_{i}\|_{2})$} (Table~\ref{tab:EXISTENCE}).
\ENSURE \textbf{Computation of the conservative component \mbox{$\nabla u$}.}
\STATE Compute the coefficient matrix in Eq.~(\ref{eq:GRAD-COEFF-MAT});  
\STATE Compute the coefficients~$\alpha$ as solution to Eq.~(\ref{eq:DIV-NORMAL-EQUATION});
\STATE Compute the potential~$u$ in Eq.~(\ref{eq:SCALAR-POTENTIAL});
\STATE Compute the conservative component \mbox{$\nabla u$} in Eq.~(\ref{eq:IMPLICIT-APPROX-GRADIENT}a).  
\ENSURE \textbf{Computation of the solenoidal component \mbox{$\nabla\wedge\mathbf{w}$}.}
\STATE Compute the anti-symmetric matrix in Eq.~(\ref{eq:ROTOR-COEFF-MAT});
\STATE Compute the coefficients~$\alpha$ as solution to Eq.~(\ref{eq:ROT-NORM-EQUATION});
\STATE Compute the potential~$\mathbf{w}$ in Eq.~(\ref{eq:VECTOR-POTENTIAL});
\STATE Compute the solenoidal component \mbox{$\nabla\wedge\mathbf{w}$} in Eq.~(\ref{eq:MESHLESS-ROTOR}).
\ENSURE \textbf{Computation of the harmonic component~$\mathbf{h}$.}
\STATE Compute \mbox{$\mathbf{h}\approx\mathbf{v}-(\nabla u+\nabla\wedge\mathbf{w})$}.
\end{algorithmic}}}
\end{algorithm} 
In the paper examples, the components of a 3D HHD are visualised by drawing the corresponding streamlines from a set of starting points, and the behaviour of the conservative potential is visualised through its iso-surfaces. 
\subsection{Meshless HHD\label{sec:APPROX-DECOMP}}
\begin{figure}[t]
\centering
\begin{tabular}{cc}
\includegraphics[height=85pt]{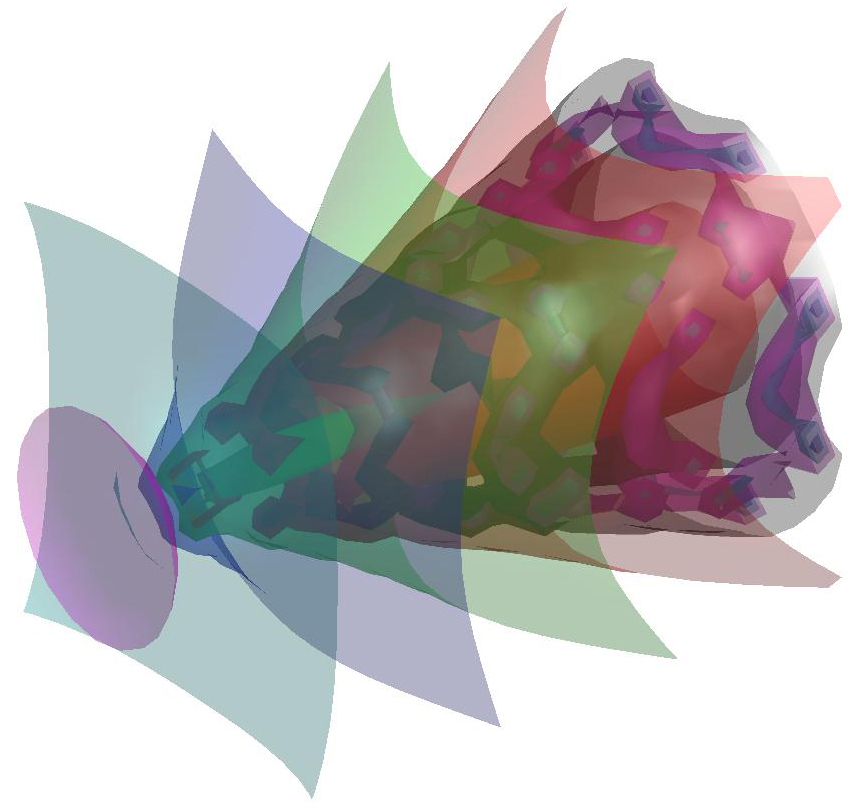}
&\includegraphics[height=85pt]{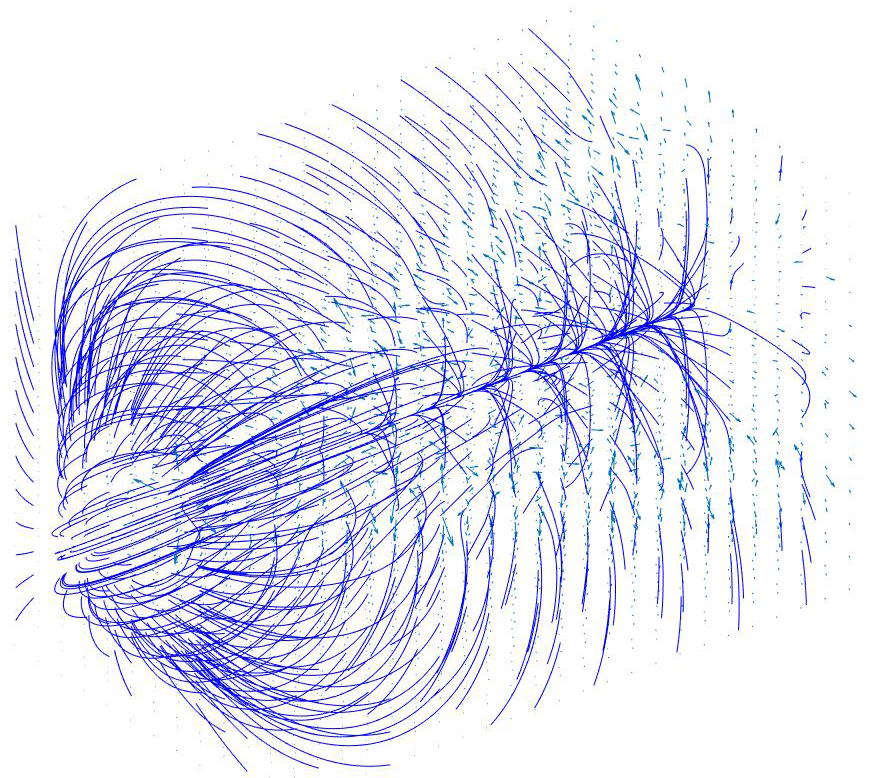}\\
(a)~$u$ &(b)~$\nabla u$\\
\includegraphics[height=85pt]{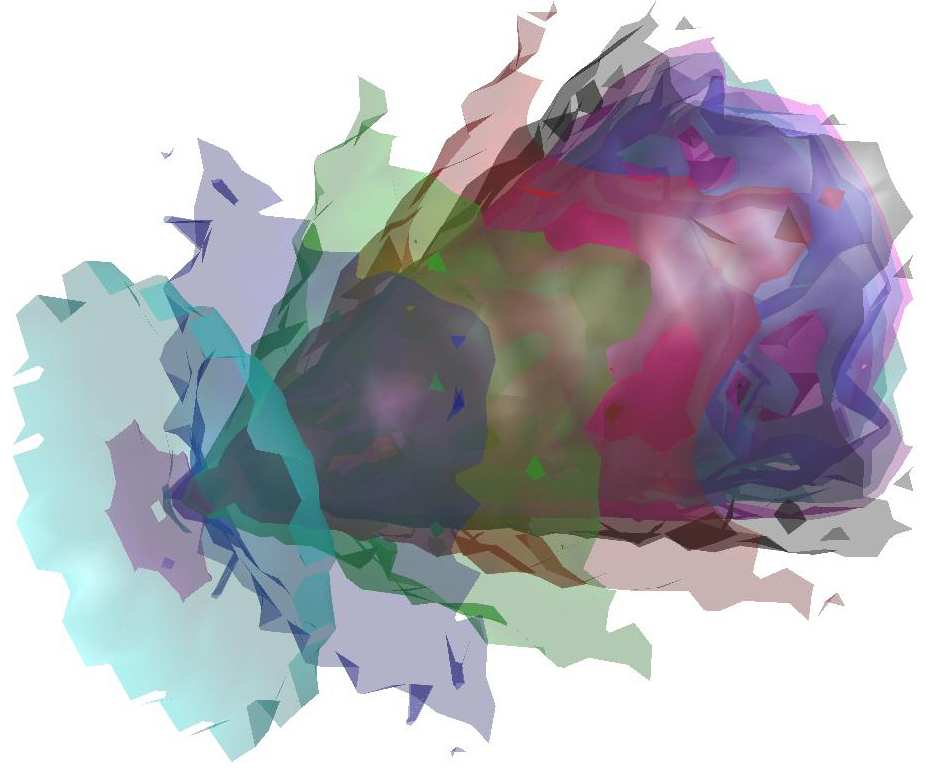}
&\includegraphics[height=85pt]{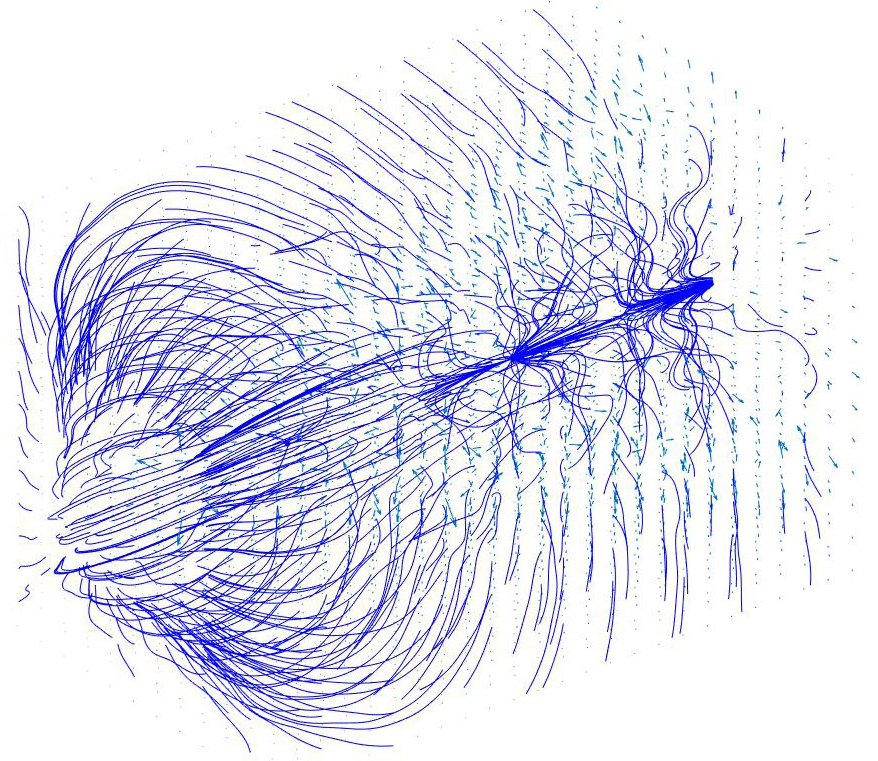}\\
(c)~$\tilde{u}$ &(d)~$\nabla\tilde{u}$
\end{tabular}
\caption{(a,c) Iso-surfaces of the meshless potential of the curl-free vector field in (b,d). Perturbing (b) with a~$25\%$ Gaussian noise, we compute the corresponding least-squares potential (c), whose gradient field is reported in (d). The~$\ell_{\infty}$ error between the ground-truth potential (a) and its approximation (c) is lower than~$1.2\%$. Different colours correspond to different iso-values.\label{fig:HH-FLOW}}
\end{figure}

\begin{figure}[t]
\centering
\begin{tabular}{cc}
\includegraphics[height=68pt]{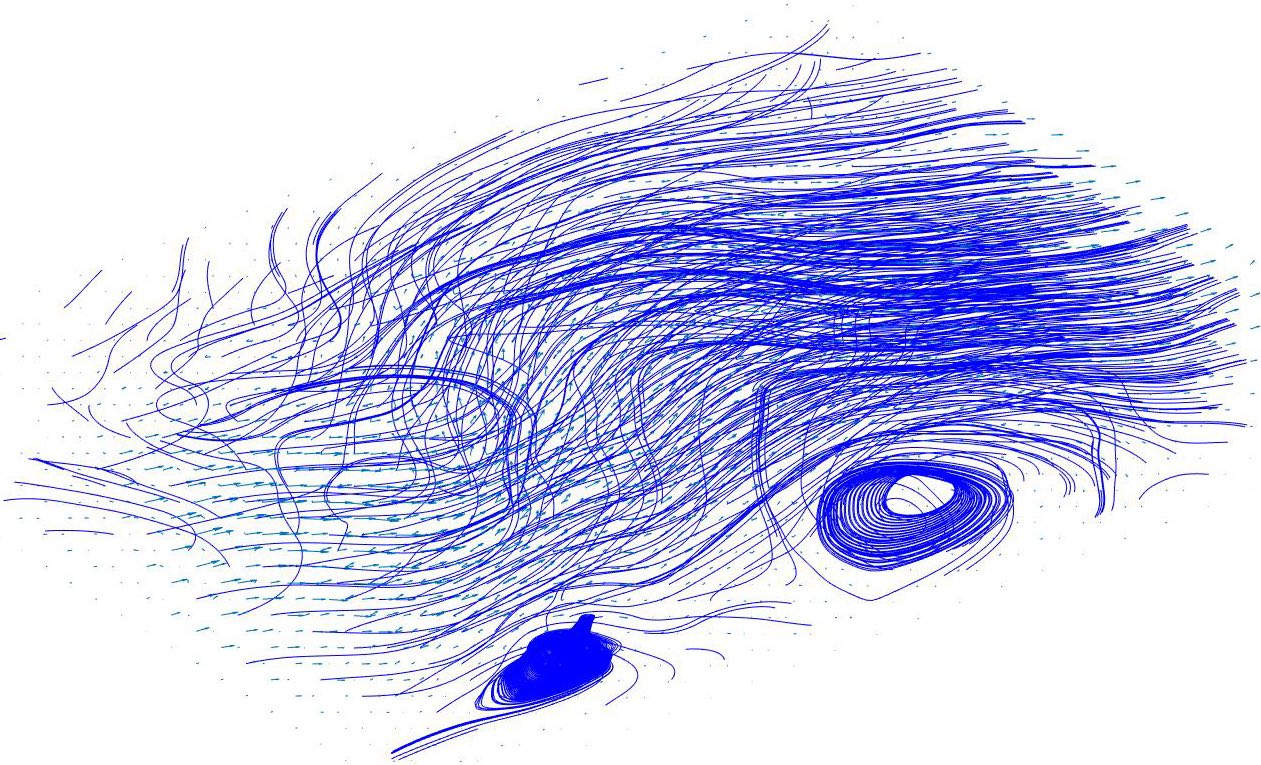}
&\includegraphics[height=68pt]{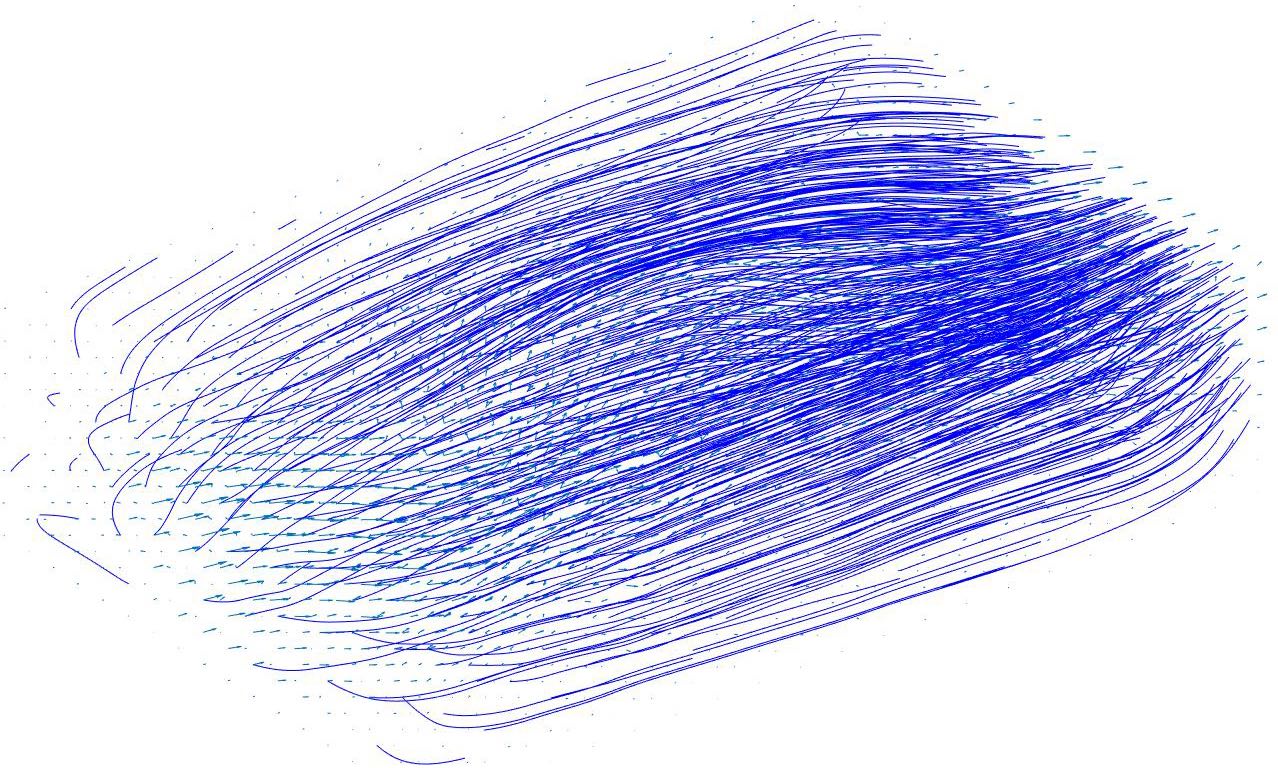}\\
(a)~$\mathbf{v}$ &(b)~$\nabla u$\\
\includegraphics[height=68pt]{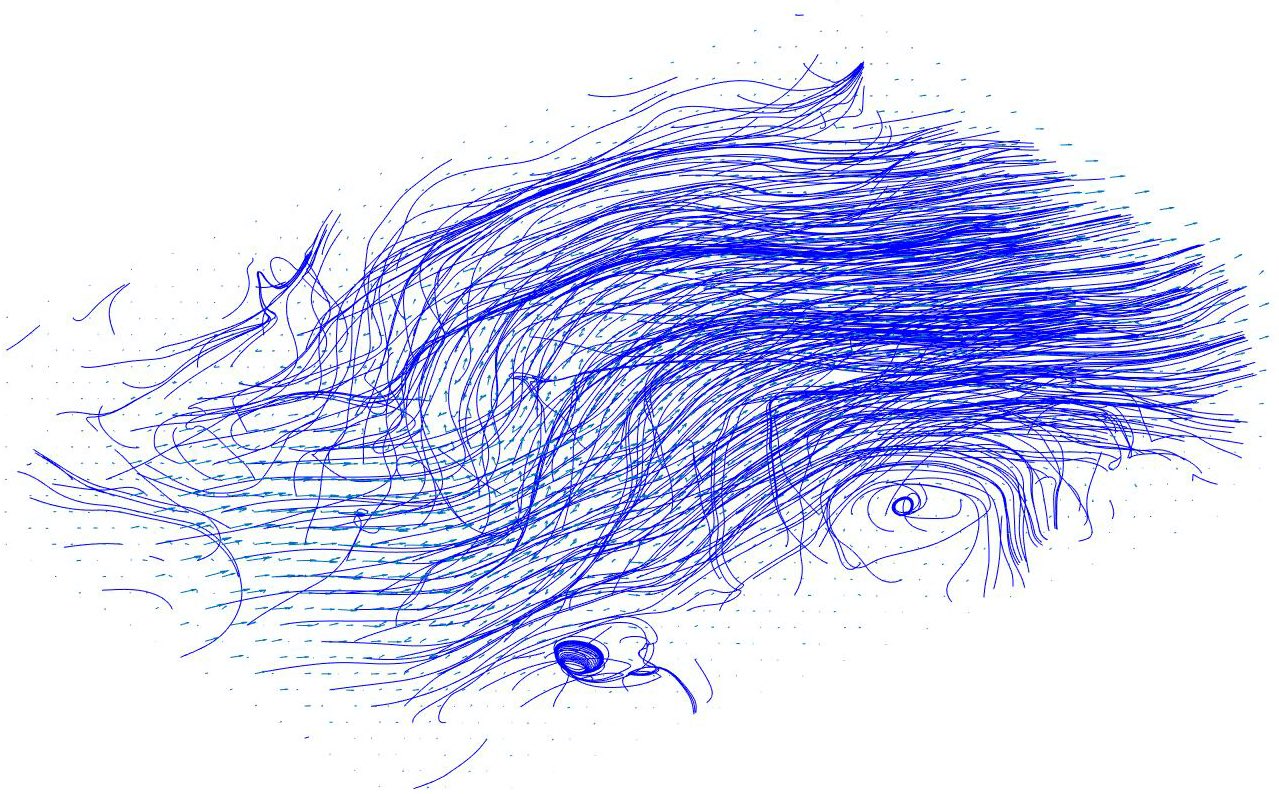}
&\includegraphics[height=68pt]{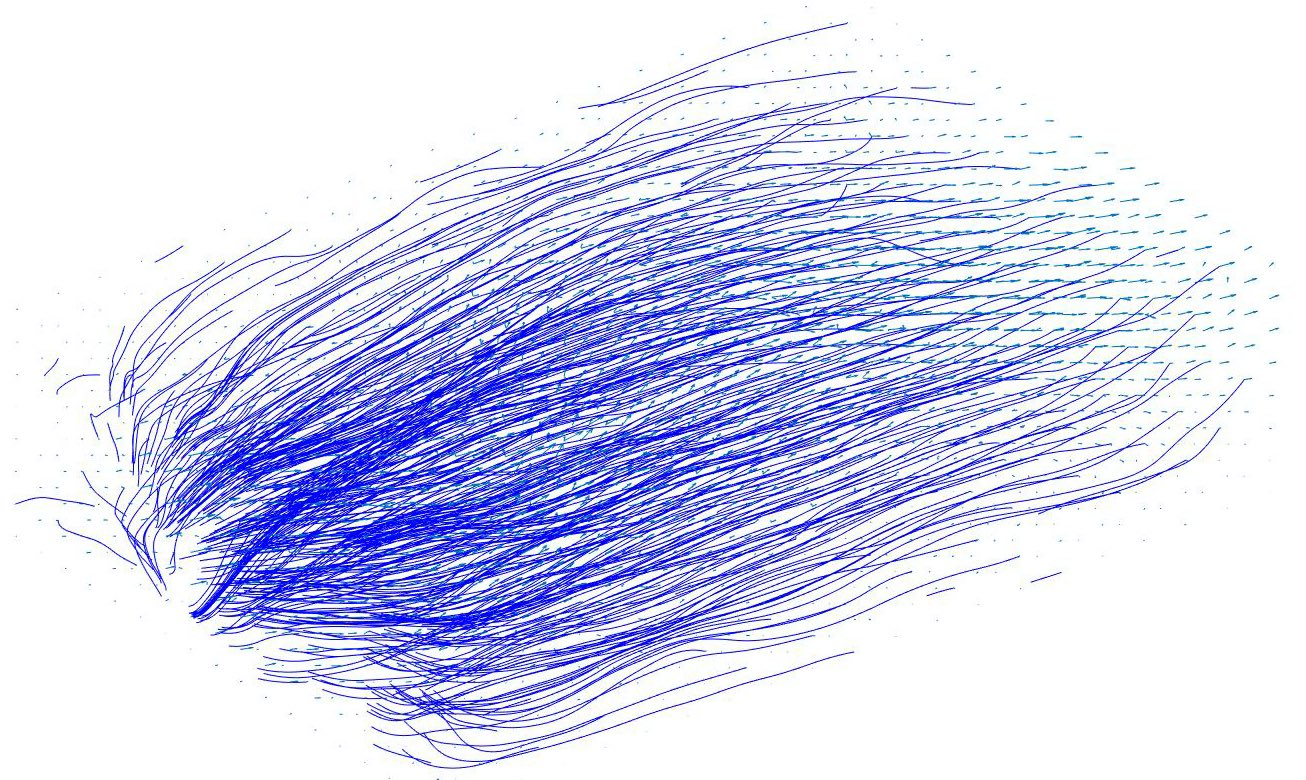}\\
(c)~$\nabla\wedge\mathbf{w}$ &(d)~$\mathbf{h}$
\end{tabular}
\caption{(a) Input field, (b-d) streamlines of the curl-free, div-free, and harmonic components of the HHD. 
\label{fig:WIND}}
\end{figure}

The \emph{meshless potential of the conservative component}
\begin{equation}\label{eq:SCALAR-POTENTIAL}
u(\mathbf{p})=\sum_{i=1}^{k}\alpha_{i}\phi_{i}(\mathbf{p}),\quad
\phi_{i}(\mathbf{p}):=\phi(\|\mathbf{p}-\mathbf{c}_{i}\|_{2}),
\end{equation}
is represented as a linear combination of RBFs \mbox{$(\phi_{i
})_{i}$} generated by a positive-definite kernel \mbox{$\phi:\mathbb{R}\rightarrow\mathbb{R}$} and centred at \mbox{$\mathcal{C}:=\{\mathbf{c}_{i}\}_{i=1}^{k}$}~\cite{DYN1986,MICCHELLI1986,WENDLAND1995}. The set~$\mathcal{C}$ is generally different from~$\mathcal{P}$ and its selection will be addressed in Sect.~\ref{sec:KERNEL-SELECTION}. Applying the linearity of the gradient operator, we get that 
\begin{equation}\label{eq:IMPLICIT-APPROX-GRADIENT}
\centering
\left\{
\begin{array}{ll}
\mathbf{v}(\mathbf{p})
\approx\nabla u(\mathbf{p})
=\sum_{i=1}^{k}\alpha_{i}\varphi_{i}(\mathbf{p}),	&(a)\\
\varphi_{i}(\mathbf{p}):=\nabla\phi_{i}(\mathbf{p})=\phi^{\prime}_{i}(\mathbf{p})\frac{\mathbf{p}-\mathbf{c}_{i}}{\|\mathbf{p}-\mathbf{c}_{i}\|_{2}},	&(b)
\end{array}
\right.
\end{equation}
where the basis field \mbox{$\{\varphi_{i}\}_{i}$} is centred at~$\mathbf{c}_{i}$, has length \mbox{$\vert\phi^{\prime}(\|\mathbf{p}-\mathbf{c}_{i}\|_{2})\vert$}, points towards the centre~$\mathbf{c}_{i}$, and is radially symmetric. Then, we impose the conditions \mbox{$\nabla u(\mathbf{b}_{i})\approx\mathbf{v}_{i}$}, \mbox{$i=1,\ldots,t$}, by minimising the corresponding least-squares error \mbox{$\sum_{i=1}^{t}\|\nabla u(\mathbf{b}_{i})-\mathbf{v}_{i}\|_{2}^{2}$}. Deriving this error with respect to the coefficients, we get the normal equation 
\begin{equation}\label{eq:DIV-NORMAL-EQUATION}
\Phi\alpha=\tilde{\mathbf{v}},
\,
\Phi:=
\left[
\begin{array}{l}
\Phi_{1}\\
\Phi_{2}\\
\Phi_{3}
\end{array}
\right]
\in\mathbb{R}^{3t\times k},
\, \tilde{\mathbf{v}}=
\left[
\begin{array}{l}
v_{x}\\
v_{y}\\
v_{z}
\end{array}
\right]\in\mathbb{R}^{3t},
\end{equation}
where the building blocks of the coefficient matrix are
\begin{equation}\label{eq:GRAD-COEFF-MAT}
\Phi_{i}:=
\left[
\begin{array}{llll}
\varphi_{1}^{x_{i}}(\mathbf{b}_{1})	&\varphi_{2}^{x_{i}}(\mathbf{b}_{1})	&\ldots		&\varphi_{k}^{x_{i}}(\mathbf{b}_{1})\\
\varphi_{1}^{x_{i}}(\mathbf{b}_{2})	&\varphi_{2}^{x_{i}}(\mathbf{b}_{2})	&\ldots		&\varphi_{k}^{x_{i}}(\mathbf{b}_{2})\\
\vdots								&\vdots									&\vdots		&\vdots\\
\varphi_{1}^{x_{i}}(\mathbf{b}_{t})	&\varphi_{2}^{x_{i}}(\mathbf{b}_{t})	&\ldots		&\varphi_{k}^{x_{i}}(\mathbf{b}_{t})\\
\end{array}
\right]
\in\mathbb{R}^{t\times k},
\end{equation}
\mbox{$i=1,2,3$},~$\tilde{\mathbf{v}}$ is the array of the input vector data, and \mbox{$\varphi_{i}^{x_{j}}(\mathbf{b}_{l})$} is the~$j$-th component of \mbox{$\varphi_{i}(\mathbf{b}_{l})$}. If the number of basis fields is lower than the number of samples (i.e., \mbox{$k<n$}), then we solve the \mbox{$k\times k$} least-squares system \mbox{$\Phi^{\top}\Phi\alpha=\Phi^{\top}\tilde{\mathbf{v}}$}. Since the generating kernel is positive-definite, the solution~$\alpha$ and the resulting potential are unique.
\begin{figure}[t]
\centering
\begin{tabular}{cc}
\includegraphics[height=90pt]{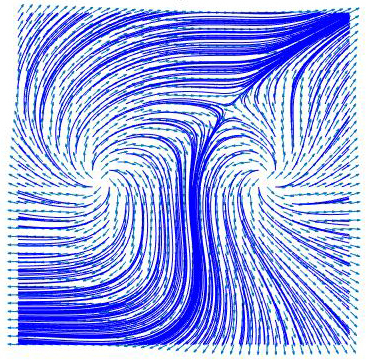}
&\includegraphics[height=90pt]{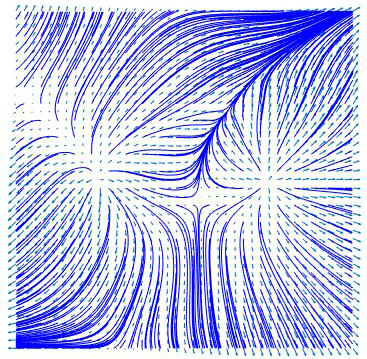}\\
(a)~$\mathbf{v}$ &(b)~$\nabla u$\\
\includegraphics[height=90pt]{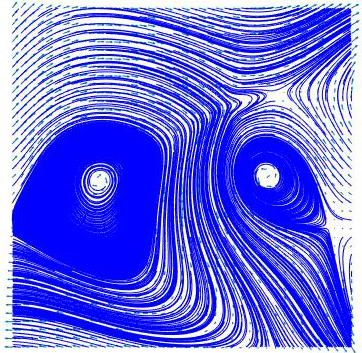}
&\includegraphics[height=90pt]{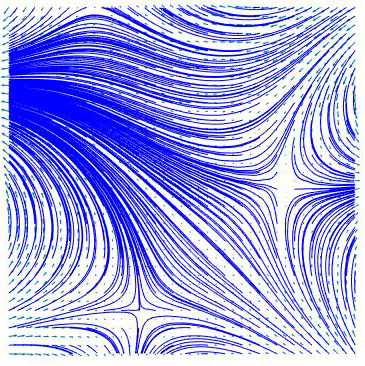}\\
(c)~$\nabla\wedge\mathbf{w}$ &(d)~$\mathbf{h}$
\end{tabular}
\caption{Meshless decomposition \mbox{$\mathbf{v}=\nabla u+\nabla\wedge\mathbf{w}+\mathbf{h}$}: (a) input field, (b) curl-free, (c) div-free, and (d) harmonic component.\label{fig:eFUN}}
\end{figure}
Each component of the potential \mbox{$\mathbf{w}(\mathbf{p})$} of~$\mathbf{v}$ is expressed in terms of the basis \mbox{$(\phi_{i}(\mathbf{p}))_{i=1}^{k}$} as
\begin{equation}\label{eq:VECTOR-POTENTIAL}
\mathbf{w}(\mathbf{p})=
\left[
\sum_{i=1}^{k}\alpha_{i}^{(1)}\phi_{i}(\mathbf{p}), \sum_{i=1}^{k}\alpha_{i}^{(2)}\phi_{i}(\mathbf{p}), \sum_{i=1}^{k}\alpha_{i}^{(3)}\phi_{i}(\mathbf{p})
\right],
\end{equation}
with~$3k$ unknowns \mbox{$\alpha^{(j)}:=(\alpha_{i}^{(j)})_{i=1}^{k}$}, \mbox{$j=1,2,3$}. Then, we minimise the least-squares error \mbox{$\mathcal{G}(\mathbf{w})=\|\nabla\wedge\-\mathbf{w}-\mathbf{v}\|_{2}^{2}$}, between the rotor
\begin{equation}\label{eq:MESHLESS-ROTOR}
\begin{split}
&\nabla\wedge\mathbf{w}
=\left[\sum_{i=1}^{k}(\alpha_{i}^{(3)}\partial_{y}\phi_{i}-\alpha_{i}^{(2)}\partial_{z}\phi_{i},\right.\\
&\sum_{i=1}^{k}(\alpha_{i}^{(1)}\partial_{z}\phi_{i}-\alpha_{i}^{(3)}\partial_{x}\phi_{i},
\left.\sum_{i=1}^{k}(\alpha_{i}^{(2)}\partial_{x}\phi_{i}-\alpha_{i}^{(1)}\partial_{y}\phi_{i}\right]
\end{split}
\end{equation}
and the input vector field. The corresponding least-squares problem is rewritten in matrix form as \mbox{$\|\mathbf{A}\alpha-\tilde{\mathbf{v}}\|_{2}$}, where the \mbox{$3t\times 3k$} anti-symmetric coefficient matrix
\begin{equation}\label{eq:ROTOR-COEFF-MAT}
\begin{array}{l}
\mathbf{A}=\left[
\begin{array}{l|l|l}
\mathbf{0}					&-\partial_{z}\phi			&\partial_{y}\phi\\
\hline
\partial_{z}\phi			&\mathbf{0}					&-\partial_{x}\phi\\
\hline
-\partial_{y}\phi			&\partial_{x}\phi			&\mathbf{0}
\end{array}
\right],\quad
\mathbf{0}\in\mathbb{R}^{t\times k},
\end{array}
\end{equation}
has \mbox{$3kt$} non-null elements and the blocks \mbox{$\partial_{y}\phi$}, \mbox{$\partial_{z}\phi$} are defined analogously to \mbox{$\partial_{x}\phi:=(\partial_{x}\phi_{j}(\mathbf{p}_{i}))_{i=1,\ldots,t}^{j=1,\ldots,k}\in\mathbb{R}^{t\times k}$}. The entries in Eq.~(\ref{eq:ROTOR-COEFF-MAT}) are computed by applying the relation in Eq.~(\ref{eq:IMPLICIT-APPROX-GRADIENT}b). Then, the coefficients solve the linear system
\begin{equation}\label{eq:ROT-NORM-EQUATION}
\mathbf{A}^{\top}\mathbf{A}\alpha=\mathbf{A}^{\top}\tilde{\mathbf{v}},\,
\alpha=\left[
\begin{array}{l}
\alpha^{(1)}\\\alpha^{(2)}\\\alpha^{(3)}
\end{array}
\right]\in\mathbb{R}^{3k},\,
\tilde{\mathbf{v}}=\left[
\begin{array}{l}
v_{x}\\ 
v_{y}\\
v_{z}
\end{array}
\right]\in\mathbb{R}^{3n},
\end{equation}
where~$v_{x}$,~$v_{y}$, and~$v_{z}$ are the components of~$\mathbf{v}$. The solution to the linear systems in Eqs.~(\ref{eq:DIV-NORMAL-EQUATION}), (\ref{eq:ROT-NORM-EQUATION}) is computed through the conjugate gradient if the input matrix is positive-definite or iterative solvers of sparse and symmetric linear systems if the coefficient matrix is positive semi-definite. To evaluate the derivatives in Eqs.~(\ref{eq:GRAD-COEFF-MAT}), (\ref{eq:ROTOR-COEFF-MAT}), we apply the derivative of composite functions, as detailed in Sect.~\ref{sec:WELL-POSEDNESS}. The meshless representation of the potentials~$u$,~$\mathbf{w}$ induces the meshless approximation \mbox{$\mathbf{v}\approx\nabla u+\nabla\wedge\mathbf{w}+\mathbf{h}$} of the input vector field and of the harmonic component \mbox{$\mathbf{h}\approx\mathbf{v}-\nabla u-\nabla\wedge\mathbf{w}$}.
\begin{figure}[t]
\centering
\begin{tabular}{cc}
\includegraphics[height=74pt,width=114pt]{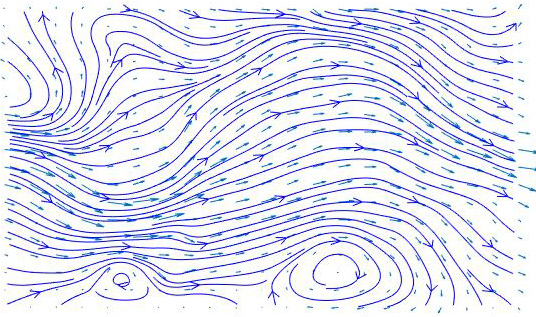}
&\includegraphics[height=74pt,width=114pt]{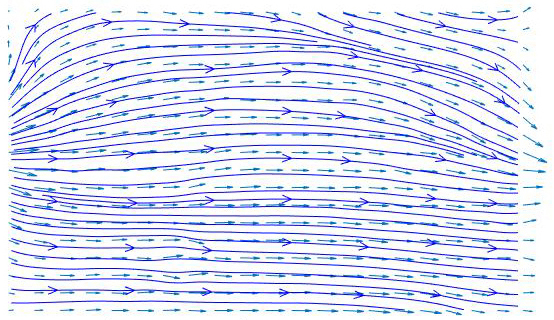}\\
(a)~$\mathbf{v}$ &(b)~$\nabla u$\\
\includegraphics[height=74pt,width=114pt]{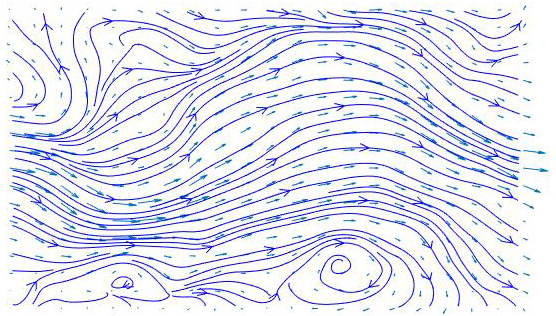}
&\includegraphics[height=74pt,width=114pt]{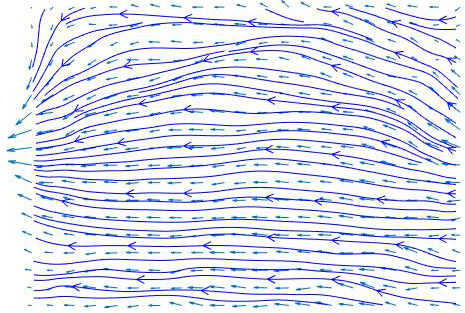}\\
(c)~$\nabla\wedge\mathbf{w}$ &(d)~$\mathbf{h}$\\
\end{tabular}
\caption{HHD: (a) input vector field, (b) curl-free, (c) div-free, and (d) harmonic component.\label{fig:WIND13}}
\end{figure}
The HHD of 2D and 3D vector fields is shown in Figs.~\ref{fig:eFUN},~\ref{fig:WIND13} and Figs.~\ref{fig:HH-FLOW},~\ref{fig:WIND}, respectively. In Fig.~\ref{fig:HH-FLOW}, a perturbation of~$25\%$ Gaussian noise of the input vector field corresponds to a~$\ell_{\infty}$ error of~$1.2\%$ between the ground-truth and the meshless potentials. The multi-quadratic, Gaussian ($1.51\%$), and inverse multi-quadratic ($1.57\%$) kernels provide analogous results. Finally, the iso-surfaces and streamlines of the meshless potential confirm that it preserves the global behaviour of the input data, removes the noise, and guarantees a good approximation accuracy.
\begin{table*}[t]
\caption{Generating kernels for RBFs, first and second order derivatives, existence of the gradient and the Hessian of the RBFs.\label{tab:EXISTENCE}}
\centering
\resizebox{1.0\linewidth}{!}{
\begin{tabular}{|l|l|l|l|l|l|}
\hline
\textbf{Global Kernels}			&$\phi(r,\sigma)$									&\textbf{First order derivative -~$\phi^{\prime}(r,\sigma)$}				&\textbf{Second order derivative -~$\phi^{\prime\prime}(r,\sigma)$}															&\textbf{$\exists$ Grad.}			&\textbf{$\exists$ Hess.}\\
\hline
Cubic					&$\sigma r^3$										&$3\sigma r^{2}$				&$6\sigma r$																&Yes ($\sigma=0$)					&Yes\\
\hline
Gaussian		&$\exp(-\sigma r^{2})$										&$-2\sigma r\exp(-\sigma r^{2})$&$-2\sigma(1-2\sigma r^{2})\exp(-\sigma r^{2})$								&Yes								&Yes\\
\hline
Thin Plate Spline		&$r^{2}\log(\sigma r)$								&$2r\log(\sigma r)+r$			&$2(\log(\sigma r)+2)$														&No									&No\\
\hline
Inv. multi-quad.		&$(r^{2}+\sigma^{2})^{-1/2}$						&$-r(r^{2}+\sigma^{2})^{-3/2}$  &$(r^{2}+\sigma^{2})^{-3/2}(3(r^{2}+\sigma^{2})^{-1}r^{2}-1)$				&Yes								&Yes ($\sigma\neq 0$)\\
\hline
Multi-quad.				&$(r^{2}+\sigma^{2})^{1/2}$	($\sigma\neq 0$)		&$r(r^{2}+\sigma^{2})^{-1/2}$	&$\sigma^{2}(r^{2}+\sigma^{2})^{-3/2}$										&Yes								&Yes\\
\hline
\textbf{Local Kernels}			&									&				&															&			&\\
\hline
Local II-deg. polyn.			&$(1-\sigma)^{2}_{+}$									&$-2(1-\sigma)_{+}$																&2																															&Yes								&Yes\\
\hline
Local IV-deg. polyn.			&$(1-\sigma)^{4}_{+}(4\sigma+1)$						&$-20\sigma(1-\sigma)_{+}^{3}$																	&$-20(1-\sigma)^{2}_{+}(1-4\sigma)$																											&Yes							&Yes\\
\hline
\end{tabular}}
\end{table*}
\subsection{Weighted meshless decomposition\label{WEIGHTED-HH-DECOMPOSIION}}
\begin{figure}[t]
\centering
\begin{tabular}{cc}
\includegraphics[height=70pt,width=110pt]{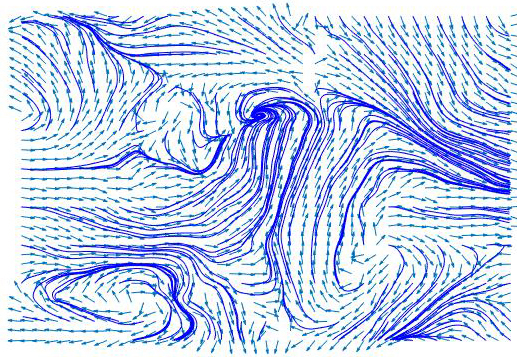}
&\includegraphics[height=70pt,width=110pt]{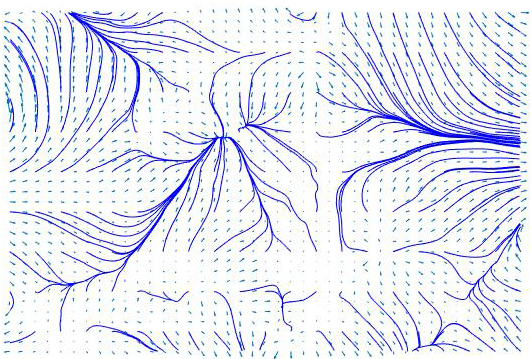}\\
(a)~$\mathbf{v}$ &(b)~$\nabla u$\\
\includegraphics[height=70pt,width=110pt]{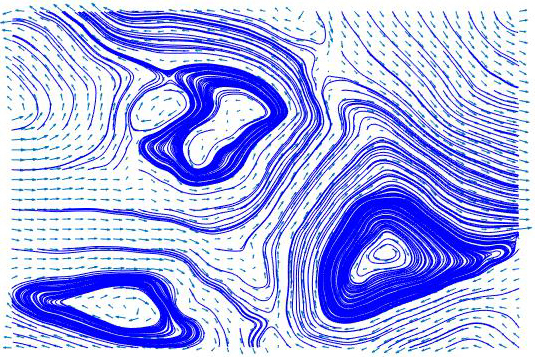}
&\includegraphics[height=70pt,width=110pt]{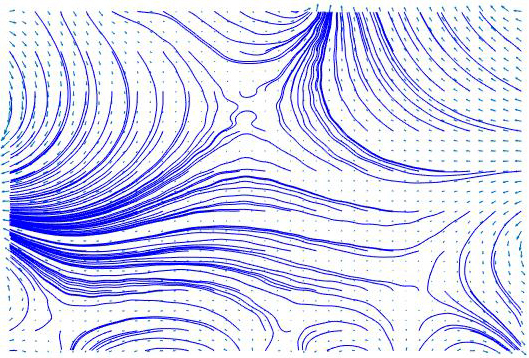}\\
(c)~$\nabla\wedge\mathbf{w}$ &(d)~$\mathbf{h}$
\end{tabular}
\caption{Least-squares HHD: (a) input field, (b) curl-free, (c) div-free, and (d) harmonic components.\label{fig:WIND2D}}
\end{figure}

\begin{figure}[t]
\centering
\begin{tabular}{cc}
\includegraphics[height=70pt]{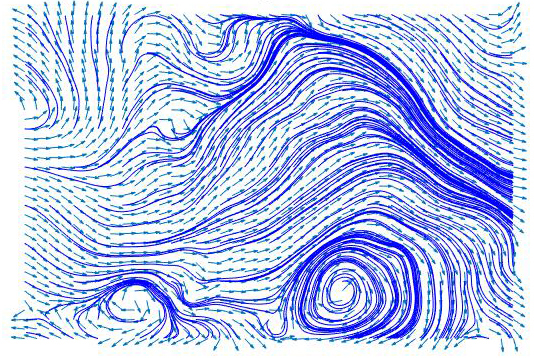}
&\includegraphics[height=70pt]{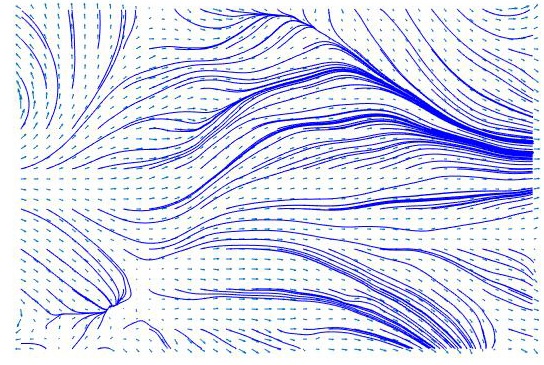}\\
(a)~$\mathbf{v}$ &(b)~$\nabla u$\\
\includegraphics[height=70pt]{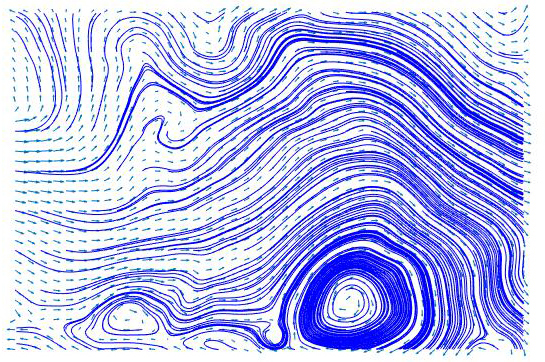}
&\includegraphics[height=70pt]{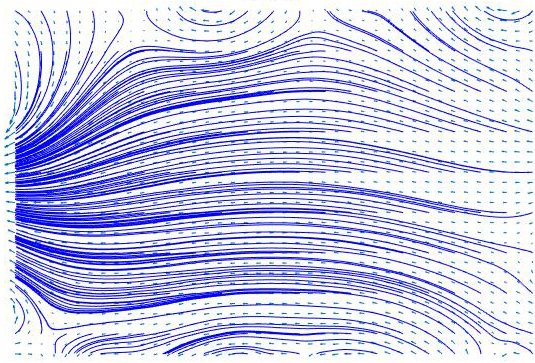}\\
(c)~$\nabla\wedge\mathbf{w}$ &(d)~$\mathbf{h}$
\end{tabular}
\caption{Least-squares HHD: (a) input field, (b) curl-free, (c) div-free, and (d) harmonic component.\label{fig:WIND2D-SECOND}}
\end{figure}

Alternatively, we consider the meshless approximation \mbox{$u(\mathbf{p})=\sum_{i=1}^{k}\alpha_{i}\phi_{i}(\mathbf{p})$} of the potential of the \emph{conservative component} \mbox{$\nabla u$} of~$\tilde{\mathbf{v}}$ by minimising the weighted least-squares~\cite{TONG2003} instead of the pointwise energy 
\begin{equation*}\label{eq:CURL-FREE}
\begin{split}
&\mathcal{F}:=\frac{1}{2}\int_{\Omega}\|\nabla u-\tilde{\mathbf{v}}\|_{2}^{2}\textrm{d}\mathbf{p}
=-\sum_{i=1}^{k}\alpha_{i}\int_{\Omega}\langle\nabla\phi_{i}(\mathbf{p}),\tilde{\mathbf{v}}\rangle_{2}\textrm{d}\mathbf{p}+\\
&+\frac{1}{2}\int_{\Omega}\|\tilde{\mathbf{v}}\|_{2}^{2}\textrm{d}\mathbf{p}+\frac{1}{2}\sum_{i,j=1}^{k}\alpha_{i}\alpha_{j}\int_{\Omega}\langle\nabla\phi_{i}(\mathbf{p}),\nabla\phi_{j}(\mathbf{p})\rangle_{2}\textrm{d}\mathbf{p}.
\end{split}
\end{equation*}
The derivatives of~$\mathcal{F}$ with respect to \mbox{$\alpha:=(\alpha_{i})_{i=1}^{k}$} are
\begin{equation*}
\partial_{\alpha}\mathcal{F}
=\sum_{i=1}^{k}\alpha_{i}\int_{\Omega}\langle\nabla\phi_{i}(\mathbf{p}),\nabla\phi_{j}(\mathbf{p})\rangle_{2}\textrm{d}\mathbf{p}
-\int_{\Omega}\langle\nabla\phi_{j}(\mathbf{p}),\tilde{\mathbf{v}}\rangle_{2}\textrm{d}\mathbf{p},
\end{equation*}
and the equation \mbox{$\partial_{\alpha}\mathcal{F}=\mathbf{0}$} reduces to the linear system
\begin{equation}\label{eq:DIV-FREE}
\left\{
\begin{array}{l}
\mathbf{A}\alpha=\mathbf{b},\quad A(i,j):=\int_{\Omega}\langle\nabla\phi_{i}(\mathbf{p}),\nabla\phi_{j}(\mathbf{p})\rangle_{2}\textrm{d}\mathbf{p};\\
b(i):=\int_{\Omega}\langle\nabla\phi_{i}(\mathbf{p}),\tilde{\mathbf{v}}\rangle_{2}\textrm{d}\mathbf{p}.
\end{array}
\right.
\end{equation}
The least-squares approximation requires only to evaluate the RBFs at the input points, while the weighted least-squares approximation needs an underlying connectivity for the evaluation of the integral. Indeed, the least-squares approximation is particularly useful for the computation of the potential of vector fields on point sets. 
For the \emph{solenoidal component}, we minimise the error \mbox{$\mathcal{H}(\mathbf{w})=\frac{1}{2}\int_{\Omega}\mathcal{G}(\mathbf{w})\textrm{d}\mathbf{p}=\frac{1}{2}\int_{\Omega}\|\nabla\wedge\mathbf{w}-\tilde{\mathbf{v}}\|_{2}^{2}\textrm{d}\mathbf{p}$}, where \mbox{$\mathcal{G}(\mathbf{w})$} and \mbox{$\nabla\wedge\mathbf{w}$} in Eq. (\ref{eq:MESHLESS-ROTOR}) are defined in Sect.~\ref{sec:APPROX-DECOMP}. Then, the minimum is achieved by solving the normal equation \mbox{$\partial_{\alpha}\mathcal{H}$}, where the derivatives are computed with respect to the coefficients \mbox{$\alpha_{k}^{(1)},\alpha_{k}^{(2)},\alpha_{k}^{(2)}$}. In particular,
\begin{equation*}
\begin{split}
\partial_{\alpha_{k}^{(1)}}\mathcal{H}
&=\int_{\Omega}\left[\left(\sum_{i=1}^{n}(\alpha_{i}^{(1)}\partial_{z}\phi_{i}-\alpha_{i}^{(3)}\partial_{x}\phi_{i})-w^{y}\right)\partial_{z}\phi_{k}+\right.\\
&+\left.\left(\sum_{i=1}^{n}(\alpha_{i}^{(2)}\partial_{x}\phi_{i}-\alpha_{i}^{(1)}\partial_{y}\phi_{i})-w^{z}\right)\partial_{y}\phi_{k}\right]\textrm{d}\mathbf{p},
\end{split}
\end{equation*}
and analogous relations apply to the other derivatives. From these relations and analogously to Eqs.~(\ref{eq:ROTOR-COEFF-MAT}),~(\ref{eq:ROT-NORM-EQUATION}), we derive the corresponding normal equation. The meshless representation of the potential of the conservative and solenoidal components of the input vector field guarantees the smoothness and stability to regular/irregular space sampling or noise, and an approach general enough to deal with 2D (Figs.~\ref{fig:WIND2D},~\ref{fig:WIND2D-SECOND}) and 3D (Fig.~\ref{fig:ROTOR-EXAMPLE}) data.
\subsection{Meshless decomposition with Laplace equation\label{eq:MESHLESS-HH-LAPLACE}}
\begin{figure}[t]
\centering
{\small{
\begin{equation*}
\left\{
\begin{array}{l}
\mathbf{v}=\mathbf{v}_{1}+\mathbf{v}_{2};\quad
\mathbf{v}_{1}=\nabla u,\quad u=x^2-2xz+yz,\\
\mathbf{v}_{2}=\nabla\wedge\mathbf{w},\quad
\mathbf{w}=(x^2yz,xy\exp(-z),x^2+y^2-z^2);\\
\end{array}
\right.
\end{equation*}}}
\begin{tabular}{ccc}
\includegraphics[height=75pt]{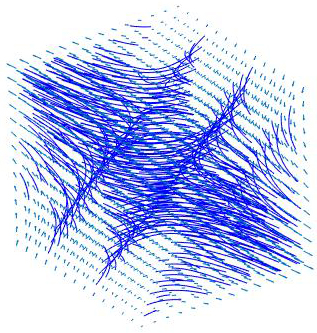}
&\includegraphics[height=75pt]{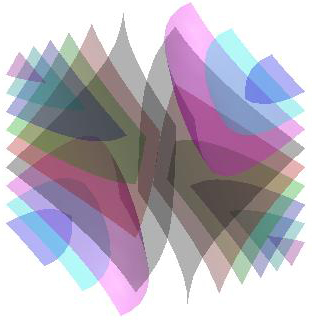}
&\includegraphics[height=75pt]{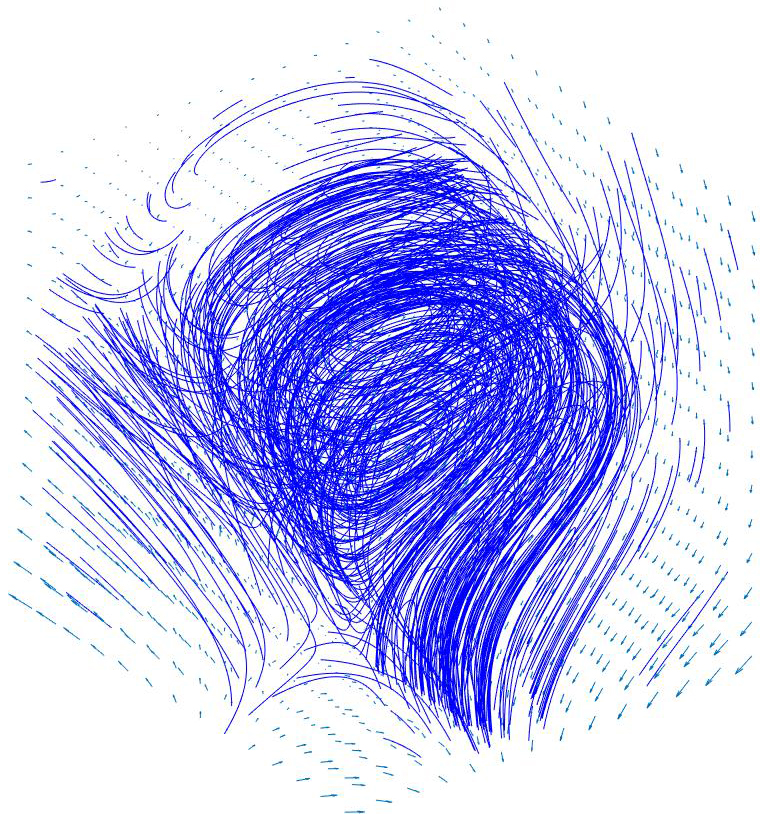}\\
(a)~$\nabla u$ &(b)~$u$ &(c)~$\nabla\wedge\mathbf{w}$\\
\end{tabular}
\caption{Meshless HHD \mbox{$\mathbf{v}=\nabla u+\nabla\wedge\mathbf{w}$} of an analytic vector field: (b) iso-surfaces of~$u$, and streamlines of the (a) irrotational and (c) div-free components. The~$\ell_{\infty}$ error between the ground-truth and the computed decomposition is lower than~$2.3$\%. 
\label{fig:ROTOR-EXAMPLE}}
\end{figure}

Recalling that the components~$u$ and~$\mathbf{w}$ of the Helmholtz-Hodge decomposition of~$\mathbf{v}$ satisfy the relations \mbox{$\Delta u=\nabla\cdot\mathbf{v}$}, \mbox{$\overline{\Delta}\mathbf{w}=\nabla\wedge\mathbf{v}$}, we approximate these functions as a linear combination of RBFs. To this end, we approximate~$\mathbf{v}$ with a meshless vector field \mbox{$\tilde{\mathbf{v}}$} such that \mbox{$\tilde{\mathbf{v}}(\mathbf{p}_{i})=\mathbf{v}_{i}$}, \mbox{$i=1,\ldots,n$}; in this way, we evaluate \mbox{$\nabla\cdot\tilde{\mathbf{v}}$} and \mbox{$\nabla\wedge\tilde{\mathbf{v}}$} analytically. Then, we define \mbox{$u(\mathbf{p}):=\sum_{j=1}^{k}\alpha_{j}\phi_{j}(\mathbf{p})$}, where the coefficient vector \mbox{$\alpha:=(\alpha_{j})_{j=1}^{k}$} satisfies the linear system
\begin{equation}\label{LAPLACE-DECOMPOSITION}
\mathbf{L}\alpha=\mathbf{b},\,
L(i,j):=\Delta\phi_{j}(\mathbf{p}_{i}),\,
b(i):=\nabla\cdot\tilde{\mathbf{v}}(\mathbf{p}_{i}),
\end{equation}
\mbox{$i=1,\ldots,n$}, \mbox{$j=1,\ldots,k$}. In Eqs.~(\ref{eq:DIV-NORMAL-EQUATION}), (\ref{LAPLACE-DECOMPOSITION}), the meshless approximations of the irrotational component of the HHD are equivalent, requiring the solution of a least-squares linear system but having a different degree of smoothness, i.e.,~$\mathcal{C}^{1}$ for Eq.~(\ref{eq:DIV-NORMAL-EQUATION}) and~$\mathcal{C}^{2}$ for Eq.~(\ref{LAPLACE-DECOMPOSITION}). The large null space of the potential of the irrotational component can lead to numerical instabilities in Eqs.~(\ref{eq:DIV-NORMAL-EQUATION}), (\ref{LAPLACE-DECOMPOSITION}), which generally happen in case of a large number of centres and are removed by adding the regularisation term \mbox{$\epsilon\mathbf{I}$}, \mbox{$\epsilon\approx 0$}, to the coefficient matrices. In our experiments, we did not face high numerical instabilities and the choice \mbox{$\epsilon:=10^{-10}$} was enough to handle them. Each~$x$,~$y$,~$z$ component of~$\mathbf{w}$ is approximated as a linear combination of RBFs of class~$\mathcal{C}^{2}$. Since~$\mathbf{w}$ is~$\mathcal{C}^{2}$, the vector Laplace-Beltrami operator \mbox{$\overline{\Delta}
=\nabla\nabla\cdot-\nabla\wedge\nabla
=(\Delta,\Delta,\Delta)$} reduces to the scalar Laplace-Beltrami operator on each component of the vector field, as a consequence of the Schwartz commutativity of the second order partial derivatives. Indeed, the equation \mbox{$\overline{\Delta}\mathbf{w}=\nabla\wedge\tilde{\mathbf{v}}$} is equivalent to solving three harmonic equations, as done for the component~$u$. Examples are shown in Fig.~\ref{fig:ROTOR-EXAMPLE}.

\section{Discussion\label{sec:DISCUSSION-GRAD}}
We derive the conditions on the kernel for the existence of the derivatives (Sect.~\ref{sec:WELL-POSEDNESS}) of the RBFs. Then, we discuss the properties of the meshless decomposition  (Sects.~\ref{sec:KERNEL-SELECTION},~\ref{sec:HH-NUMERICAL-STABILITY}). 
\subsection{Well-posedness of the meshless decomposition\label{sec:WELL-POSEDNESS}}
Expressing the potential of the conservative and solenoidal components as a linear combination of RBFs, the evaluation of differential operators reduces to 1D derivatives. To this end, we identify the conditions on the kernel that guarantee the existence of the gradient and of the rotor of the potential of the conservative and solenoidal component, respectively. These conditions are easily evaluated for an arbitrary kernel by checking the existence of its first and second derivatives at zero. In Table~\ref{tab:EXISTENCE}, we discuss the existence of the first and second order derivatives of the RBFs commonly used for the approximation and HHD of vector fields.

Assuming that~$\phi$ is \mbox{$\mathcal{C}^{1}(\mathbb{R}^{+})$}, we study the well-posedness, continuity, and differentiability of the potential at any point of~$\mathbb{R}^{3}$ in terms of the generating kernel. Since the potential is a linear combination of RBFs, it is enough to study the existence of the gradient in Eq.~(\ref{eq:IMPLICIT-APPROX-GRADIENT}b) of these basis functions, which is well-defined, continuous, and differentiable for any kernel and at any point of~$\mathbb{R}^{3}$ except the centre~$\mathbf{c}_{i}$. To analyse the regularity of the meshless function in Eq.~(\ref{eq:IMPLICIT-APPROX-GRADIENT}b) at~$\mathbf{c}_{i}$, we evaluate this expression in a neighbourhood of~$\mathbf{c}_{i}$, i.e., on the points \mbox{$\mathbf{p}:=\mathbf{c}_{i}+\beta\mathbf{a}$}, \mbox{$\|\mathbf{a}\|_{2}=1$}, of the sphere~$\mathcal{S}$ of centre~$\mathbf{c}_{i}$ and radius~$\beta$. Then, we have that
\begin{equation*}
\lim_{\beta\rightarrow 0^{\pm}}\nabla \phi_{i}(\mathbf{p})
=\lim_{\beta\rightarrow 0^{\pm}}\textrm{sgn}(\beta)\phi^{\prime}(\vert\beta\vert)\mathbf{a}
=\pm\phi^{\prime}(0)\mathbf{a},\quad
\mathbf{p}\in\mathcal{S},
\end{equation*}
where \mbox{$\textrm{sgn}(\cdot)$} is the sign function. Indeed, the gradient of~$\phi_{i}$ is well-defined and continuous at~$\mathbf{c}_{i}$ if and only if \mbox{$\phi^{\prime}(0)=0$}. Since the rotor involves the first order partial derivatives of the components of~$\mathbf{w}$, previous conditions also guarantee the well-posednees of the solenoidal potential. Deriving the function in Eq.~(\ref{eq:IMPLICIT-APPROX-GRADIENT}b), the entries of the Hessian matrix of the RBFs are
\begin{equation}\label{eq:IMPLICIT-HESSIAN-MATRIX-SINGLE}
\begin{split}
&\partial_{x_{k}x_{j}}^{2}\phi_{i}(\mathbf{p})
=\frac{\phi^{\prime\prime}(\|\mathbf{p}-\mathbf{c}_{i}\|_{2})}{\|\mathbf{p}-\mathbf{c}_{i}\|_{2}}(x_{k}-c_{i}^{(k)})(x_{j}-c_{i}^{(j)})\\
&+\frac{\phi^{\prime}(\|\mathbf{p}-\mathbf{c}_{i}\|_{2})}{\|\mathbf{p}-\mathbf{c}_{i}\|_{2}}\delta_{kj}
-\frac{\phi^{\prime}(\|\mathbf{p}-\mathbf{c}_{i}\|_{2})}{\|\mathbf{p}-\mathbf{c}_{i}\|_{2}^{3}}(x_{k}-c_{i}^{(k)})(x_{j}-c_{i}^{(j)}),
\end{split}
\end{equation}
\mbox{$i,j=1,\ldots,d$}, where~$c_{i}^{(j)}$ is the~$j$-th component of~$\mathbf{c}_{i}$. To study the continuity of these derivatives at~$\mathbf{c}_{i}$, we consider their restriction on the sphere~$\mathcal{S}$, i.e.,
\begin{equation*}
\partial^{2}_{x_{k}x_{j}}\phi_{i}
=\phi^{\prime\prime}(\vert\beta\vert)\vert\beta\vert\eta^{(k)}\eta^{(j)}
+\frac{\phi^{\prime}(\vert\beta\vert)}{\vert\beta\vert}\delta_{kj}+
-\frac{\phi^{\prime}(\vert\beta\vert)}{\vert\beta\vert}\eta^{(k)}\eta^{(j)}.
\end{equation*}
Applying the following identities (derivative of a composite function) \mbox{$\lim_{\vert\beta\vert\rightarrow 0^{\pm}}\frac{\phi^{\prime}(\vert\beta\vert)}{\vert\beta\vert}
=\lim_{\vert\beta\vert\rightarrow 0^{\pm}}\phi^{\prime\prime}(\vert\beta\vert)=\phi^{\prime\prime}(0)$}, we get that \mbox{$\lim_{\vert\beta\vert\rightarrow 0^{\pm}}\partial^{2}_{x_{k}x_{j}}\phi_{i}(\mathbf{p})
=(\delta_{kj}-\eta^{(k)}\eta^{(j)})\phi^{\prime\prime}(0)$}. Indeed, the existence of \mbox{$\phi^{\prime\prime}(0)$} is enough to compute the second order derivatives of the RBFs.
\begin{table}
\caption{Timings of the meshless approximation and HHD with respect to the number of samples and centres, induced by globally- (Gl.) and locally- (Loc.) supported RBFs.\label{tab:HHD-COST}}
\centering
\begin{tabular}{|l|l|l|l|}
\hline
\textbf{Tests}										&\textbf{$\#$Input Sampl./}			&\textbf{Meshless}		&\textbf{Meshless}\\
													&\textbf{$\#$ Sel. Centr.}			&\textbf{Approx.}		&\textbf{H.H.D.}\\
\hline
\multicolumn{4}{|l|}{\textbf{2D Data~$\&$ Regular sampl./Sel. Centres}}\\
\hline
Fig.~\ref{fig:NONANALYTIC-EXAMPLE}					&1.3K/13.K									&4.2s (Gl. RBFs)			&7.2s\\
Fig.~\ref{fig:eFUN}									&1.4K/1.4K									&8.9s (Gl. RBFs)			&15.1s\\
Fig.~\ref{fig:WIND13}								&1.4K/1.4K									&5.4s (Loc. RBFs)			&7.2s\\
Fig.~\ref{fig:WIND2D}								&1.4K/1.4K									&4.1s (Loc. RBFs)			&7.9s\\
Fig.~\ref{fig:WIND2D-SECOND}						&1.4K/1.4K									&3.5s (Loc. RBFs)			&6.1s\\
\hline
\multicolumn{4}{|l|}{\textbf{2D Data~$\&$ Adaptive sampl./Sel. Centres -  Global RBFs}}\\
\hline
Fig.~\ref{FIG:vectorvalued}							&0.6K/0.3K									&5.1s						&0.45s\\
Fig.~\ref{FIG:vectorvaluedsamples}					&16K/0.5K, 1K, 2K							&3.2s/4.7s/4.9s				&5.1s/9.1s/8.9s\\
\hline
\multicolumn{4}{|l|}{\textbf{3D Data~$\&$ Regular sampl.~$\&$ 1.6K centres -  Global RBFs}}\\
\hline
Fig.~\ref{fig:HH-FLOW}								&150K										&3.3s						&5.9s\\
Fig.~\ref{fig:WIND}									&22K										&2.8s						&4.1s\\
Fig.~\ref{fig:ROTOR-EXAMPLE}						&125K										&3.1s						&5.2s\\
\hline
\end{tabular}
\end{table}
\subsection{Kernel/centres' selection and computational cost\label{sec:KERNEL-SELECTION}}
The kernel and centres' selection is guided by the order of smoothness of the resulting approximation (Table~\ref{tab:EXISTENCE}), the computational cost and storage overhead.

\textbf{Kernel selection}
The first and second order derivatives of the cubic, Gaussian, II- and IV-degree polynomial kernels are well defined; indeed, they are valid choices for the approximation of vector fields, the computation of the HHD, the classification of the critical points of the potential of the conservative and solenoidal components. The derivatives of the thin-plate, inverse multi-quadratic, and multi-quadratic kernels might be not well defined as the radial distance~$r$ and/or the parameter~$\sigma$ of the generating kernels in Table~\ref{tab:EXISTENCE} becomes close to zero. Indeed, these last three kernels are valid choices for the approximation of a vector field but the components of the HHD might be not defined at the centres of the RBFs or might have spurious critical points. For sparse samples, we select globally-supported kernels (e.g., the Gaussian kernel in our tests); for dense data, we select locally-supported kernels (e.g., the II-order polynomial, in our experiments), whose support is computed according to geometric~\cite{PAULY2003} or functional properties~\cite{RIPPA1999}.

\textbf{Centres' selection}
In case of a dense point set, we select locally-supported RBFs centred at any input point or globally-supported RBFs centred at a subset of the input points through clustering, kernel-based sampling, or sparsification, and with a maximum number of centres determined according to the available computational resources (e.g., 5K points in our experiments). Indeed, the location and number of centres of the RBFs are adapted to the behaviour of the input vector field by increasing the number of centres at each iteration until the residual least-squares error is lower than a given threshold (e.g.,~$5\%$, in our experiments). However, the centres of the RBFs remain fixed (i.e., they are not optimised further) and are used for the computation of the coefficient matrices of the linear systems in Eq.~(\ref{eq:DIV-NORMAL-EQUATION}).
\begin{figure}[t]
\centering
\begin{tabular}{c}
Input vector field~$\mathbf{v}$:~$25\times 25$\\
\includegraphics[width=.28\columnwidth]{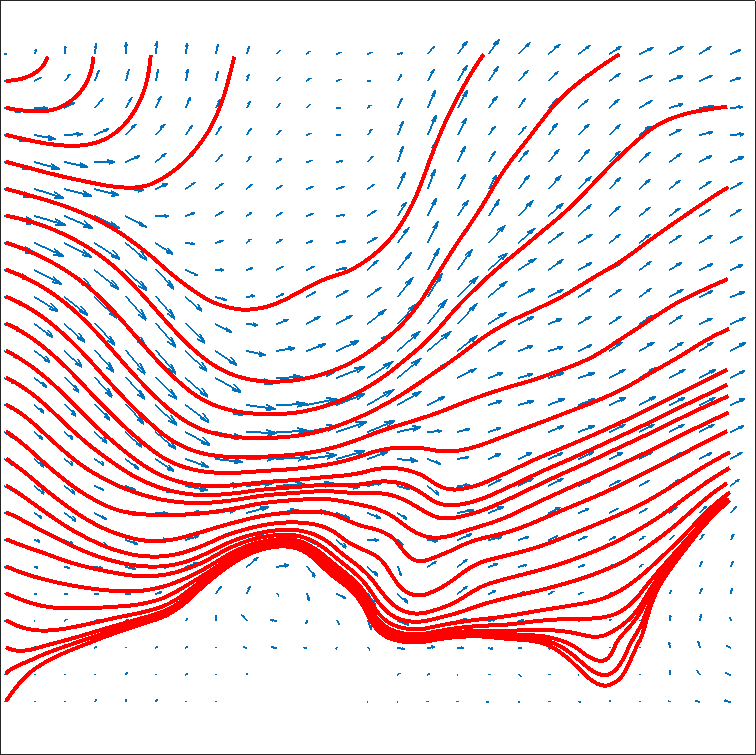}
\end{tabular}
\begin{tabular}{ccc}
\hline
168 RBFs' centres &Gradient of	&Angle \mbox{$\angle(\nabla u,\mathbf{v})$}\\
&meshless pot.~$u$ &\\
\includegraphics[width=.26\columnwidth]{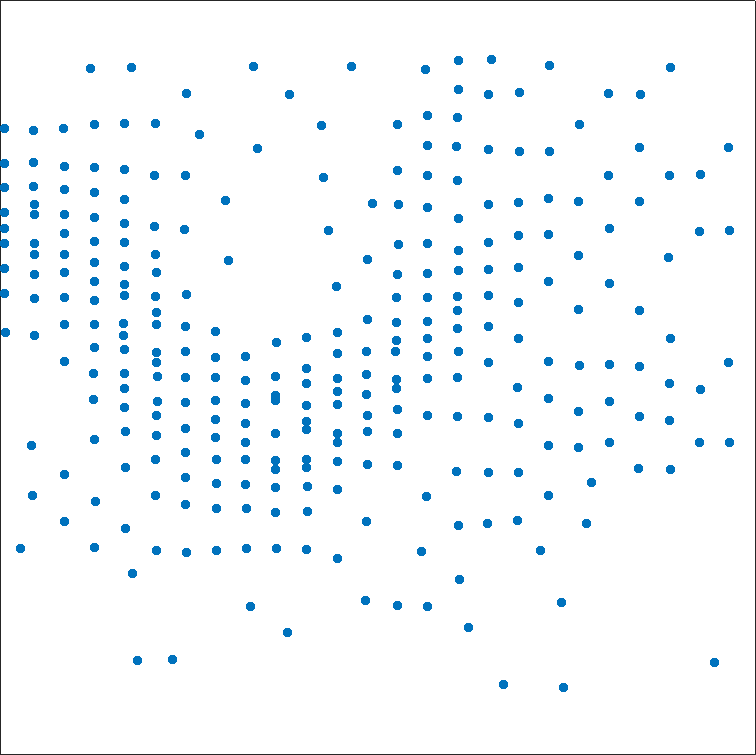}
&\includegraphics[width=.26\columnwidth]{HHD-IMAGES/v_inputStream.png}
&\includegraphics[width=.26\columnwidth]{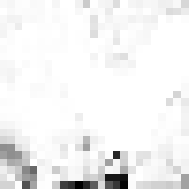}\\
\includegraphics[width=.26\columnwidth]{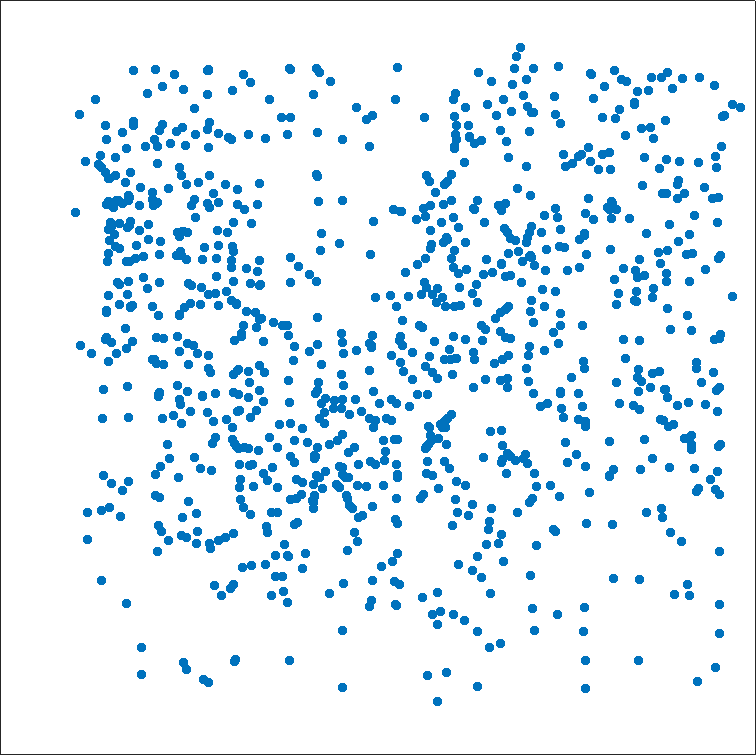}
&\includegraphics[width=.26\columnwidth]{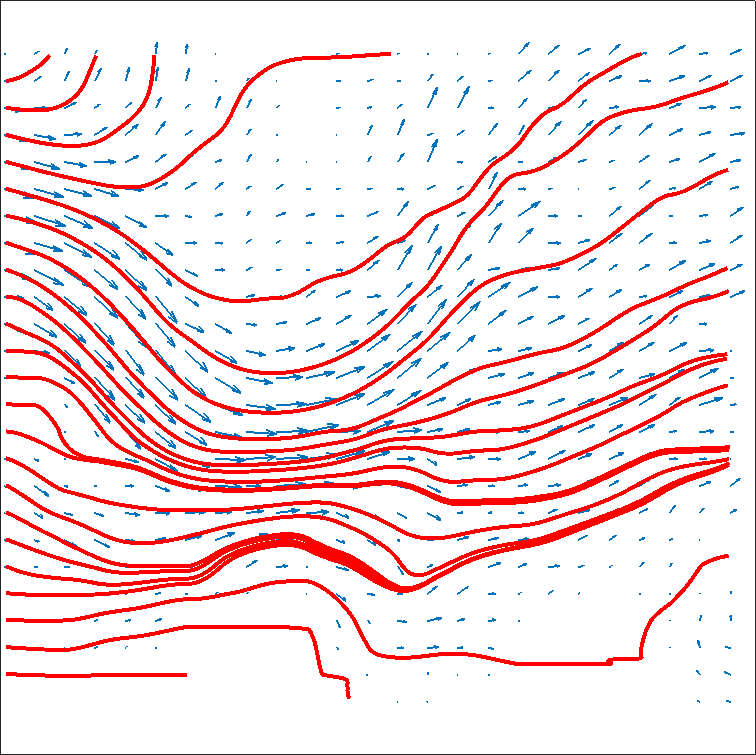} 
&\includegraphics[width=.26\columnwidth]{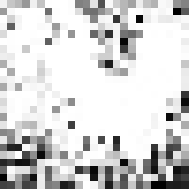}\\
\includegraphics[width=.26\columnwidth,height=.26\columnwidth]{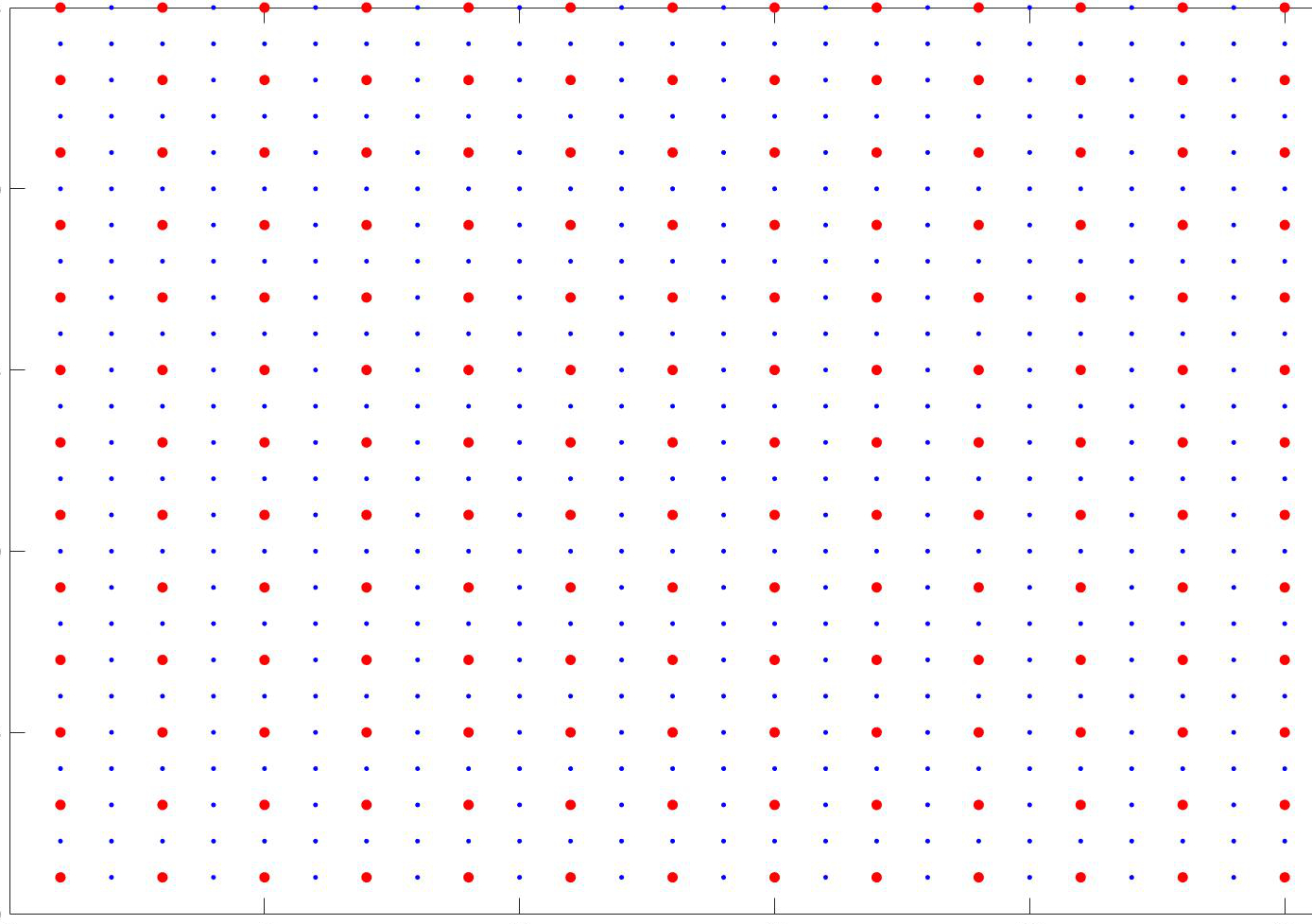}
&\includegraphics[width=.26\columnwidth]{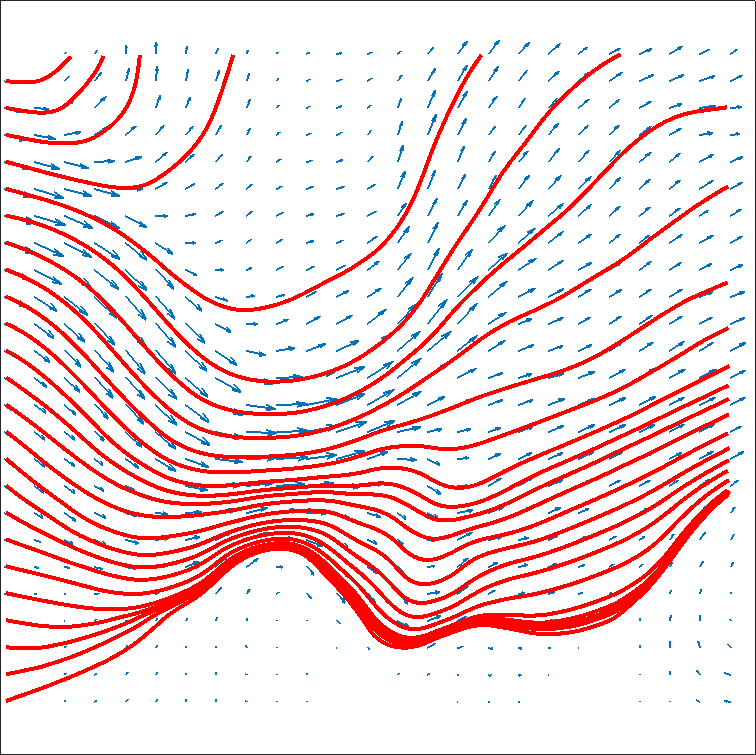}
&\includegraphics[width=.26\columnwidth]{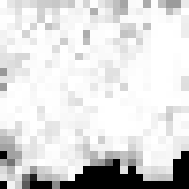}\\
(a) &(b) &(c)
\end{tabular}
\caption{1-st Row: wind field and streamlines on a \mbox{$25\times 25$} regular grid. Centres computed with the kernel-based sampling applied to the field magnitude \mbox{$\|\mathbf{v}\|_{2}$} (2nd row) and to the potential (3rd row), or uniformly distributed (4th row). (a) Selected centres of the RBFs, (b) streamlines of gradient of the meshless potential, and (c) angle distribution \mbox{$\angle(\nabla u,\mathbf{v})$}: white represents a null angle and black corresponds to~$\pi$. See also Table~\ref{TAB:vectorerror}.\label{FIG:vectorvalued}}
\end{figure}

\textbf{Computational cost} 
For the computation of the potential functions of the conservative and solenoidal components with~$k$ \emph{globally-supported RBFs}~\cite{CARR2001,TURK2002}, the allocation of the coefficient matrices in Eqs.~(\ref{eq:GRAD-COEFF-MAT}),~(\ref{eq:ROTOR-COEFF-MAT}) takes \mbox{$\mathcal{O}(kt)$} memory and the computation of the coefficients of the  solution to the linear system in Eqs.~(\ref{eq:DIV-NORMAL-EQUATION}),~(\ref{eq:ROT-NORM-EQUATION}) takes \mbox{$\mathcal{O}(k^3)$} time with direct solvers and \mbox{$\mathcal{O}(kt)$} time with iterative solvers. Finally, the computation of the harmonic component takes linear time. For \emph{locally-supported RBFs}~\cite{WENDLAND1995,MORSE2001}, the memory allocation reduces to \mbox{$\mathcal{O}(rk)$} and computational cost is \mbox{$\mathcal{O}(rk\log k)$}, where~$r$ is the average number of points in the neighbourhood of each center (e.g., \mbox{$r=10,20$} for the~$k$-nearest neighbourhood). In Table~\ref{tab:HHD-COST}, we report the performances of the meshless HHD in terms of computation time and memory consumptions.
\begin{figure}[t]
\centering
\begin{tabular}{c|cc}
\hline
Input domain &\multicolumn{2}{|c}{Input vector field~$\mathbf{v}$}\\
\includegraphics[width=.24\columnwidth]{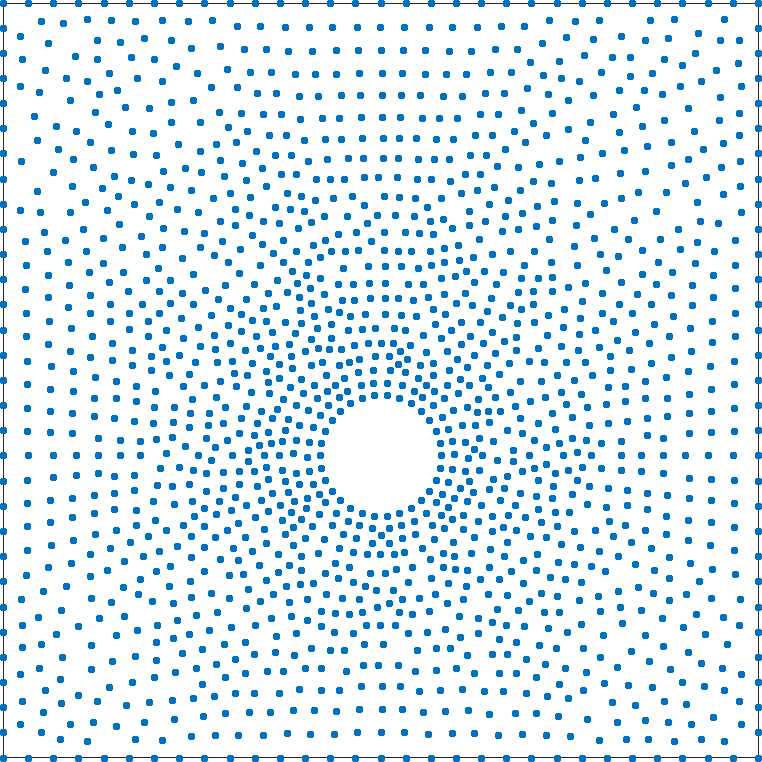}
&\includegraphics[width=.24\columnwidth]{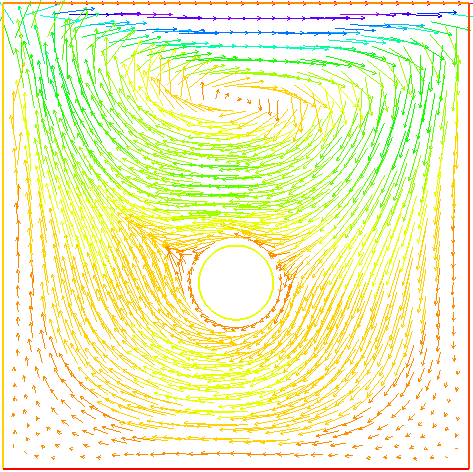}
&\includegraphics[width=.24\columnwidth]{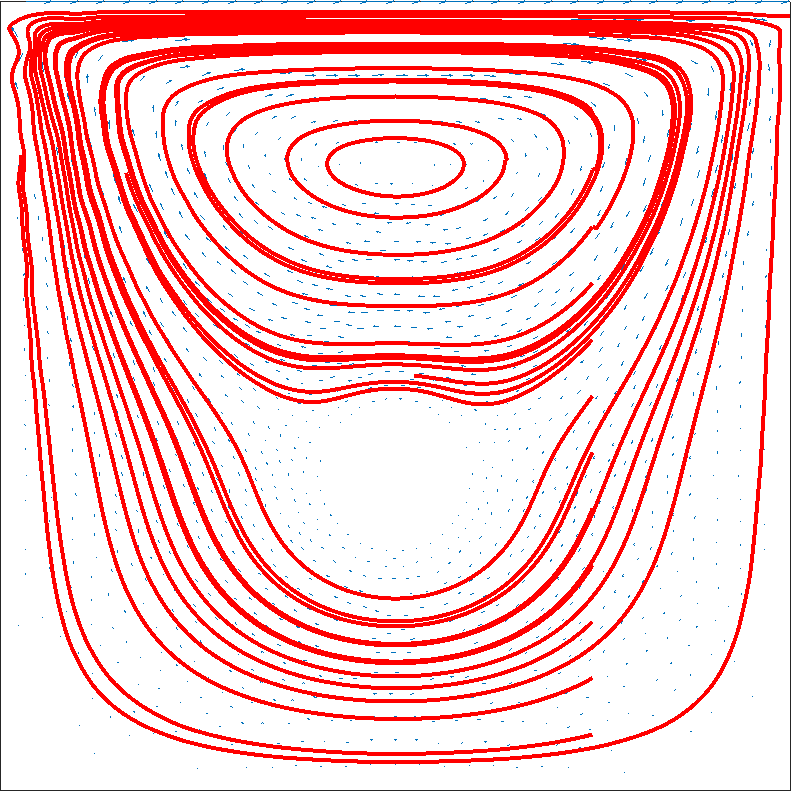}\\
\hline
Sampling & Reconstruction & Error \\ 
\includegraphics[width=.24\columnwidth]{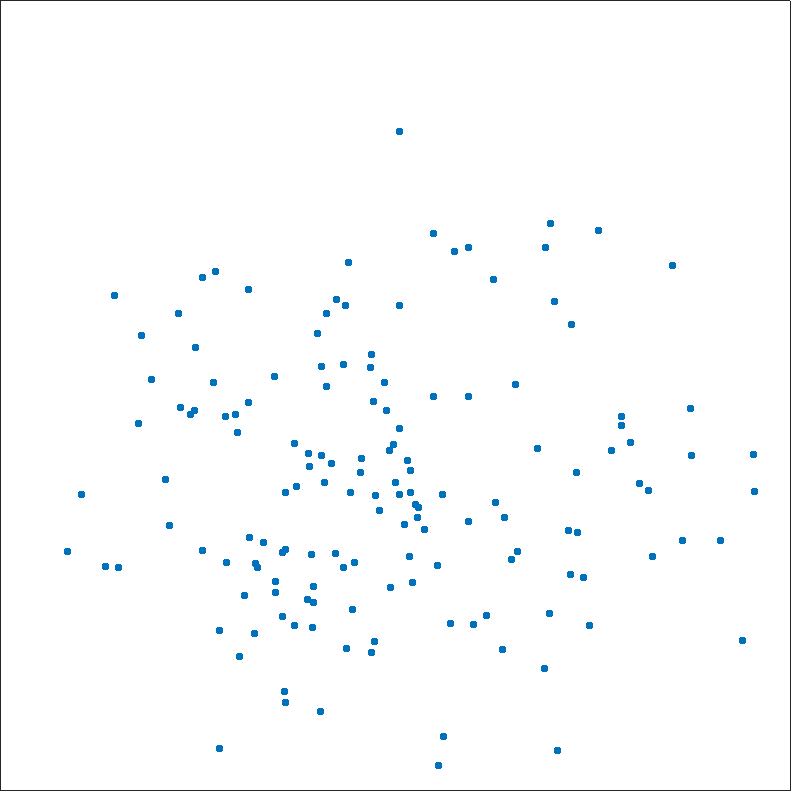} &
\includegraphics[width=.24\columnwidth]{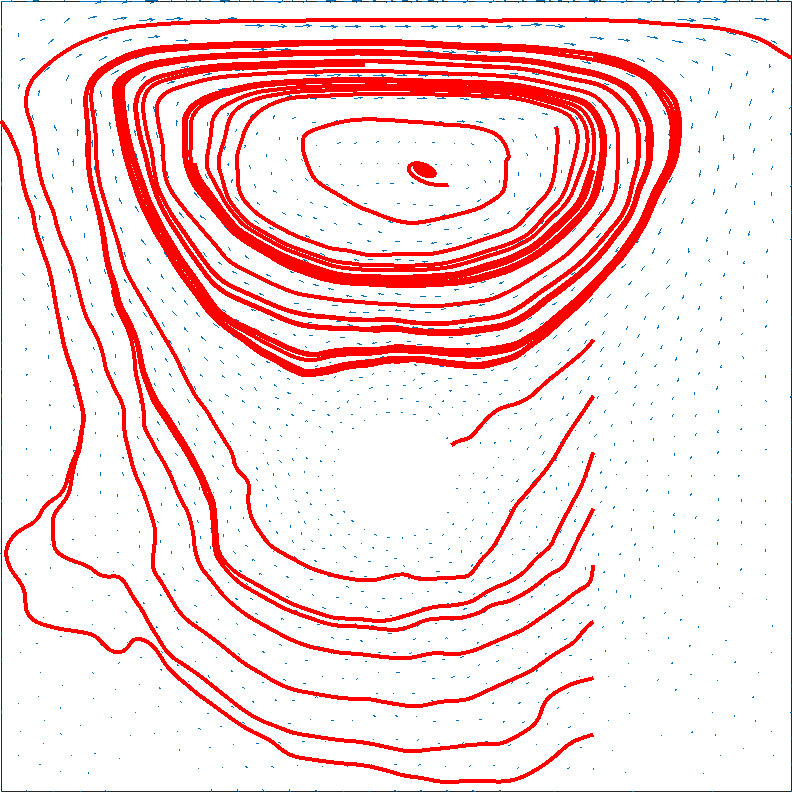} &
\includegraphics[width=.24\columnwidth]{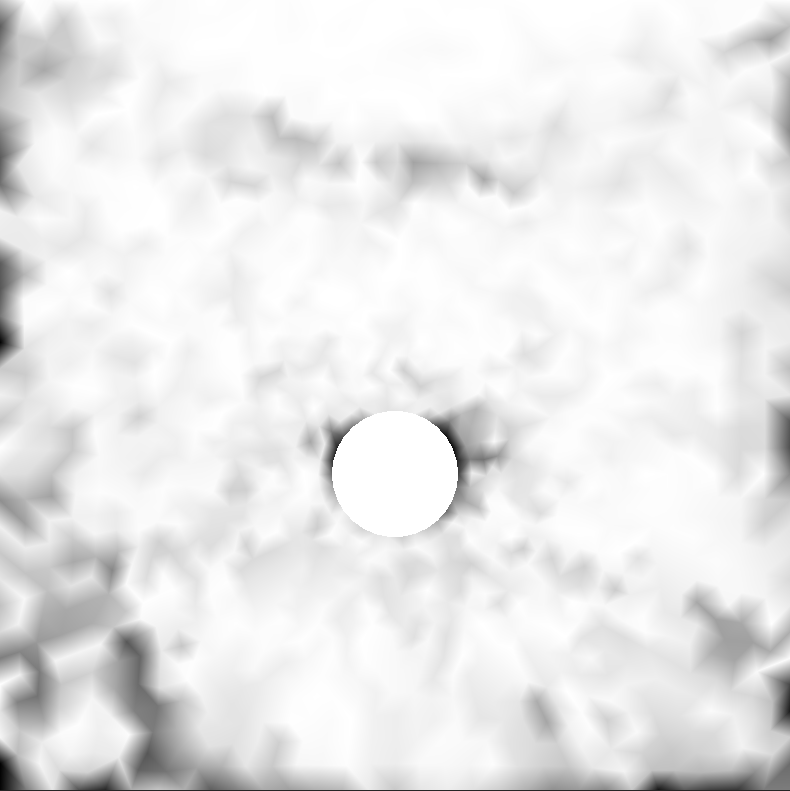} (a) \\
\includegraphics[width=.24\columnwidth]{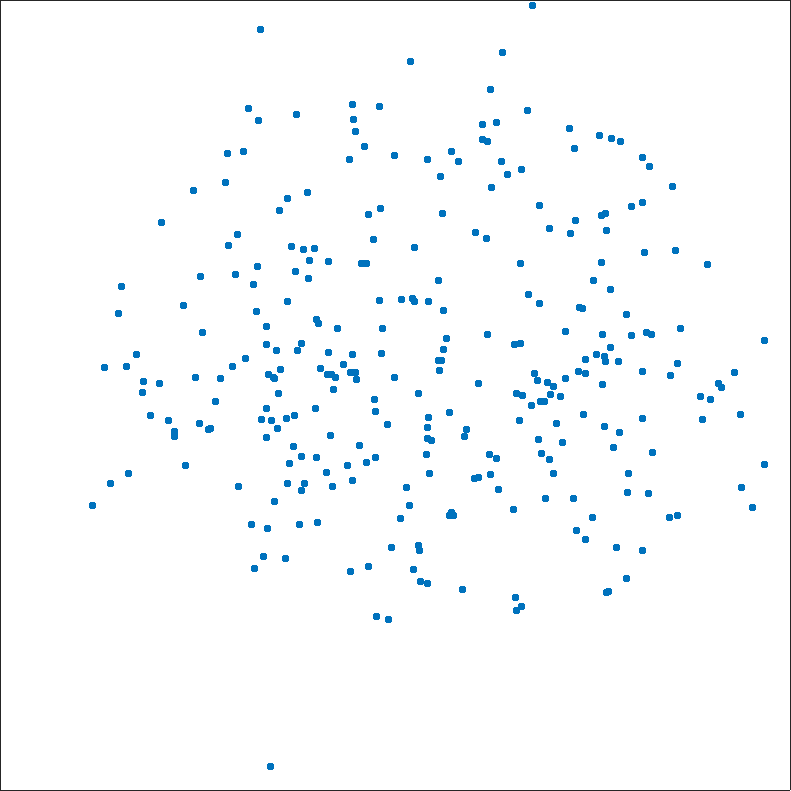} &
\includegraphics[width=.24\columnwidth]{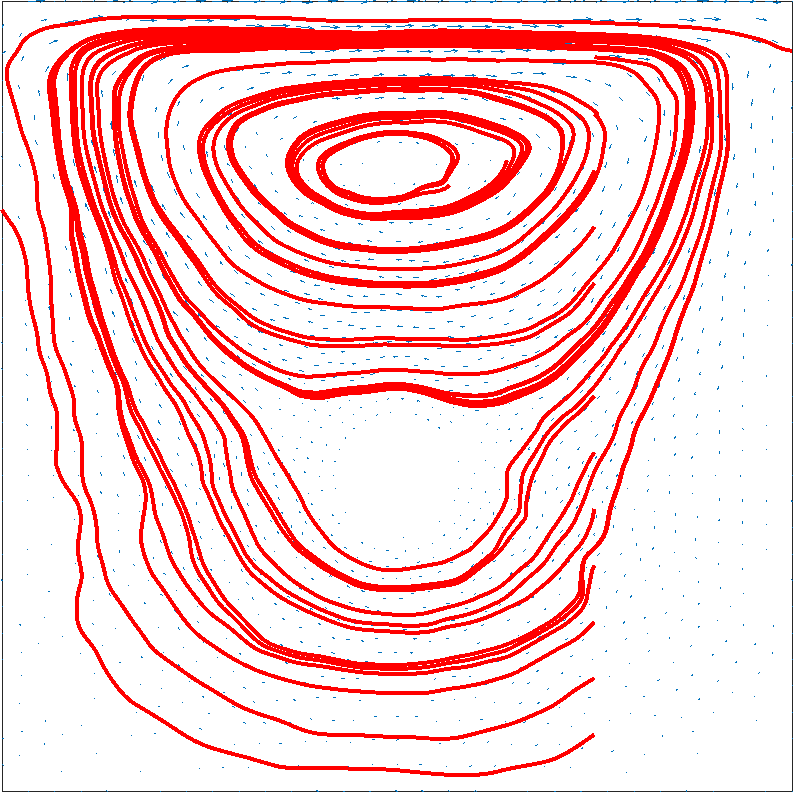} &
\includegraphics[width=.24\columnwidth]{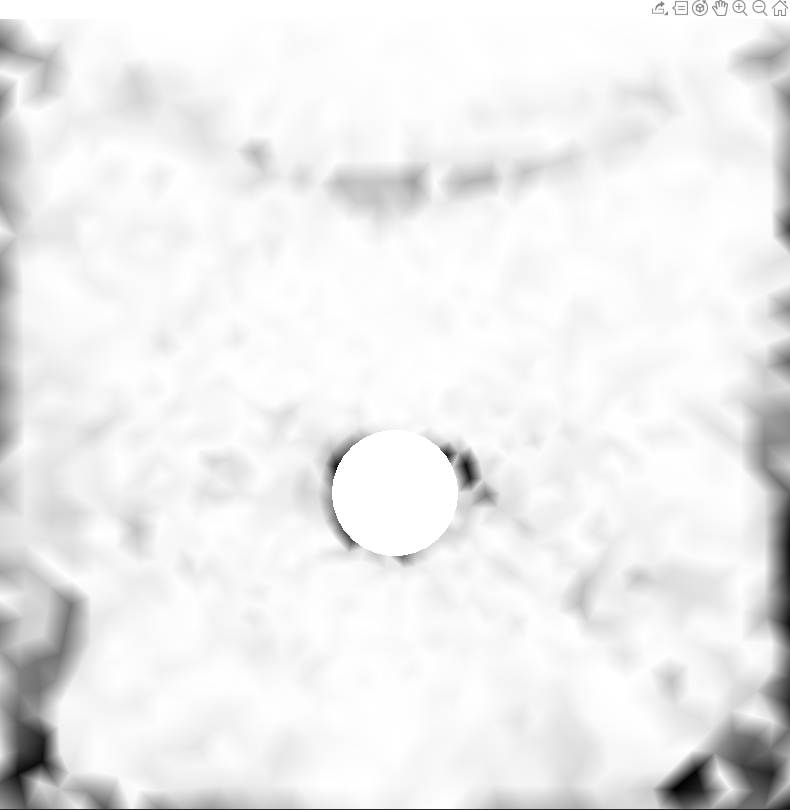} (b) \\
\includegraphics[width=.24\columnwidth]{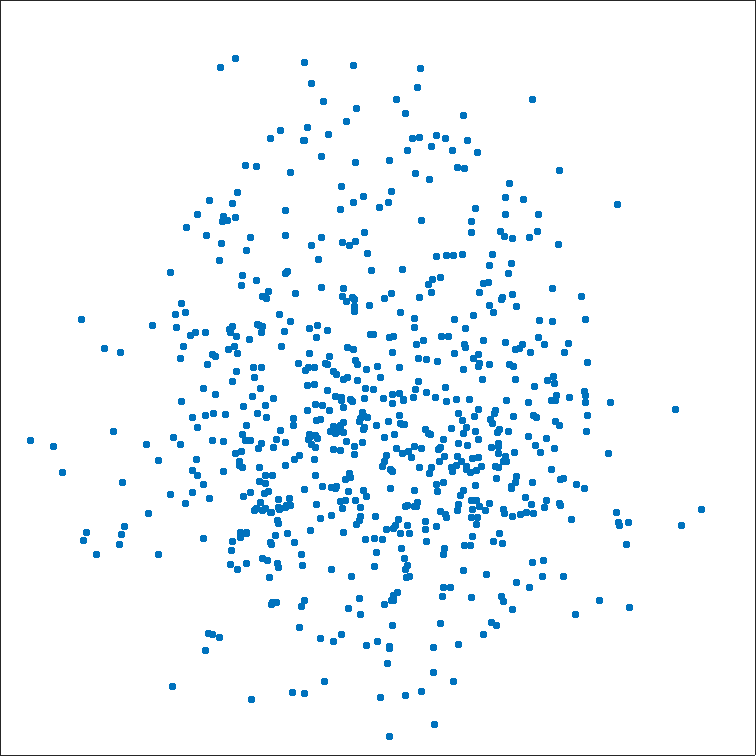} &
\includegraphics[width=.24\columnwidth]{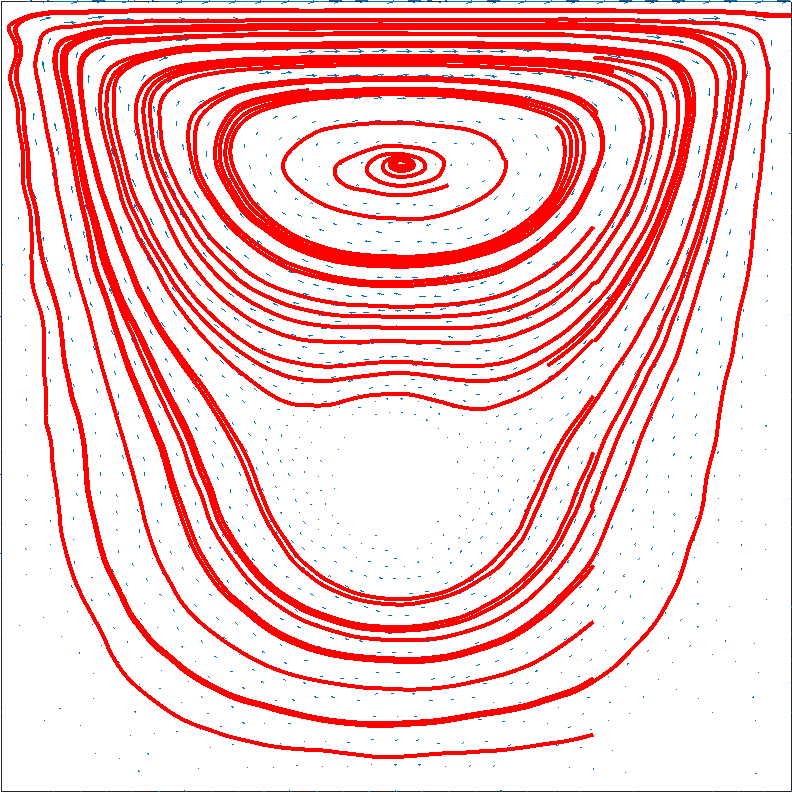} &
\includegraphics[width=.24\columnwidth]{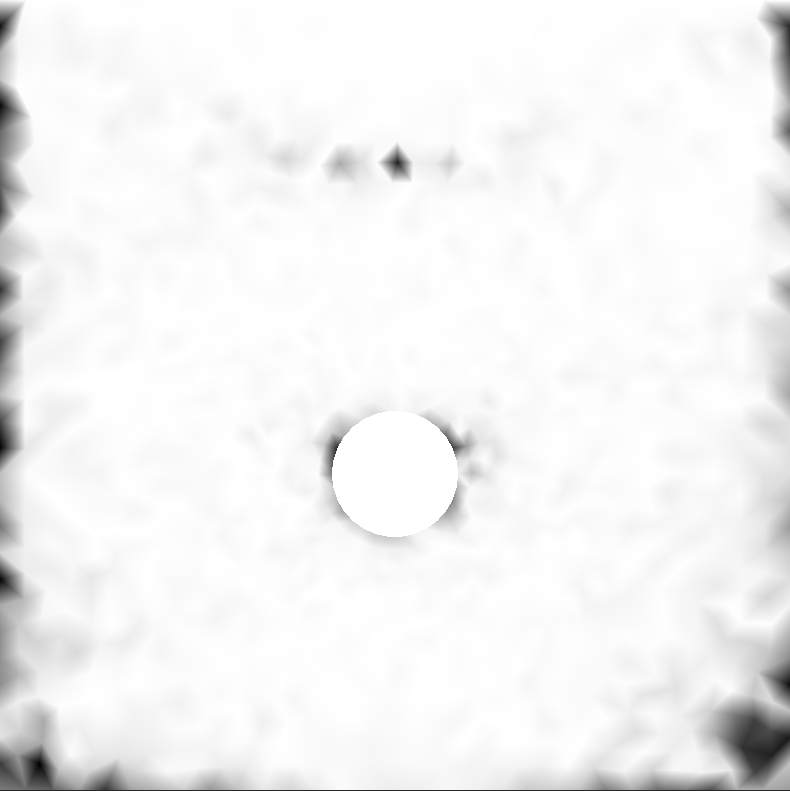} (c)
\end{tabular}
\caption{Sampling, reconstruction, and error of the kernel-based sampling applied to the field magnitude \mbox{$\|\mathbf{v}\|_{2}$} (first row) with (a) \mbox{$n=150$}, (b) \mbox{$n=300$}; (c) \mbox{$n=700$} samples. See also Table~\ref{TAB:methodvector-NEW}.\label{FIG:vectorvaluedsamples-NEW}}
\end{figure}

\textbf{Experimental results}
For the selection of the centres of the meshless approximation of a 2D vector field defined on a regular (Fig.~\ref{FIG:vectorvalued}, wind field) and irregular sampling (Fig.~\ref{FIG:vectorvaluedsamples-NEW}, a fluid flow), we apply the kernel-based sampling, a uniform sampling, and a random sampling. The error is measured as the angle between the input and the reconstruction vector fields; the white colour corresponds to a null angle and black identifies the maximum angle, which is lower or equal to~$\pi$ in our experiments. For all the methods, the centres of the RBFs are denser in those regions where the magnitude of the vector field is higher (e.g., in the central region of the input domain); here, the reconstruction is very accurate and preserves the streamlines of the vector field. The error is localised mainly in the boundary regions, where we have only a partial information on the behaviour of the input vector field. The kernel-based sampling generally provides the best results in terms of minimum angle between the input and meshless vector fields, shape and distribution of the streamlines.
\begin{table}[t]
\centering
\caption{Approximation accuracy of the meshless vector field in Fig.~\ref{FIG:vectorvalued} with centres selected through kernel-based and uniform centres. Best results in bold.\label{TAB:vectorerror}}
\begin{tabular}{l|ccc}
\hline
\textbf{Method} 											&Kernel sampl.~$\|\mathbf{v}\|_{2}$ 		&Kernel sampl.~$u$			&Unif. sampl. 	\\
\hline
\textbf{NC} 												&\textbf{0.994} 							&0.355  					&0.929\\
\textbf{NRMSE} 												&\textbf{0.084} 							&0.938						&0.311  \\
\textbf{$P_{0.05}$} 										&\textbf{99.9\%} 							&$35.5\%$					&88.7\%  \\
\textbf{$P_{0.10}$} 										&\textbf{100\%} 							&$62.7\%$					&96.9\%
\end{tabular}
\end{table}

We analyse the quality of the meshless approximation and decomposition in terms of the selected centres and kernels. To this end, we focus on 2D vector fields and on the computation of the conservative component. 
%
%
%
%
%
%
%
\begin{figure}[t]
\centering
\begin{tabular}{c}
Input vector field~$\mathbf{v}=\nabla u$:~$128 \times 128$\\
(a)\includegraphics[width=.28\columnwidth]{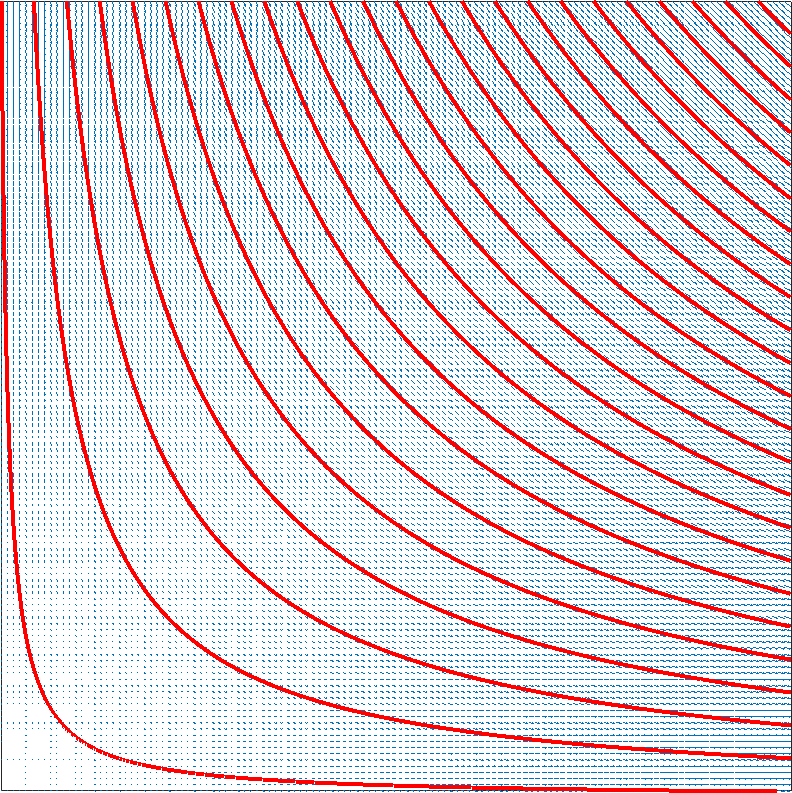}
\end{tabular}
\begin{tabular}{ccc}
\hline
RBFs' Centres &Gradient of &Angle \mbox{$\angle(\nabla\tilde{u},\mathbf{v})$}\\
		&meshless pot.~$\tilde{u}$					&\\
\includegraphics[width=.25\columnwidth]{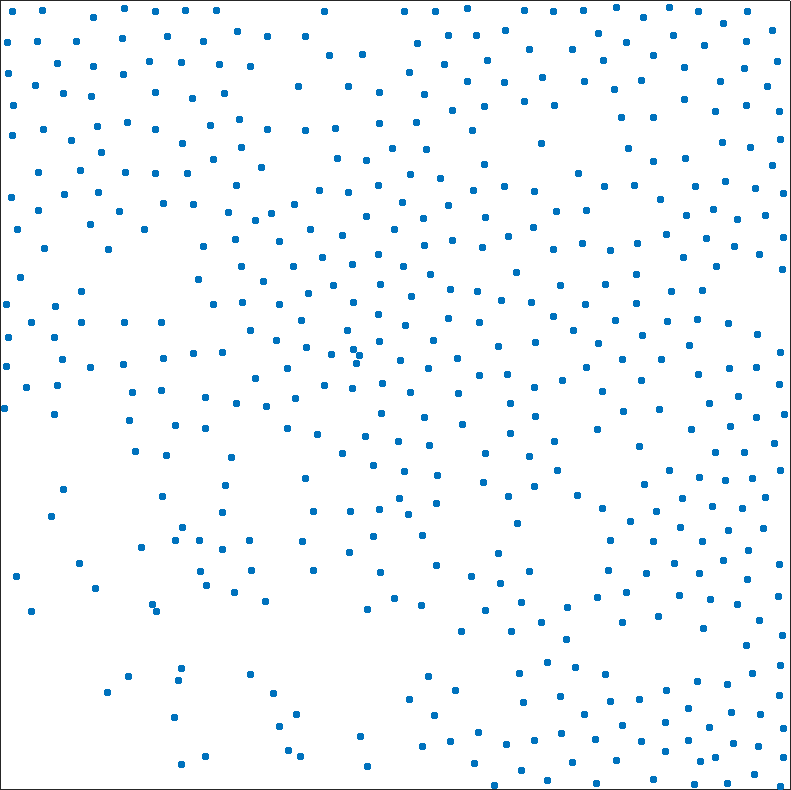}
&\includegraphics[width=.25\columnwidth]{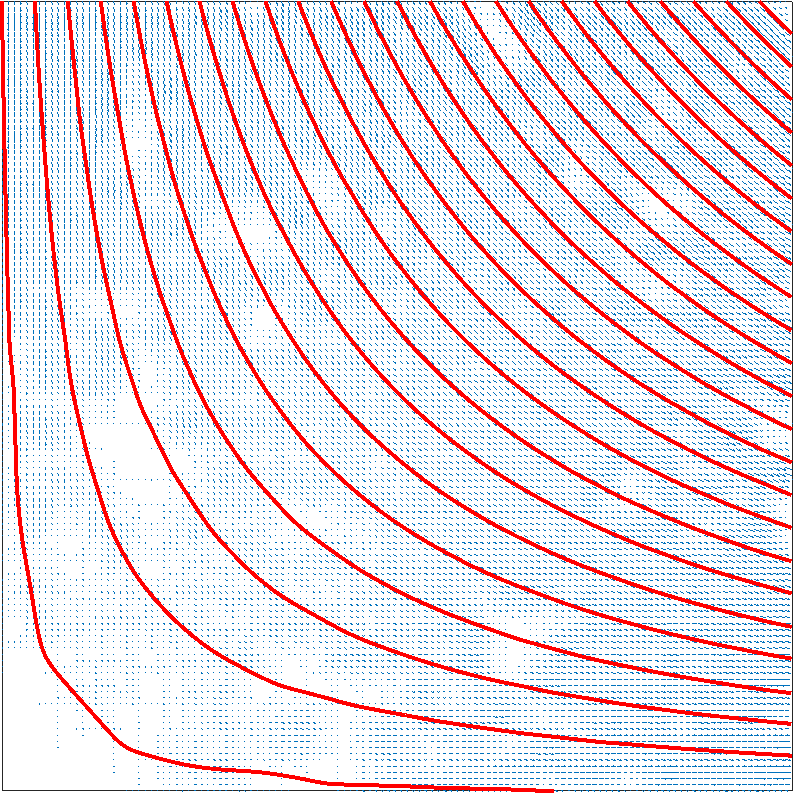}
&\includegraphics[width=.25\columnwidth]{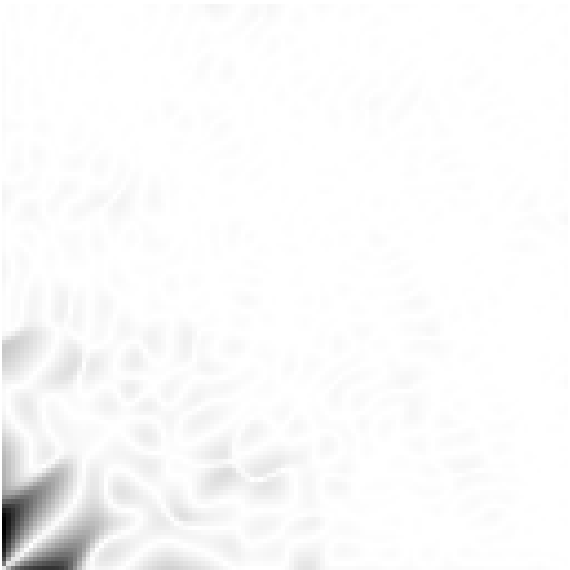}\\
\includegraphics[width=.25\columnwidth]{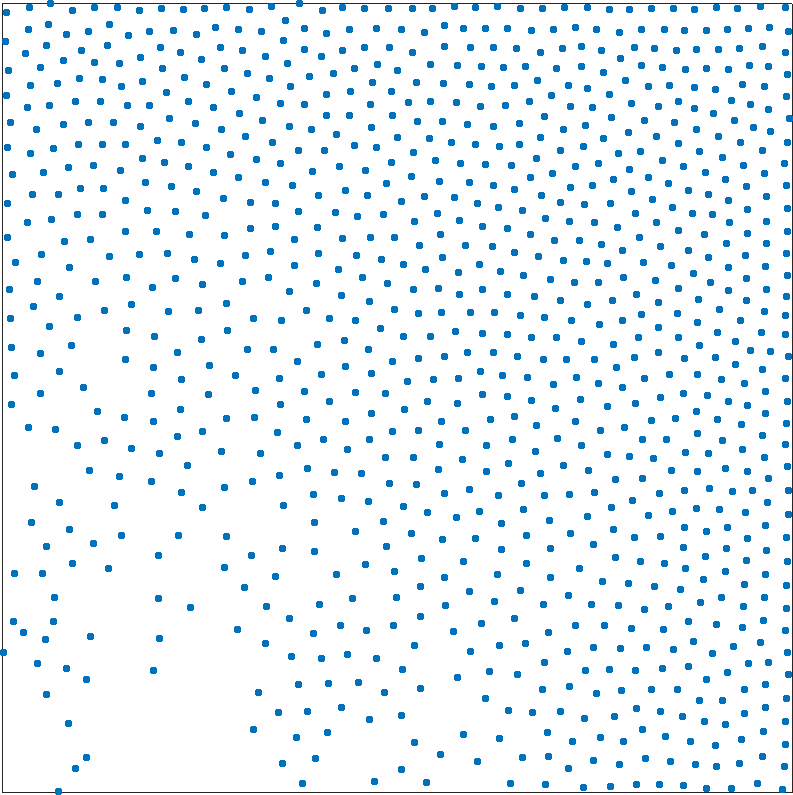}
&\includegraphics[width=.25\columnwidth]{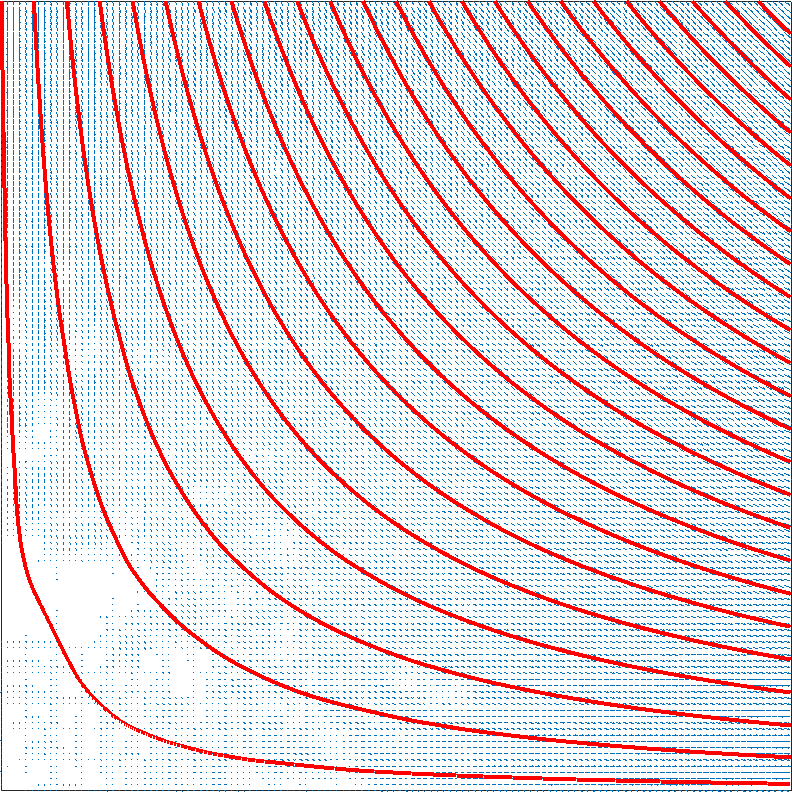}
&\includegraphics[width=.25\columnwidth]{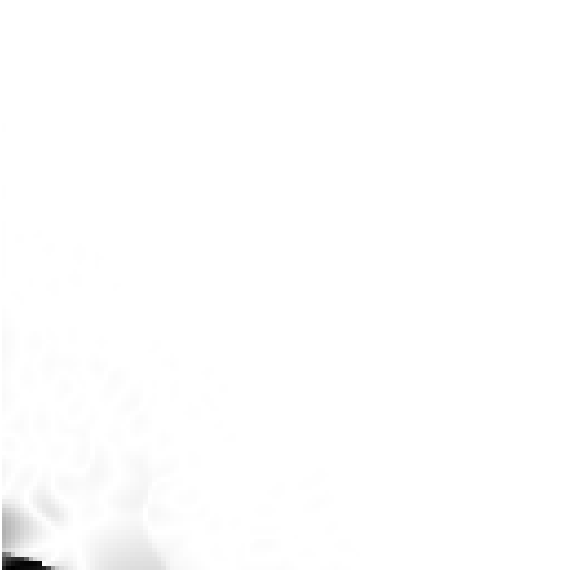}\\
\includegraphics[width=.25\columnwidth]{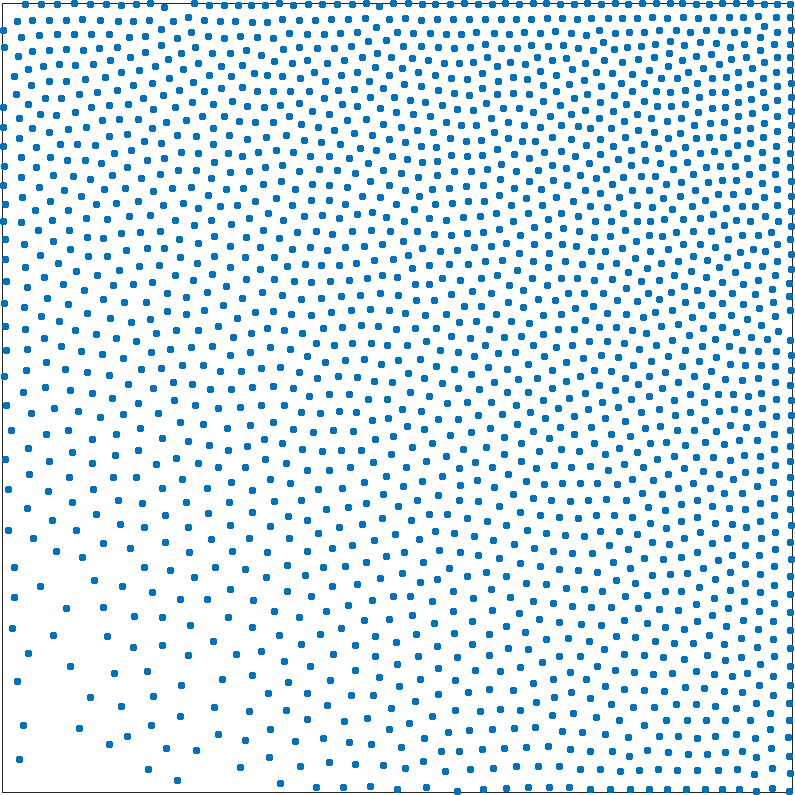} 
&\includegraphics[width=.25\columnwidth]{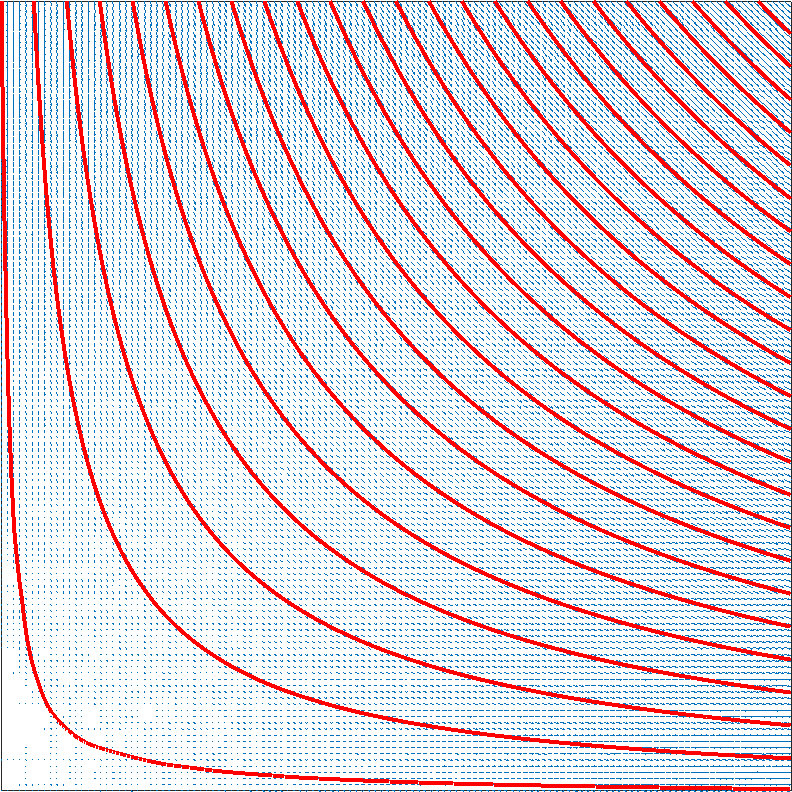}
&\includegraphics[width=.25\columnwidth]{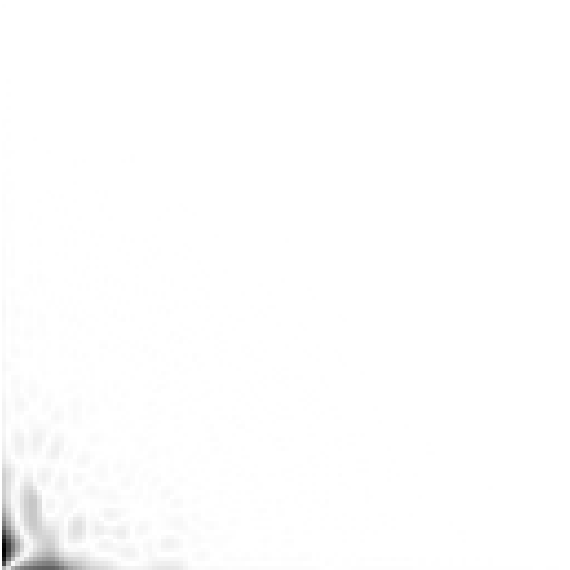}\\
(b) &(c) &(d)
\end{tabular}
\caption{(a) Streamlines of a conservative vector field. (b, 2nd row) 0.5K, (b, 3rd row) 1K, (d, 4th row) 2K kernel-based samples of the field magnitude \mbox{$\|\mathbf{v}\|_{2}$}. (c) Streamlines of the gradient of the meshless potential and (d) angles between the input vector field and its meshless approximation. See also Table~\ref{TAB:methodvector}.\label{FIG:vectorvaluedsamples}} 
\end{figure}
To this end, we evaluate the quality of the reconstruction at~$m$ input points in terms of the \emph{normalised cross correlation} NC, the \emph{normalised root mean square error} (NRMSE), and the \emph{$P_{k}$-percentile}, defined as the percentage of points whose reconstruction error is lower than~$k$. Each metric is computed as the average of its value on the components of the field. In out tests (Table~\ref{TAB:vectorerror}), the kernel-based sampling has the best results with a NC value of 0.994 and a~$P_{0.05}$ value of 0.99.
\begin{table}[t]
\centering
\caption{With reference to Fig.~\ref{FIG:vectorvaluedsamples}, we report the approximation accuracy with respect to a larger number of centres of the RBFs, selected with the kernel-based sampling. Best results in bold.\label{TAB:methodvector}}
\begin{tabular}{l|ccccc}
\hline
\textbf{Samples} & 4K & 3K & 2K & 1K & 0.5K  \\ \hline
\textbf{NC} & \textbf{0.999} & 0.998 & 0.995 & 0.981 & 0.913 \\
\textbf{NRMSE} & \textbf{0.020} & 0.029 & 0.048 & 0.102 & 0.226 \\
\textbf{$P_{0.05}$} & \textbf{100\%} & 99.9\% & 98.4\% & 88.2\% & 67.1\% \\ 
\textbf{$P_{0.10}$} & \textbf{100\%} & 100\% & 99.9\% & 97.1\% & 85.1\%
\end{tabular}
\end{table}
Since the kernel-based sampling provides the best approximation accuracy, we further analyse its accuracy with respect to the number of samples. Selecting a larger numbers of samples (i.e., from 500 to 4K), the reconstruction of the input vector field improves. The error is mainly localised at the bottom left corner, where the samples are less dense, and it reduces where the number of samples increases (Fig.~\ref{FIG:vectorvaluedsamples}). Comparing the error metrics with a larger number of samples (Table~\ref{TAB:methodvector}), the kernel-based sampling is very accurate, even when we use only 1K samples (i.e., the~$6\%$ of the input points). In fact, the~$97.1\%$ of the points have a reconstruction error lower than 0.10. Selecting 4K samples, the reconstruction error is lower than 0.02.

\textbf{Comparison with respect to previous work\label{sec:PW-COMPARISON}}
For the comparison of the proposed approach with previous work, we notice that our main goals (i.e., meshless interpolation of scalar/vector values, meshless HHD of arbitrary 3D fields) are still an open problem in Computer Graphics and Visualisation, which have been focused mainly on 2D vector fields. To this end, we compare its main properties with respect to~\cite{PETRONETTO2010,RIBEIRO2016}, which have been applied only to 2D SPH flows. According to Table~\ref{tab:COMPARISON}, the proposed meshless decomposition has a higher approximation accuracy for the computation of the potential of the conservative and irrotational components. Considering an irregularly sampled vector field (Fig.~\ref{FIG:vectorvaluedsamples-NEW}) and comparing the error metrics with a larger number of samples (Table~\ref{TAB:methodvector-NEW}), the kernel-based sampling is very accurate, even when we use only 150 samples (i.e., the~$10\%$ of the input points). In fact, the~$96.1\%$ of the points have a reconstruction error lower than 0.10. Selecting 700 samples, the reconstruction error is lower than 0.07. In Table~\ref{TAB:methodvector-NEW}, we report the execution time and the number of iterations with respect the number of selected centres.
\begin{table}[t]
\caption{For the analytic fields in Fig.~\ref{fig:ANALYTIC-STATISTICS},~\ref{fig:ROTOR-EXAMPLE},~\ref{FIG:vectorvaluedsamples}, we compare the proposed approach with~\cite{PETRONETTO2010}, where~$\ell_{\infty}$ error ($\ell_{\infty}$-$u$) between the ground-truth~$u$ and the computed meshless potential of the conservative component (i.e., \mbox{$\max_{i}\vert u(\mathbf{p}_{i})-\tilde{u}(\mathbf{p}_{i})\vert$}). We measure the angle between (i) the ground-truth and the meshless conservative fields, and (ii) the ground-truth~$\mathbf{w}$ and meshless~$\tilde{\mathbf{w}}$ irrotational component (i.e., \mbox{$\max_{i}\angle ((\nabla\wedge \mathbf{w})(\mathbf{p}_{i}),(\nabla\wedge \tilde{\mathbf{w}})(\mathbf{p}_{i})$}).\label{tab:COMPARISON}}
\centering
\begin{tabular}{|l||l|l|l|}
\hline
\textbf{Tests} &\multicolumn{3}{|c|}{\textbf{Our method}}\\
\hline
\textbf{Ex.}										&$\ell_{\infty}$-$u$			&$\angle\nabla u$	&\textbf{$\angle\nabla\wedge\mathbf{w}$}\\
\hline
Fig.~\ref{fig:ANALYTIC-STATISTICS}(a)				&$1.2\times 10^{-6}$			&$1.2^{\circ}$ 		&$3.3^{\circ}$\\			
Fig.~\ref{fig:ANALYTIC-STATISTICS}(b)				&$2.5\times 10^{-7}$			&$2.4^{\circ}$ 		&$4.6^{\circ}$\\
Fig.~\ref{fig:ROTOR-EXAMPLE}						&$1.2\times 10^{-6}$			&$1.9^{\circ}$		&$2.1^{\circ}$\\
Fig.~\ref{FIG:vectorvaluedsamples}					&$6.2\times 10^{-8}$			&$3.2^{\circ}$		&$4.1^{\circ}$\\
\hline\hline
&\multicolumn{3}{|c|}{\textbf{Previous work: ~\cite{PETRONETTO2010}}}\\
\hline
													&$\ell_{\infty}$-$u$			&$\angle\nabla u$	&\textbf{$\angle\nabla\wedge\mathbf{w}$}\\
\hline
Fig.~\ref{fig:ANALYTIC-STATISTICS}(a)				&$5.2\times 10^{-3}$			&$5.1^{\circ}$ 		&$7.3^{\circ}$\\			
Fig.~\ref{fig:ANALYTIC-STATISTICS}(b)				&$3.5\times 10^{-2}$			&$8.5^{\circ}$ 		&$7.5^{\circ}$\\
Fig.~\ref{fig:ROTOR-EXAMPLE}						&$1.2\times 10^{-3}$			&$7.6^{\circ}$		&$5.2^{\circ}$\\
Fig.~\ref{FIG:vectorvaluedsamples}					&$4.2\times 10^{-3}$			&$10.8^{\circ}$		&$4.6^{\circ}$\\
\hline
\end{tabular}
\end{table}
\subsection{Properties of the decomposition\label{sec:HH-NUMERICAL-STABILITY}}
%
\textbf{Unicity and exactness}
The proposed approach implements the natural HHD~\cite{BHATIA2014,BHATIA2014-HODGE}, without imposing additional boundary conditions to guarantee a unique decomposition. However, the approximations (\ref{eq:SCALAR-POTENTIAL}), (\ref{eq:VECTOR-POTENTIAL}) of the potential of the conservative and irrotational components are uniquely defined in terms of the RBFs. In fact, the coefficients of their representations solve the corresponding least-squares systems in Eqs.~(\ref{eq:DIV-NORMAL-EQUATION}), (\ref{eq:ROT-NORM-EQUATION}). Since our meshless HHD is based on the evaluation of differential operators in the continuous setting, the residual divergence and the residual rotor are null, i.e., the relations \mbox{$\nabla\wedge \nabla u=\mathbf{0}$}, \mbox{$\nabla\cdot\nabla\wedge\mathbf{w}=0$}, \mbox{$\nabla\cdot\nabla f=\Delta f$} apply in an exact way. On the contrary, these relations apply in an approximate way for~\cite{PETRONETTO2010,RIBEIRO2016}, as differential operators are discretised as finite differences, and are affected by a residual divergence and rotor.
\begin{figure}[t]
\centering
\begin{tabular}{cc}
$\Omega:=[-2,2]^{2}$ &$u(x,y):=x\exp\left[-(x^{2}+y^{2})\right]$\\
\includegraphics[height=75pt]{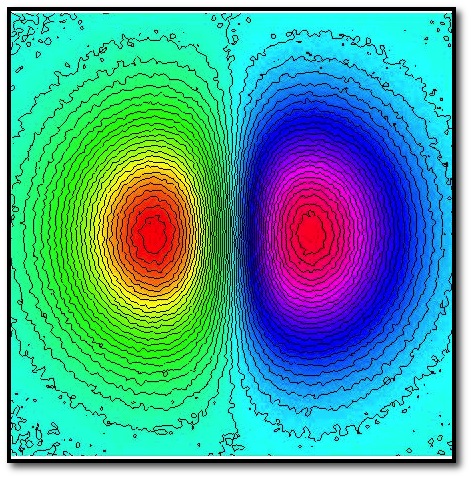}
&\includegraphics[height=75pt]{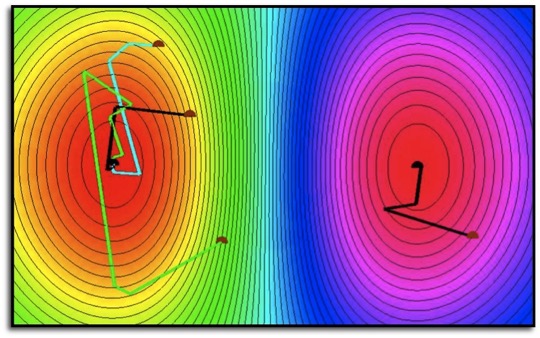}
\end{tabular}
\begin{tabular}{ccc}
\multicolumn{3}{c}{$\Omega=[-2\pi,2\pi]^{2},\,
u(x,y):=\sin x\cos y,\,
\mathcal{A}:=(\pi i,\pi j)_{i,j\in\mathbb{Z}}$}\\
\includegraphics[height=68pt]{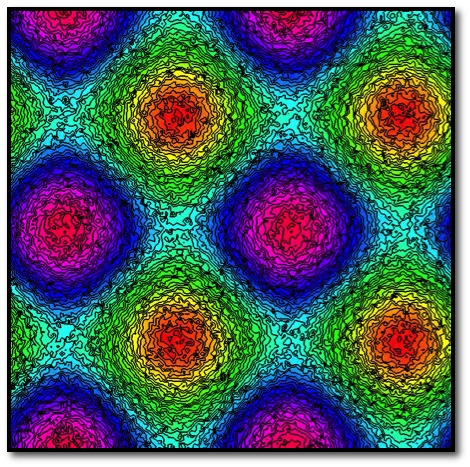}
&\includegraphics[height=68pt]{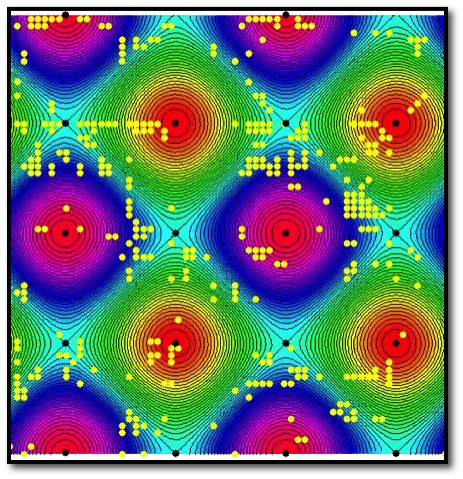}
&\includegraphics[height=68pt]{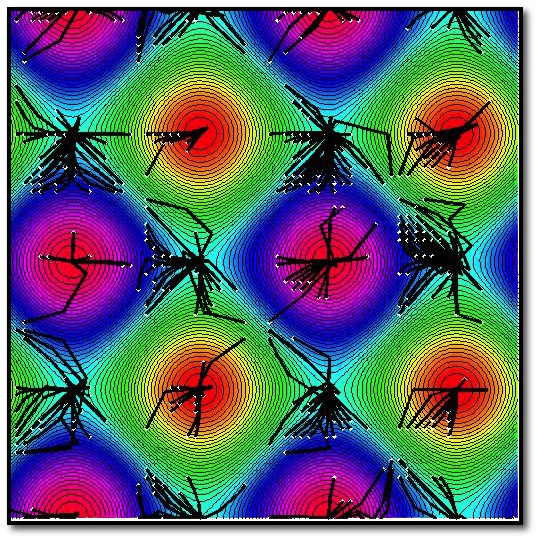}
\end{tabular}
\caption{(Left) Noisy potential, (middle) meshless approximation, (right) critical points (black dots) with paths computed by the iterative scheme from random guesses (yellow dots).\label{fig:ALL-CP-NOISE}}
\end{figure}

\textbf{Approximation of derivatives}
Let~$\phi$ be the Gaussian or multi-quadratic kernel and let us assume that~$\phi$ is conditionally positive of order~$m$. Let~$\Omega$ be a bounded set of~$\mathbb{R}^{d}$ that satisfies the interior cone condition and let~$u$ be the interpolant of~$f$ on~$\mathcal{P}$ with respect to the RBFs centred at~$\mathcal{P}$. Then, for any \mbox{$l\in\mathbb{N}$} with \mbox{$l\geq\max\{\vert\alpha\vert,m-1\}$}, there exist constants \mbox{$h_{0}(l)$}, \mbox{$f_{l}>0$} such that \mbox{$\vert D^{\alpha} f-D^{\alpha}u\vert\leq f_{l}h_{\Omega} \vert f\vert$}, with \mbox{$h_{\Omega}<h_{0}(l)$}. According to this last relation, we accurately approximate the derivatives of a given function through the derivatives the RBFs of its approximation. For quantitative error bounds, we refer the reader to~\cite{NARCOWICH2006,WENDLAND2005}.


\textbf{Numerical stability}
Let \mbox{$u_{\mathbf{e}}(\mathbf{p})=\sum_{i=1}^{k}\alpha_{\mathbf{e},i}\phi_{i}(\mathbf{p})$} be the potential of the perturbed vector field \mbox{$\tilde{\mathbf{v}}:=\mathbf{v}+\mathbf{e}$}. Then, the variation on the corresponding gradient is estimated as
\begin{equation}\label{eq:GARD-UPPER-BOUND}
\begin{split}
&\|\nabla u-\nabla u_{\mathbf{e}}\|_{2}
=\|\sum_{i=1}^{k}(\alpha_{i}-\alpha_{\mathbf{e},i})\nabla\phi_{i}\|_{2}\\
&\leq\sqrt{k}\max_{i=1,\ldots,k}\{\|\nabla\phi_{i}\|_{2}\}\|\alpha-\alpha_{\mathbf{e}}\|_{2},\quad
\alpha_{\mathbf{e}}=(\alpha_{\mathbf{e},i})_{i=1}^{k},\\
&\leq_{\textrm{Eq.~(\ref{eq:DIV-NORMAL-EQUATION})}} \sqrt{k}\|\phi^{\prime}\|_{\infty}\|(\Phi^{\top}\Phi)^{-1}\Phi^{\top})\|_{2}\|\mathbf{e}\|_{2}\\
\end{split}
\end{equation}
Indeed, the stability of the meshless computation of the potential of the conservative component is controlled by the maximum variation the derivative of the generating kernel and the inverse of the minimum eigenvalue of the least-squares matrix. Let~$\mathbf{w}_{\mathbf{e}}$ be the potential of~$\tilde{\mathbf{v}}$. Recalling that~$\alpha$ satisfies Eq.~(\ref{eq:ROT-NORM-EQUATION}) and analogously to Eq.~(\ref{eq:GARD-UPPER-BOUND}), we get that
\begin{equation}\label{eq:VECTOR-UPPER-BOUND}
\|\nabla\wedge\mathbf{w}-\nabla\wedge\mathbf{w}_{\mathbf{e}}\|_{2}\leq\sqrt{6k}\|\phi^{\prime}\|_{\infty}\lambda_{\min}^{-1}(\tilde{\mathbf{A}})\|\mathbf{e}\|_{2},
\end{equation}
with \mbox{$\tilde{\mathbf{A}}:=(\mathbf{A}^{\top}\mathbf{A})^{-1}\mathbf{A}$}. Our experiments confirm the robustness of the meshless approximation and HHD to sampling, a good independence of the selection of the shape parameters and kernels. An ill-conditioned coefficient matrix is associated with almost coincident points or a badly scaled coefficient matrix. Indeed, it is useful to check and remove almost coincident points from the input data set and select the shape parameter of the RBFs through optimised criteria, such as ``trial and error'' procedure, or an adaptive leave-one-out cross-validation~\cite{RIPPA1999}, or optimality constraints~\cite{DAVYDOV2011} with respect to the selected (e.g., Gaussian) RBFs. For a fixed number of centres, a smaller shape parameter generally produces a more accurate approximation, but is associated with a poorly conditioned coefficient matrix. For a fixed shape parameter, the conditioning number also grows with the number of centres. The upper bounds in Eqs.~(\ref{eq:GARD-UPPER-BOUND}) and (\ref{eq:VECTOR-UPPER-BOUND}) highlight the case when numerical instabilities might happen. In these cases, which we have not encountered in our tests, it is generally enough to regularise the linear systems in Eq.~(\ref{eq:ROT-NORM-EQUATION}) and~(\ref{eq:DIV-FREE}) by adding the term \mbox{$\epsilon\mathbf{I}$} to the coefficient matrix, with \mbox{$\epsilon\rightarrow 0$} and~$\mathbf{I}$ identity matrix. 

\textbf{Memory footprint of the meshless decomposition}
While the input vector field is stored as a matrix of doubles, whose dimension is equal to the number of input nodes/vertices, the meshless potentials are represented as a set of coefficients and corresponding centres, which are the nodes of a grid with a lower resolution. Indeed, the meshless HHD (Sect.~\ref{sec:MESHLESS-HH}) allows us to achieve a strong reduction of the memory footprint of the input data (Figs.~\ref{FIG:vectorvalued},~\ref{FIG:vectorvaluedsamples-NEW},~\ref{FIG:vectorvaluedsamples}, Tables~\ref{TAB:methodvector},~\ref{TAB:methodvector-NEW},~\ref{table:TAB-SPACE-SAVE}) and to decouple the representation of the potential from the discretisation of the input domain. This last aspect is important to distinguish the complexity of the domain geometry from the complexity of the vector field.
\begin{table}[t]
\centering
\caption{With reference to Fig.~\ref{FIG:vectorvaluedsamples-NEW}, we report (i) the error metrics when varying the number of samples, the reduction of the execution time and the variation of the number of iterations for the kernel-based sampling with a larger number of samples, with respect to the execution time~$T=58$ sec. with 700 samples.\label{TAB:methodvector-NEW}}
\begin{tabular}{|l|cccc|}
\hline
\textbf{Samples} & 700 & 500 & 300 & 150   \\ \hline
\textbf{Objective function} & \textbf{1.49} & 1.81 & 4.6 & 8.3  \\
\textbf{NC} & \textbf{0.997} & 0.996 & 0.983 & 0.959  \\
\textbf{NRMSE} & \textbf{0.075} & 0.152 & 0.181 & 0.279  \\
\textbf{$P_{0.05}$} & \textbf{99.9\%} & 98.7\% & 98.6\% & 96.1\%  \\ 
\textbf{$P_{0.10}$} & \textbf{100\%} & 99.7\% & 99.6\% & 99.1\% \\
\hline
\textbf{Execution Time} [s] &~$T$  &~$0.4 T$ &~$0.23 T$ &~$0.22 T$  \\
\textbf{Iterations} & 1634 & 842 & 631 & 1236  \\
\hline
\end{tabular}
\end{table}
\section{Conclusions and future work\label{sec:FUTURE-WORK}}
This paper has addressed the approximation and analysis of an arbitrary vector field through a meshless representation of the HHD with RBFs. To the best of our knowledge, we introduce the first work that addresses the meshless computation of the HHD of~$n$D instead of 2D vector fields, which is based entirely on a continuous approach. This new HHD framework for meshless vector fields is also aimed to set-up the foundation for other tasks for the analysis of meshless vector fields, such as the detection of critical points, topology construction, and the analysis of time-depending vector fields. For the meshless classification of the critical points of the potential \mbox{$u:\Omega\rightarrow\mathbb{R}$}~\cite{SKRABA2016}, we solve the equation \mbox{$\nabla u(\mathbf{p})=\mathbf{0}$} in Eq.~(\ref{eq:IMPLICIT-APPROX-GRADIENT}a) through a trust-region iterative solver, initialised as \mbox{$\mathbf{p}_{j_{1}}:=\mathbf{p}^{\star}$}. Then (Fig.~\ref{fig:ALL-CP-NOISE}, Table~\ref{tab:ITER-ERROR}), a critical point is classified according to the sign of the eigenvalues of the Hessian matrix of~$u$ (c.f., Eq.~(\ref{eq:IMPLICIT-HESSIAN-MATRIX-SINGLE})). For the potential~$\mathbf{w}$ of the solenoidal component, we can apply a similar procedure to the symmetric Jacobian matrix \mbox{$\mathbf{J}=\nabla\wedge\mathbf{w}$} (for a~$\mathcal{C}^{2}$ kernel). Each RBF has at most one critical point at its centre and the critical points of the potential are determined uniquely by its meshless representation as a linear combination of the RBFs, and not by the single RBF. Even though the order of convergence of \mbox{$\nabla u(\mathbf{p})$} to zero is high, in case of critical points of higher order we might experience a larger approximation error around the critical points. Indeed, the classification of the critical points of the conservative potential and of the singularities of the solenoidal component will be addressed in future work.
\begin{table}[t]
\caption{Reduction of the memory footprint~$\%$ ($p:=k/n$) of the input data with the meshless approximation, in terms of the original grid size \mbox{$n\times n$} and the number~$k$ of RBFs in the meshless approximation. The value~$\epsilon_{\infty}$ is the~$\ell_{\infty}$ error between the input field and its meshless approximation, evaluated at the grid nodes.\label{table:TAB-SPACE-SAVE}}
\centering
\small{
\begin{tabular}{|l|l||l|l|}
\hline
\textbf{2D Tests}								&$n=256^{2}$								&\textbf{3D Tests}						&$n=128^{3}$\\
$k=64^{2}$										&$p=0.6\%$									&$k=11K$								&$p=0.5\%$\\				
\hline
Fig.~\ref{fig:eFUN}								&$1.2\times 10^{-9}$						&Fig.~\ref{fig:HH-FLOW}					&$7.3\times 10^{-8}$\\
\hline	
Fig.~\ref{fig:WIND13}							&$2.5\times 10^{-10}$						&Fig.~\ref{fig:WIND}					&$2.1\times 10^{-8}$\\
\hline
Fig.~\ref{fig:WIND2D}							&$1.9\times 10^{-9}$						&										&\\
\hline					
Fig.~\ref{fig:WIND2D-SECOND}					&$2.9\times 10^{-11}$						&										&\\
\hline
\end{tabular}}
\end{table}
Finally, the computation of the optimal centres with respect to the target accuracy and computational resources can be achieved by minimising the least-squares energy in Eqs.~(\ref{eq:DIV-NORMAL-EQUATION}), (\ref{eq:ROT-NORM-EQUATION}) with respect to the unknown coefficients and to the coordinates of the centres. The minimum of the normal equation is computed through iterative solvers of non-linear systems or the iterative optimisation method L-BFGS (\emph{Limited-memory Broyden, Fletcher, Goldfarb, Shanno})~\cite{ZHU1997}.


\textbf{Acknowledgements}
We thank the Reviewers for their thorough review and constructive comments, which helped us to improve the technical part and presentation of the paper, and Dr. Simone Cammarasana (CNR-IMATI) for his tests on the kernel-based sampling. This work is partially supported by the ERC Advanced Grant CHANGE, Nr. 694515.
\begin{table}[t]
\caption{'\emph{It}' iterations, '\emph{Ev}' function evaluations, convergence of \mbox{$\|\nabla u(\mathbf{p})\|_{2}$} to zero of the iterative trust-region method (Fig.~\ref{fig:ALL-CP-NOISE}).\label{tab:ITER-ERROR}}
\centering
\begin{tabular}{|l|l|l||l|l|l|}
\hline
It.	&Ev.	&$\|\nabla u(\mathbf{p})\|_{2}$	&It.	&Ev.		&$\|\nabla u(\mathbf{p})\|_{2}$\\
\hline 
 	~$0$          &$4$      &$0.0494952$ 			&$0$         &$4$         &$0.172086$\\             
    ~$1$          &$8$      &$0.0400076$    		&$1$         &$8$        &$0.015144$\\            
    ~$2$         &$12$      &$0.0231691$      		&$2$         &$12$       &$0.002548$\\                
    ~$3$         &$16$      &$0.0005562$   			&$3$         &$16$       &$5.0169\,\,e-05$\\                 %
	.			&.			&.						&.			 &.			&.\\	
   ~$28$        &$116$      &$1.1088\,\,e-16$    		&$18$         &$76$     &$4.7386\,\,e-17$\\      
\hline
\end{tabular}
\end{table}  
%

%
\begin{IEEEbiography}
[{\includegraphics[width=1in,height=1.10in,clip,keepaspectratio]{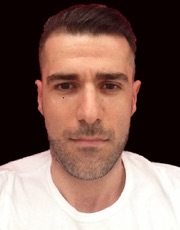}}]{Giuseppe Patan\`e}
Giuseppe Patan\`e is senior researcher at CNR-IMATI. Since 2001, his research is focused on Computer Graphics and Applied Mathematics. He obtained the National Scientific Qualification as Full Professor of Computer Science. He is author of journal and conference papers in the field of Computer Graphics and Visualisation. He is  tutor of Ph.D. students and Post.Docs and responsible for R$\&$D activities in national and EU projects.
\end{IEEEbiography}
\end{document}